**The Lifecycle of Hollows on Mercury: An Evaluation of Candidate Volatile Phases and a Novel Model of Formation.**

M. S. Phillips[1], J. E. Moersch[1], C. E. Viviano[2], J. P. Emery[3]
[1]Department of Earth and Planetary Sciences, University of Tennessee, Knoxville
[2]Planetary Exploration Group, Johns Hopkins University Applied Physics Laboratory
[3]Department of Astronomy and Planetary Sciences, Northern Arizona University
Corresponding author: Michael Phillips (mphill58@vols.utk.edu)

## Abstract

On Mercury, high-reflectance, flat-floored depressions called hollows are observed nearly globally within low-reflectance material, one of Mercury's major color units. Hollows are thought to be young, or even currently active, features that form via sublimation, or a "sublimation-like" process. The apparent abundance of sulfides within LRM combined with spectral detections of sulfides associated with hollows suggests that sulfides may be the phase responsible for hollow formation. Despite the association of sulfides with hollows, it is still not clear whether sulfides are the hollow-forming phase. To better understand which phase(s) might be responsible for hollow formation, we calculated sublimation rates for 57 candidate hollow-forming volatile phases from the surface of Mercury and as a function of depth beneath regolith lag deposits of various thicknesses. We found that stearic acid ($C_{18}H_{36}O_2$), fullerenes ($C_{60}$, $C_{70}$), and elemental sulfur (S) have the appropriate thermophysical properties to explain hollow formation. Stearic acid and fullerenes are implausible hollow-forming phases because they are unlikely to have been delivered to or generated on Mercury in high enough volume to account for hollows. We suggest that S is most likely the phase responsible for hollow formation based on its abundance on Mercury and its thermophysical properties. We discuss the possibility that S is the phase responsible for hollow formation within the hollow-formation model framework proposed by Blewett et al. (2013). However, several potential limitations with that model lead us to suggest an alternative hollow-formation model: a subsurface heat source (most often impact-induced) generates thermal systems that drive sulfur-rich fumaroles in which S and other phases accumulate on and within the surface at night and sublimate during the day to create hollows. We call this hollow-formation model "Sublimation Cycling Around Fumarole Systems" (SCArFS). We suggest that thermal decomposition of sulfides within LRM is a main contributor to S and S-bearing gases within the proposed fumarole systems and that (re-)precipitation of sulfides may occur at the surface along hollow floors and rims.

## Highlights

- A thermophysical model was developed to test the viability of 57 candidate hollow-forming volatiles within the hollow-formation model framework proposed in Blewett et al. (2013).
- We find that the thermophysical properties of elemental sulfur (S) combined with the abundance of S on Mercury, make it the most likely hollow-forming volatile explored in our study.
- We propose a novel model for hollow formation in which a subsurface heat source drives sulfur-rich fumarole systems that deposit volatiles (importantly, S) in the near-surface at night within a "sulfur permafrost zone", and daytime solar heating drives sublimation to form hollows.

## 1. Introduction

Mercury hosts features unique in the solar system that were first observed in Mariner 10 images and described as bright crater floor deposits (BCFDs, Dzurisin, 1977). Images from instruments onboard the MErcury Surface, Space ENvironment, GEochemistry, and Ranging (MESSENGER, Solomon et al., 2007) spacecraft have revealed that many BCFDs host sub-kilometer-scale, flat-floored, rimless depressions with a typical depth of around 20-50 m (Fig. 1) and lateral extents that range from meters to kms (Blewett et al., 2011; Blewett et al., 2013; Thomas et al., 2014). These features have become known as hollows to distinguish them from structures such as pyroclastic vents and pits (Head et al., 2009; Kerber et al., 2011; Goudge et al., 2014).

### 1.1. Distribution of Hollows on Mercury and Their Association with Mercury's Low Reflectance Material

Hollows are nearly global in extent, though some spatial variability has been noted (Fig. 2). Thomas et al. (2014) showed that hollows are more abundant and cover more area at equatorial latitudes and on slopes with equator-facing aspects than at higher latitudes and on slopes that face away from the equator. This result was in agreement with Blewett et al. (2011, 2013), who first noted some examples of hollows at mid- to high-latitudes on equator-facing slopes. Blewett et al. (2016) measured the depth of hollows using shadow length measurements on 565 high-resolution images (< 20 m/pixel) of Mercury's northern hemisphere and found that hollows have a mean depth of 24 m, a minimum depth of 1.6 m and a maximum of 110.5 m. These depth measurements indicate a lower average depth than those from Thomas et al. (2014), but the uncertainty is smaller because the image resolution was much higher (< 20 m/pixel versus < 180 m/pixel). We used the data collected by Blewett et al. (2016) to show that hollow depth decreases as latitude increases (Fig. 2d). The apparent preference of hollows to cover more area at equatorial latitudes and to preferentially form on slopes with equator-facing aspects at mid- to high-latitudes has led previous workers to suggest that solar heating drives a sublimation process responsible for hollow formation (Blewett et al., 2011; Blewett et al., 2013; Thomas et al., 2014). The trend of decreasing hollow depth with increasing latitude further supports this suggestion (Fig. 2d).

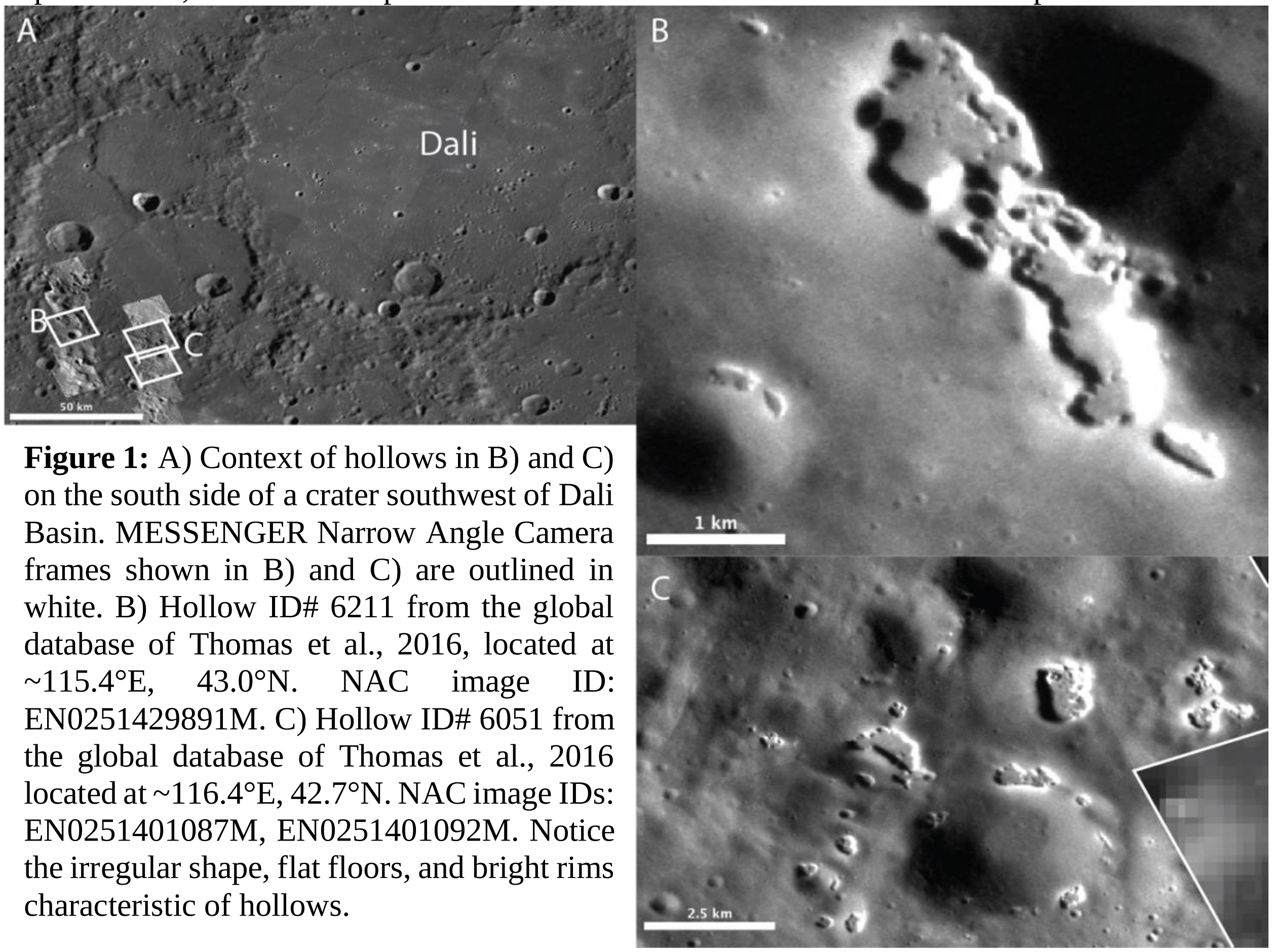

**Figure 1:** A) Context of hollows in B) and C) on the south side of a crater southwest of Dali Basin. MESSENGER Narrow Angle Camera frames shown in B) and C) are outlined in white. B) Hollow ID# 6211 from the global database of Thomas et al., 2016, located at ~115.4°E, 43.0°N. NAC image ID: EN0251429891M. C) Hollow ID# 6051 from the global database of Thomas et al., 2016 located at ~116.4°E, 42.7°N. NAC image IDs: EN0251401087M, EN0251401092M. Notice the irregular shape, flat floors, and bright rims characteristic of hollows.

Mercury's surface has an extreme thermal environment due to the planet's high orbital eccentricity (0.206) and proximity to the Sun (0.308-0.467 AU). Mercury is also in a 3:2 spin-orbit resonance, which means it has 3 sidereal days (and 1 synodic day) every 2 hermean years (Colombo, 1965; Pettengill and Dyce, 1965). The combined effect of the 3:2 resonance and high eccentricity is that (i) two meridians (0°E and 180°E) experience solar noon at alternate perihelia and receive maximum solar insolation ("hot meridians"), whereas (ii) two meridians (90°E and 270°E) experience solar noon at alternate aphelia and receive minimum solar insolation ("warm meridians"). Because Mercury also has an obliquity near zero, the intersections of the hot meridians with the equator are known as Mercury's "hot poles" and the intersections of the warm meridians with the equator are Mercury's "warm poles" (e.g., Soter and Ulrichs, 1967). An additional oddity resulting from Mercury's orbit is that for a portion of its orbit surrounding perihelion the orbital angular velocity exceeds the rotational angular velocity. Thus, for an observer on Mercury, the apparent motion of the sun across the sky is retrograde. This effectively lengthens noon-time heating for the hot poles, because the sun passes through its apex three times, and causes a double sunset/sunrise at the warm poles that imparts a signature on the diurnal temperature curve (Fig. S1). Mercury has most likely been in its present-day 3:2 spin-orbit resonance since near the end of the Late Heavy Bombardment ~3.8 Ga (Knibbe and van Westrenen, 2017). Prior to capture into its present-day resonance, Mercury may have been in either a 1:1 or 2:1 spin-orbit resonance (Correia and Laskar, 2009; Correia and Laskar, 2010; Wieczorek et al., 2012; Knibbe and van Westrenen, 2017). Because hollows are generally considered the youngest features on Mercury (and potentially currently active), the ancestral resonance is not likely to have affected hollow formation, but it may have implications for the sequestration of a hypothesized hollow-sourcing volatile layer between ~4.6 Ga and ~3.8 Ga (Wieczorek et al., 2012; Knibbe and van Westrenen, 2017), as will be discussed in more detail in *Section 4.1.1*.

Despite Mercury's longitudinal asymmetry in insolation and the apparent dependence of hollow formation on thermal environment, there does not appear to be a clear correlation between Mercury's hot meridians and hollow occurrence. Thomas et al. (2014) report a lower areal extent of hollows near the cold poles between 30°S and 30°N and an uptick in areal extent of hollows near the 0°E hot pole in this same latitudinal band, indicating that solar insolation may impart some control on hollow formation. This has led previous authors to suggest that equatorial temperatures on Mercury are sufficient to create hollows at all longitudes (Thomas et al., 2014). The higher concentration of hollows around Mercury's 180°E hot meridian at latitudes above 50°N than at other longitudes can be explained by the distribution of the northern smooth plains in this region, which are anticorrelated with hollow distribution (Fig. 2a,b) as suggested by (Blewett et al., 2013).

Hollows are not clearly correlated with Mercury's hot poles, but they are also not uniformly distributed across all longitudes. An anomalously high concentration of hollows between longitudes 300°E and 320°E (Blewett et al., 2013; Thomas et al., 2014), could be considered to extend between ~280°E and 340°E (Fig. 2b,c; Blewett et al., 2013). Possible causes for this nonuniform longitudinal distribution of hollows are formation-age differences and lateral differences in surface material composition. If there are processes erasing hollows, such as impact gardening and space weathering, then areas with the most hollows may host the youngest hollows; however, a preliminary investigation of hollow reflectance across Mercury as a proxy for surface exposure age showed no geographic trend in reflectance, potentially indicating that hollows do not differ in age across Mercury (Phillips et al., 2016). The distribution of hollows may also depend on Mercury's low-reflectance material (LRM), which is one of Mercury's major color units (Robinson et al., 2008; Denevi et al., 2009) and assumed by association to be the source of hollow-forming volatiles (Blewett et al., 2013; Thomas et al., 2014). There is a high concentration of LRM between ~280°E and 360°E in the northern hemisphere (Fig. 2c). The coexistence of hollows and the LRM in this region may indicate that surface composition controls the longitudinal distribution of hollows (Blewett et al., 2013).

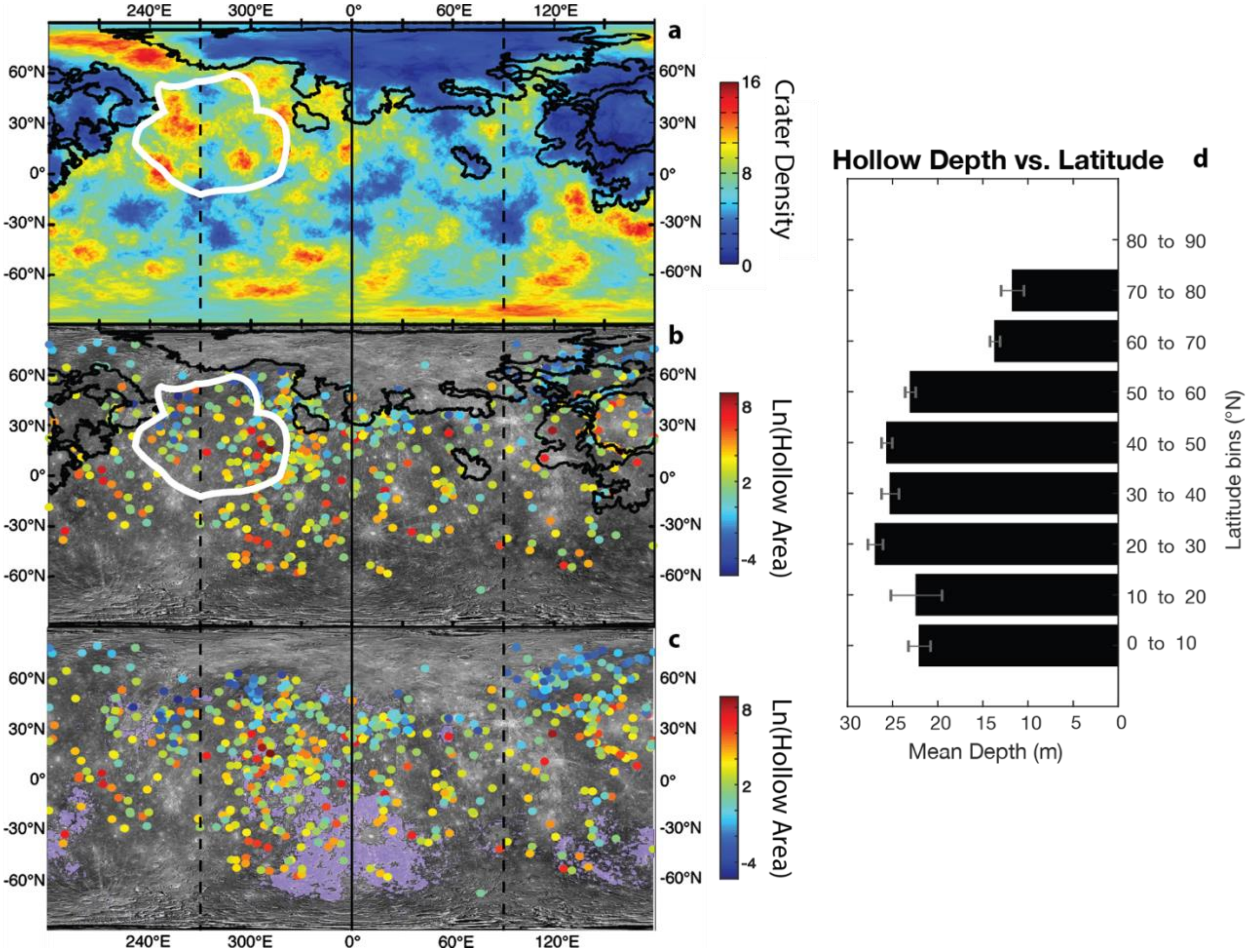

**Figure 2.** Maps of **a)** density of impact craters > 25 km diameter per $10^5$ km. Data from Marchi et al. (2013), map reproduced from Weider et al. (2015); **b)** distribution of hollows relative to the northern smooth plains (solid black line) and high-Mg region (solid white line); **c)** distribution of hollows relative to the low-reflectance material (LRM). LRM is highlighted in lavender. It should be noted that many small patches of LRM are covered by dots indicating hollow locations. In **b)** & **c)**, the areas ($km^2$) of individual hollows/hollow clusters are represented by color on a natural log scale. Hollow location and area data are from Thomas et al. (2014, 2016). Base map is MESSENGER MDIS LOI mosaic (astrogeology.usgs.gov); Vertical dashed lines are the Warm Meridians and the vertical solid line is the Hot Meridian (the other Hot Meridian is at the map edges). **d)** Hollow mean depth binned by 10° latitudinal bins in the northern hemisphere. Error bars are the standard error for hollow depth in each latitudinal bin. Hollow depth data are taken from Blewett et al. (2016).

The hollow-forming volatile is suggested to be sourced from the LRM because hollows almost exclusively form in this unit (96%, Thomas et al., 2014; Thomas et al., 2016). The distributions of hollows and the LRM are shown in Figure 2c. Murchie et al. (2015) explored several hypotheses for the nature and origin of the LRM and concluded that it is a stratigraphically deep component of Mercury's crust (estimated to be ~30 km below the surface, Denevi et al., 2009; Rivera-Valentin and Barr, 2014; Ernst et al., 2015). The LRM is exhumed and emplaced on the surface via impact cratering (Klima et al., 2018). The darkening agent within LRM is likely endogenic to Mercury and may be some form of carbon (Riner et al., 2009; Nittler et al., 2011; Xiao et al., 2013; Murchie et al., 2015; Peplowski et al., 2015; Trang et al., 2017). The hypothesis that carbon is the darkening agent is consistent with elevated carbon content in LRM and the idea that Mercury had an early carbon-bearing crust (Peplowski et al., 2016), such as a primary graphite flotation crust (Vander Kaaden and McCubbin, 2015). Spectral modeling of MESSENGER Visible and Infrared Spectrograph (VIRS) data by Trang et al. (2017) supports the idea that carbon is a plausible darkening agent for the LRM.

Additionally, Weider et al. (2015) identified a region with MESSENGER's X-Ray Spectrometer in the northern hemisphere between ~230°E and 320°E that is anomalously high in Mg called the high magnesium region (HMR; Fig. 2a & b). The HMR was proposed by Weider et al. (2015) to be either an ancient impact basin or related to large degrees of partial melting in the mantle. The HMR is also high in Ca and S, and as such is thought to be rich in (Ca, Mg) sulfides (e.g., Zolotov et al., 2013). Magma ocean models by Boukaré et al. (2019) suggest that Mercury's magma ocean may have hosted sulfide layers that, depending on the magma ocean composition, could have floated to form a sulfide crust or risen through the crust as sulfide-rich plumes. The sulfur-rich terrains on Mercury are also correlated with the LRM (Nittler et al., 2011), and have caused some to suggest that sulfides may be a darkening agent within the LRM (Nittler et al., 2011; Weider et al., 2012; Blewett et al., 2013; Helbert et al., 2013).

### 1.2. Relationship to Craters

Approximately 97% of hollows (by surface area) are within impact craters or impact crater-related material such as rims and ejecta (Thomas et al., 2014). Though the close association of hollows with craters implies a genetic link between the two (Blewett et al., 2011, 2013), the precise role that craters play in hollow formation is unclear. Impacts may simply be the mechanism by which hollow-forming material is exhumed (Blewett et al., 2011), or the heat and/or melt generated by impacts may play an active role in hollow formation (Vaughan et al., 2012; Blewett et al., 2016).

The distribution of hollows within craters differs between simple and complex craters. In simple craters, hollows are almost exclusively found on or proximal to the rims (Thomas et al., 2014). However, in complex craters hollows are predominantly observed on terraced walls and central complexes, such as central peaks and peak rings (Thomas et al., 2014). There are also instances of hollows extensively covering the floors of some complex craters, such as Tyagaraja and Sander craters (Blewett et al., 2011). Depth measurements made by Blewett et al. (2016) show that hollows near/on the central complexes of these craters tend to be deeper than the hollows along the floors (Blewett et al. 2016, supplementary materials). Our preliminary visual inspection in JMARS of the depth measurements made by Blewett et al. 2016 within complex craters suggests it may be true more generally that within complex craters deeper hollows are found nearer the central complexes. However, more work is necessary to support this suggestion.

### 1.3 The Apparent Youth of Hollows

Several lines of evidence point to the apparent youth of hollows: cross-cutting relationships with rayed craters, “crisp” morphology and delicate structures, and lack of superposed craters (Blewett et al., 2011; Blewett et al., 2013; Blewett et al., 2018). Perhaps the most direct line of evidence in support of a young age for some hollows is that at least 6 rayed craters with ages estimated at ≤ ~270 Ma (Xiao et al., 2012) are host to hollows. Because hollows occur within the craters, hollows must be younger than the craters. Hollows also show little evidence of modification by impacts or topographic muting through space weathering, although some examples of morphologically muted and dark (i.e., “old”) hollows exist (Blewett et al., 2018). There is only one published example of a small (~30 m diameter) crater within a hollow (Wang et al., 2020). Additionally, the bright floors and halos of hollows would presumably not survive long periods of exposure to the hermean space weathering environment (Domingue et al., 2014), further indicating a young age for at least the bright surfaces of hollows, if not the hollows themselves. However, as noted by Blewett et al. (2018), there is likely an observational bias toward imaging bright (i.e., “young”) hollows. This is because hollows are generally small features and the brightness of ostensibly young hollows is what drew MESSENGER team members to target these features (Blewett et al., 2018).

### 1.4. Hollow Formation Models and Candidate Volatiles

Several models have been proposed to explain hollow formation. Blewett et al. (2011) first posited a sublimation origin for hollows, noting their resemblance to martian “swiss cheese” terrain formed by sublimation of polar carbon dioxide ice (Malin et al., 2001). Blewett et al. (2013), further developed this idea into a three-stage model for hollow formation (Fig. 3a): 1) hollow-forming volatiles are deposited on the cold, night-side of Mercury from volcanic eruptions and/or magmatic outgassing events; 2) hollow-forming volatiles are rapidly buried and sequestered by lava flows and pyroclastic deposits; and 3) hollow-forming materials are exhumed by impact craters and emplaced on the surface where they are unstable and sublimate to form hollows. Thomas et al. (2014) suggested that endogenic processes, such as heating by magma in the subsurface, may drive volatiles to the surface, but ultimately concluded that the most probable mechanism for the delivery of volatiles to the surface is by exhumation from an impact. In general, the model of Blewett et al. (2013) proposes that hollows initially grow downward and laterally, but downward progression halts once a sufficient lag is built-up or when the volatile layer is exhausted. Lateral growth progresses through the extent of the volatile-bearing unit because the steep sides of hollows do not allow for a lag to build in the lateral direction. Blewett et al. (2016) suggest that because hollows have a relatively constant depth that is shallow compared to the thickness of the LRM, the downward growth of hollows is limited by the build-up of a lag deposit rather than limited by the thickness of the volatile-bearing unit. Additionally, Thomas et al. (2014) noted that some hollows are hosted in small craters and on thrust faults nested within larger impact craters that have hollows. From this observation, the authors suggest that some hollow-forming volatile material remains in the subsurface after hollow formation, within the original larger crater, has ceased, further indicating that a lag deposit suppresses hollow formation. If a lag deposit halts downward growth, then the thickness of the lag deposit would be a function of the temperature-dependent sublimation rate of the hollow-forming volatile, the physical characteristics of the lag deposit (e.g., porosity and tortuosity), and the thermal insulating properties of the lag deposit.

Blewett et al. (2013) further proposed that sulfur-bearing volatiles ($H_2S$, $SO_2$), or sulfides (that decompose to produce elemental sulfur) are likely candidate hollow-forming materials because of the volatility of sulfur-bearing phases on Mercury and the high abundance of sulfur within the LRM (Nittler et al., 2011; Weider et al., 2011). The suggestion that sulfides play a role in hollow formation has also been made by several other workers (Vaughan et al., 2012; Helbert et al., 2013; Xiao et al., 2013; Thomas et al., 2014; Thomas et al., 2016; Vilas et al., 2016; Lucchetti et al., 2018; Pajola et al., 2021). Thomas et al. (2016) investigated the spectral reflectance characteristics of hollows, their bright halos, and the surrounding LRM and concluded that formation of hollows leaves behind a lag that is darker and bluer-sloped (i.e., relatively high reflectance values at shorter, bluer, wavelengths than at longer, redder, wavelengths in the UV/VIS) compared to the volatile material that is lost, which is brighter and redder-sloped than the parent material (i.e., LRM). The authors suggest that (Ca, Mg) sulfides are plausible candidate hollow-forming materials and that graphite may be the darkening agent of the LRM based on the relative spectral characteristics. A direct comparison of hollow spectra to sulfide spectra was made by Vilas et al. (2016). The authors showed that MESSENGER Mercury Dual Imaging System (MDIS) Wide-Angle Camera (WAC) (Hawkins et al., 2007) 8-band spectra of hollows from Dominici and Hopper craters are similar to spectra of MgS and, to a lesser extent, CaS samples measured in the lab. However, Lucchetti et al. (2018) suggest that sulfides alone do not explain the spectra of hollows and that pyroxenes could contribute to the spectra.

Blewett et al. (2016) offered another mechanism for hollow formation in which carbon-rich LRM starting materials are lost through ion sputtering or converted to methane (or other carbon-bearing volatile species) through ion bombardment. From this model, we infer that the subsequent vaporization of carbon or the carbon-bearing volatiles would create hollows. If ion bombardment is responsible for the conversion of carbon to a volatile gas, such as methane, or if ion sputtering causes direct loss of carbon, then hollows should occur anywhere there is carbon-bearing material on the surface and perhaps not nearly exclusively within craters. Additionally, the processes of ion sputtering and ion-bombardment act on the uppermost layer of the surface, so unless the carbon-bearing material is pure, one might expect this process to self-limit after a very thin lag develops. Impact gardening may allow for the processes to continue as fresh material is brought to the surface, but the rate of turnover due to impact gardening is slow relative to plausible hollow growth rates (e.g., lateral growth rate of at least 1 cm per 10 kyr was inferred by Blewett et al., 2016). Killen et al. (2007) estimate that 1 cm of regolith will turn over every 150 kyr with 50% probability, although new preliminary models for impact gardening rates on Mercury indicate that the rate might be much slower (e.g., Costello et al., 2019). Blewett et al. (2016) estimate the rate of lateral hollow growth at Balanchine crater to be at least 1 cm per 10 kyr. Because downward growth of hollows is inferred to happen first, and more quickly (Blewett et al., 2018), this lateral growth rate can be considered conservative. Therefore, it seems unlikely that impact gardening can replenish volatiles to the surface quickly enough to account for the rate of hollow formation, perhaps casting doubt on whether surficial space weathering processes can account for hollow formation. However, Blewett et al. (2016) also propose that carbon-rich starting material heated and mixed with solar-wind-saturated material, for example by impact melting, could also produce methane and other carbon-bearing volatile species. In this scenario, crater-related heating is necessary for hollow formation because it generates new volatile phases that were not already present on the surface or in the subsurface. These volatile phases would ostensibly be lost as they are created within or adjacent to impact melt at the surface and hollows would subsequently form through collapse of the void space. It is not clear how the latitudinal dependence on hollow depth and areal extent is explained in this model.

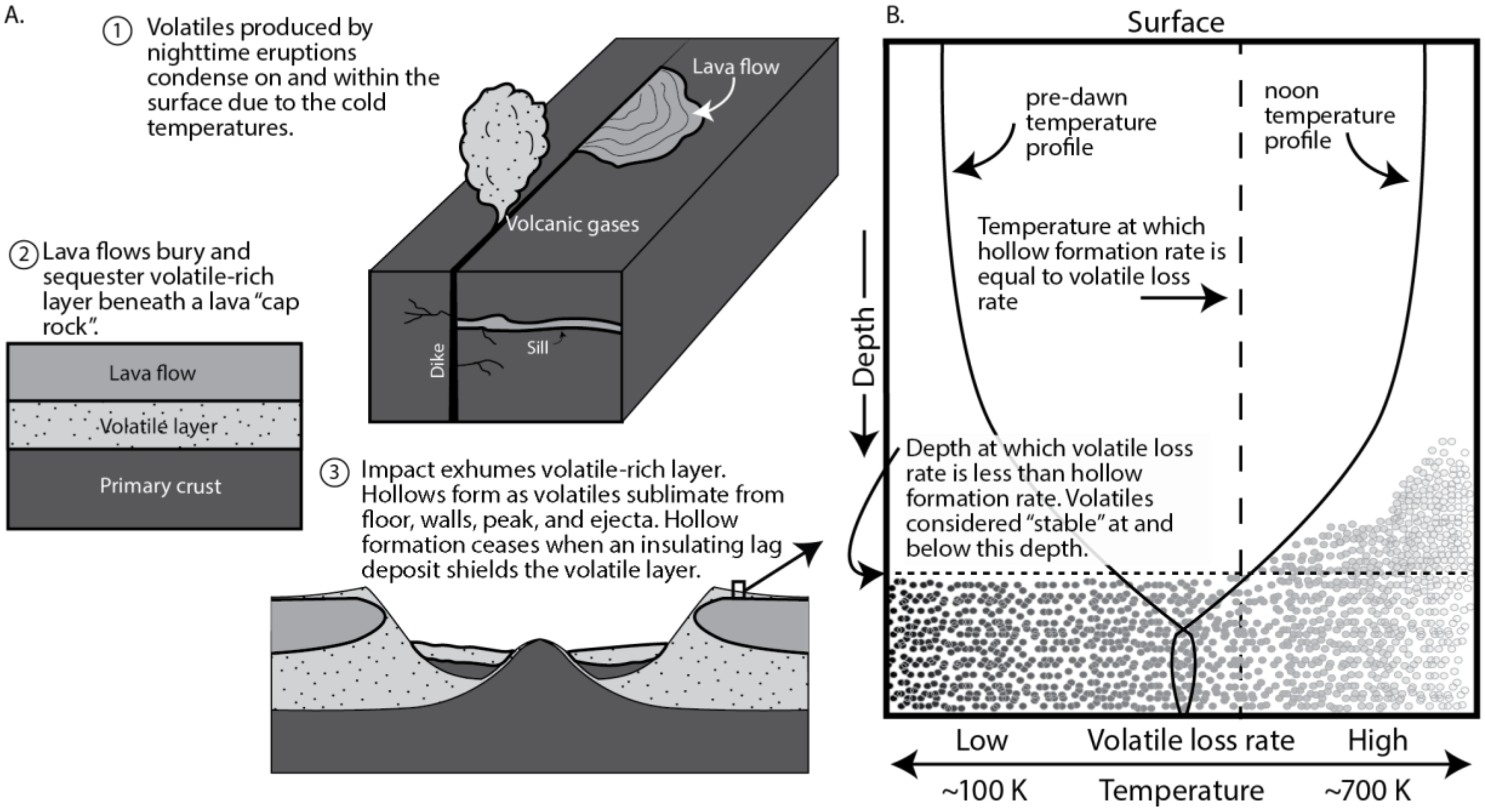

**Figure 3.** Conceptual model for hollow formation in a Surface Exposure And Lag Sequestration (SEALS) model (see text in *Section 1.4*). A) Three-stage hollow-formation model proposed in Blewett et al. (2013). The basic stages of this model are perhaps the most widely accepted of any hollow-formation model (see, e.g., Thomas et al., 2016). The mechanisms differ between models for how a volatile layer forms (e.g., volcanic sequestration, fractionation in volcanic flows, slag formation in impact melts and lavas, etc.), but the processes proposed by Blewett et al. (2011, 2013) are shown in part A (Stages 1 and 2 ) of this Figure. In Stage 3, and in all SEALS models, hollows form by loss of a volatile phase upon exposure to the surface via solar heating (or space weathering). Hollow formation ceases after a lag deposit develops and shields the volatile phase. B) A more detailed depiction of the process by which hollows form in a SEALS hollow formation model. The depth scale of the cartoon represents ~0.5 to 1 m of regolith. Dots represent a volatile phase and its loss rate at different temperatures: cold and stable (dark dots, left, low volatile loss rate) to hot and unstable (light dots, right, high volatile loss rate). If the loss rate for a given volatile is low at the equilibrium temperature but fast at daytime surface/near-surface temperatures (as the example volatile loss rate shown with the vertical dashed line), then such a volatile could be responsible for hollow formation within a SEALS model framework. However, if a compound's loss rate at subsurface temperatures is fast at or below equilibrium temperatures, then such a volatile would be unstable under a lag deposit and inconsistent with a hollow formation model that includes sequestration beneath a regolith lag deposit.

Vaughan et al., (2012) proposed a model in which hollows form via differentiation of impact melt. In this model, hollow-forming materials, sulfides or chlorides as suggested by the authors, rise to the top of impact melt where they form a "slag" layer susceptible to photodissociation and subsequent vaporization in a "sublimation-like" process. This model has some limitations. Photodissociation of sulfides is a surficial process, so a relatively pure and thick (≥ depth of the hollow) sulfide slag is necessary. Otherwise, hollow growth would be limited by development of a thin lag (such as in the case of ion bombardment and ion sputtering as described above). Additionally, hollows forming on steep slopes, on Caloris knobs (Wright et al., 2020), and proximal to the northern smooth plains unit (Denevi et al., 2013a; Thomas et al., 2014) are not easily explained by differentiated impact melt.

Helbert et al. (2013) provided a slightly different model from that of Vaughan et al. (2012), in which sulfides form as a slag product in Mg, Ca, and S-rich lavas instead of impact melts. Helbert et al. (2013) proposed thermal decomposition as the mechanism to dissociate sulfides rather than photodissociation from surficial space-weathering processes. The mechanism of thermal decomposition may play an important role in hollow formation if sulfides are the volatile substance that drives hollowing. This model, however, does not explain the 97% association of hollows with craters or the 96% association with the LRM (Thomas et al., 2014), because hollows would be expected to form in any sufficiently sulfide-rich lava.

The processes by which volatiles are delivered to or generated at the surface vary across hollow formation models, but most models in the literature have two aspects in common: 1) a volatile phase sublimates, or is otherwise decomposed via space weathering processes, when exposed at the surface, and 2) an insulating lag deposit is the mechanism by which downward hollow growth is halted. In this work, we will call models that have these characteristics Surface Exposure And Lag Sequestration (SEALS) models. The most complete version of a SEALS hollow-formation model is the three-stage model proposed in Blewett et al. (2013), and there appears to be a consensus in the literature that the basic framework of this model best explains hollow formation (see, e.g., Thomas et al., 2016).

The viability of all three stages of the Blewett et al. (2013) model (concentration and deposition of a volatile phase into a volatile layer, sequestration of that layer, and exhumation and sublimation of that layer to form hollows, see Fig. 3) depend on the composition of the hollow-forming phase. Although sulfides are the phase most commonly proposed as the hollow-forming volatile phase (Vaughan et al., 2012; Helbert et al., 2013; Xiao et al., 2013; Thomas et al., 2014; Thomas et al., 2016; Vilas et al., 2016), there has been no detailed exploration of how sulfides are expected to behave at hermean surface and subsurface temperatures. Additionally, no exhaustive exploration of the thermophysical properties of other possible hollow-forming volatile phases has been done. Identifying the likely phase responsible for hollow formation will allow for a more informed evaluation of the plausibility of the three-stage model of Blewett et al. (2013) and of SEALS models for hollow formation more generally.

### 1.5. Hypothesis

Our null hypothesis is that no volatile phase exists that is unstable at hermean surface/near-surface temperatures but is stable under a lag deposit. The alternative hypothesis is that one or more such volatile phases exist that are unstable at hermean surface/near-surface temperatures and stable under a lag deposit. We would reject our null hypothesis in favor of the alternative if a volatile phase can be found with the appropriate characteristics. For any volatile phases that show the appropriate thermophysical characteristics for hollow formation, we will then address their geological plausibility on Mercury. To address our null hypothesis, we solve the 1D time-dependent heat equation to calculate hermean diurnal temperature profiles across Mercury and then use these temperatures to calculate the loss rate of candidate volatiles at different depths using standard temperature-dependent sublimation and diffusion rate equations. In this way, we will narrow down the field of possible volatile phases involved in hollow formation, which will allow for a more informed evaluation of Stages 1 and 2 of the hollow-formation model of Blewett et al. (2013) and of the viability of SEALS models for hollow formation more generally. We describe the thermophysical model and discuss our choice of volatile phases and the calculation of volatile loss rates in *Section 2*. We report our results and discuss the implications in *Sections 3* and *4*, respectively. Additionally, we propose an alternative to a SEALS hollow-formation model in *Section 4.4*.

## 2. Method

For a host of possible hollow-forming phases, we will calculate volatile loss rates, hollow-formation timescales, and an expected latitudinal range of hollow formation and compare these to hollow-formation rates, timescales, and the latitudinal range that have been derived from observations of hollows on Mercury. We consider 57 volatile phases that can be generally categorized as inorganics, simple organics, aromatic hydrocarbons, linear amides, carboxylic acids, carbon (i.e., fullerenes and graphite), and sulfides (Table 1). We compiled this list based on volatile phases commonly found on solid solar system bodies (e.g., comets, asteroids) and in hermean volcanic eruptions (see Table S1 for information on volatile sources). It should be noted that while these phases are *possible* not all are geologically *plausible* candidate hollow-forming phases. We consider this wide range of phases to explore the thermophysical parameter space and so as not to discount phases *a priori* that may lead to interesting considerations for hollow formation and hermean geology more generally.

| Class | Molecular form | Name | Mass (amu) | Density (g/cc) | A | B | Ref |
|---|---|---|---|---|---|---|---|
| Inorganics | $NH_3$ | Ammonia | 17.031 | 0.82 | 19.6831 | -3548.4791 | [1] |
| | $H_2O$ | Water | 18.02 | 1.00 | 20.1454 | -5709.7220 | [1] |
| | CO | Carbon monoxide | 28.01 | 0.99 | 15.8039 | -901.2444 | [1] |
| | $N_2$ | Nitrogen | 28.02 | 0.95 | 15.5330 | -829.0949 | [1] |
| | S | Sulfur | 32.07 | 2.07 | 17.5020 | -9028.6679 | [1] |
| | $H_2S$ | Hydrogen sulfide | 34.09 | 1.20 | 17.7081 | -2751.9756 | [1] |
| | Ar | Argon | 39.948 | 1.62 | 15.2978 | -930.0365 | [1] |
| | $CO_2$ | Carbon Dioxide | 44.01 | 1.51 | 20.9568 | -3176.1315 | [1] |
| | $SO_2$ | Sulfur Dioxide | 64.07 | 1.90 | 18.4501 | -3600.7217 | [1] |
| | Kr | Krypton | 83.798 | 2.72 | 15.8999 | -1342.6834 | [1] |
| | Xe | Xenon | 131.129 | 3.54 | 16.4629 | -1939.6153 | [1] |
| Simple Organics | $CH_4$ | Methane | 16.04 | 0.49 | 15.3208 | -1174.8802 | [1] |
| | HCN | Hydrogen Cyanide | 27.026 | 1.03 | 18.0122 | -3973.2133 | [1] |
| | COS | Carbonyl Sulfide | 60.08 | 1.56 | 15.7027 | -2461.6665 | [1] |
| | $C_5H_{12}$ | Pentane | 72.151 | 0.91 | 16.9266 | -3632.1135 | [1] |
| | $CS_2$ | Carbon disulfide | 76.15 | 1.55 | 16.2058 | -3662.1778 | [1] |
| | $C_5H_{10}O$ | Pentanal | 86.134 | 1.06 | 16.1700 | -4166.2888 | [1] |
| | $C_7H_8$ | Toluene | 92.141 | 1.03 | 15.8398 | -4082.7779 | [1] |
| | $C_5H_{10}O_2$ | Ethyl Propanoate | 102.133 | 1.15 | 17.2577 | -4562.1276 | [1] |
| Aromatic Hydrocarbons | $C_6H_7N$ | Aniline | 93.129 | 1.22 | 16.4401 | -4961.9649 | [1] |
| | $C_6H_6O$ | Phenol | 94.119 | 1.13 | 20.7498 | -7182.3973 | [1] |
| | $C_7H_6O$ | Benzaldehyde | 106.13 | 1.04 | 17.7545 | -5870.3926 | [1] |
| | $C_7H_8O$ | Benzyl alcohol | 108.146 | 1.10 | 18.0859 | -5675.4731 | [1] |
| | $C_7H_6O_2$ | Salicylaldehyde | 122.129 | 1.37 | 17.3999 | -5965.0204 | [1] |
| | $C_6H_5NO_2$ | Nitrobenzene | 123.111 | 1.34 | 17.7701 | -6297.2056 | [1] |
| | $C_{10}H_8$ | Naphthalene | 128.174 | 1.16 | 19.9823 | -7289.6827 | [1] |
| | $C_{10}H_8O$ | 1-Naphthol | 144.173 | 1.10 | 17.6694 | -7301.9245 | [1] |
| | $C_{10}H_8O$ | 2-Naphthol | 144.173 | 1.28 | 17.6694 | -7301.9245 | [1] |
| | $C_{12}H_{10}$ | Biphenyl | 154.212 | 1.18 | 16.9013 | -6739.6483 | [1] |
| | $C_{12}H_{18}$ | Hexamethylbenzene | 162.276 | 0.98 | 17.4820 | -6439.3143 | [1] |
| | $C_{14}H_{10}$ | Anthracene | 178.22 | 1.25 | 22.4380 | -10494.5425 | [1] |
| | $C_{14}H_{10}$ | Phenanthrene | 178.234 | 1.15 | 22.4380 | -10494.5425 | [1] |
| Linear Amides | $CH_3NO$ | Formamide | 45.041 | 1.26 | 18.5481 | -6812.7311 | [1] |
| | $C_2H_5NO$ | N-Methylformamide | 59.068 | 1.20 | 18.8778 | -6676.9254 | [1] |
| | $C_2H_5NO$ | Acetamide | 59.068 | 1.16 | 18.8778 | -6676.9254 | [1] |
| | $C_3H_7NO$ | Dimethylformamide | 70.1 | 0.95 | 18.9430 | -5974.6286 | [1] |
| | $C_3H_7NO$ | N-Methylacetamide | 70.1 | 1.00 | 18.9430 | -5974.6286 | [1] |
| | $C_4H_9NO$ | Dimethylacetamide | 87.12 | 1.14 | 17.2279 | -5058.8581 | [1] |
| | $C_4H_9NO$ | Methylpropanamide | 87.12 | 1.01 | 17.2279 | -5058.8581 | [1] |
| Carboxylic Acids | $C_5H_{10}O_2$ | Valeric Acid | 102.13 | 1.11 | 17.2577 | -4562.1276 | [1] |
| | $C_6H_{12}O_2$ | Caproic Acid | 116.172 | 1.07 | 20.1922 | -5939.0959 | [1] |
| | $C_7H_{14}O_2$ | Enanthic Acid | 130.187 | 1.02 | 16.8860 | -5040.5091 | [1] |
| | $C_8H_{16}O_2$ | Octanoic Acid | 144.214 | 1.03 | 16.7891 | -5271.9032 | [1] |
| | $C_9H_{18}O_2$ | Nonanoic Acid | 158.241 | 1.05 | 16.8661 | -5528.8777 | [1] |
| | $C_{10}H_{20}O_2$ | Capric Acid | 172.268 | 0.87 | 19.6859 | -6746.3687 | [1] |
| | $C_{12}H_{24}O_2$ | Lauric Acid | 200.346 | 1.01 | 18.6443 | -7174.2935 | [1] |
| | $C_{16}H_{32}O_2$ | Palmitic Acid | 256.43 | 0.98 | 22.4353 | -10900.4317 | [1] |
| | $C_{18}H_{36}O_2$ | Stearic Acid | 256.43 | 0.88 | 22.6120 | -11419.0224 | [1] |
| Carbon | $C_{60}$ | Buckminster Fullerene | 720.6 | 1.57 | 26.6685 | -22512.3745 | [2,3] |
| | $C_{70}$ | Fullerene C70 | 840.7 | 1.57 | 26.7008 | -23530.1171 | [2,3] |
| | C | Graphite | 12.011 | 2.27 | 29.0706 | -95618.4913 | [1] |
| Sulfides | $K_2S$ | Potassium Sulfide | 110.262 | 1.80 | 27.6517 | -29972.0000 | [1] |
| | $Na_2S$ | Sodium Sulfide | 78.0452 | 1.86 | 28.0546 | -30987.0000 | [1] |
| | MgS | Magnesium Sulfide | 56.38 | 2.68 | 28.7618 | -44693.0000 | [1] |
| | MnS | Manganese Sulfide | 87.003 | 4.00 | 28.5099 | -44838.0000 | [1] |
| | FeS | Iron Sulfide | 87.92 | 4.84 | 31.2236 | -46553.0000 | [1] |
| | CaS | Calcium Sulfide | 72.143 | 2.59 | 28.7698 | -58092.0000 | [1] |

[1] Rumble (2018); [2] Pan (1992); [3] Dresselhaus et al. (1996)

**Table 1:** Chemical formulas, names, molecular masses, densities, and constants A and B (used in **Eq. 9)** for candidate hollow-forming volatile phases explored in this study.

2.1. Thermophysical Model

2.1.1. Heat Diffusion Equation

There have been several thermophysical models for Mercury formulated in past studies (e.g., Ulrichs and Campbell, 1969; Morrison, 1970; Chase et al., 1976; Mitchell and de Pater, 1994; Salvail and Fanale, 1994; Emery et al., 1998; Vasavada et al., 1999; Hale and Hapke, 2002; Yan et al., 2006; Paige et al., 2013; Bandfield et al., 2019).

Similar to past models, we consider heat conduction along the vertical dimension, and use the 1D time-dependent heat equation (Fourier, 1822):

**Eq. 1**

$$\rho c_p \frac{\partial T}{\partial t} = \frac{\partial}{\partial \mathrm{z}} k \frac{\partial T}{\partial z}$$

Where $\rho$ is density, $c_p$ is the specific heat capacity, and $k$ is the effective thermal conductivity. We allow $\rho$ to vary with depth ($z$), $c_p$ to vary with temperature ($T$), and $k$ to vary with $z$ and $T$. For $\rho$, we use Carrier et al.'s (1991) empirically derived formula for lunar cores (**Eq. 2**) because these data serve as the best available analog to a hermean regolith density profile due to likely similarities (at least in gross properties) between the physical properties of hermean and lunar regolith (e.g., Chase et al., 1976; Hale and Hapke, 2002):

**Eq. 2**

$$\rho(z) = \rho_b \left(\frac{z + 0.122}{z + 0.18}\right)$$

Where $\rho(z)$ is density (kg m$^{-3}$) as a function of depth ($z$, in meters) and the subscript $b$ indicates evaluation at the bottom boundary. We adopt a larger value for $\rho_b$ than is used by Carrier et al. (1991) to account for Mercury's higher surface gravity (Tables 2 and 3, see Carrier et al., 1991 for details). $c_p$ is calculated using a fourth degree polynomial (**Eq. 3**) fit to data from Hemingway et al. (1973), Hemingway et al. (1981), Ledlow et al. (1992), and Wakabayashi and Matsumoto (2006).

**Eq. 3**

$$c_p = c_0 + c_1 T + c_2 T^2 + c_3 T^3 + c_4 T^4$$

This polynomial fit is valid for regolith and a range of temperatures between ~30 < T < ~1000 K (Fig. S2) and is thus appropriate for our model. Polynomial coefficients are presented in Tables 2 and 3. We allow $k$ to vary with temperature using an expression for effective thermal conductivity used in previous modeling that incorporates an empirical term for radiative heat transfer (e.g., Whipple, 1950; Watson, 1964; Cuzzi, 1974; Mitchell and de Pater, 1994):

**Eq. 4**

$$k = k_c + \beta T^3$$

Where $k_c$ is the contact conductivity and $\beta$, defined as $4\sigma\epsilon l$, is a factor that incorporates the effects of inter-grain radiative transfer (Whipple, 1950). The values $\sigma$, $\epsilon$, and $l$ are the Stefan-Boltzmann constant, thermal emissivity, and the inter-grain length scale respectively. Empirical calculations by Fountain and West (1970) show that contact conductivity varies linearly with regolith density according to:

**Eq. 5**

$$k_c = k_b - (k_b - k_s) \frac{\rho_b - \rho(z)}{\rho_b - \rho_s}$$

where the subscript $s$ indicates evaluation at the surface boundary. Because of the dependence of $k$ and $c_p$ on temperature, the thermal inertia ($TI = \sqrt{k\rho c_p}$) varies with time and depth. These dependencies will be important when comparing our model results to others where $TI$ is an input parameter to the model (*Section 2.1.3*).

| Table 2: Range of parameter values explored and useful model outputs ($dw, TI, dt$) | | | | $dw$ (m) | $TI$ (time-averaged) | $dt$ (s) |
|---|---|---|---|---|---|---|
| median values | | | | 0.1247 | 70.1 | 410.6 |
| Albedo, $A$ | Lower $A$ | | 0.06 | 0.1247 | 70.7 | 410.6 |
| Albedo, $A$ | Upper $A$ | | 0.14 | 0.1247 | 69.6 | 410.6 |
| Emissivity, $\epsilon$ | Lower $\epsilon$ | | 0.9 | 0.1226 | 70.2 | 455.3 |
| Emissivity, $\epsilon$ | Upper $\epsilon$ | | 1 | 0.1268 | 70.1 | 401.0 |
| inter-grain length, l (µm) $\beta = 4\sigma\epsilon l$ | Lower $l$ | | 20 | 0.0903 | 58.7 | 675.5 |
| inter-grain length, l (µm) $\beta = 4\sigma\epsilon l$ | Upper $l$ | | 130 | 0.1485 | 79.9 | 807.5 |
| Density (kg m$^{-3}$) $\rho(z) = \rho_b \left(\frac{z+0.122}{z+0.18}\right)$ | Lower $\rho$ | $\rho_b$ | 1920 | 0.1454 | 59.9 | 807.5 |
| Density (kg m$^{-3}$) $\rho(z) = \rho_b \left(\frac{z+0.122}{z+0.18}\right)$ | Upper $\rho$ | $\rho_b$ | 3300 | 0.1109 | 79.2 | 746.2 |
| Contact conductivity ($W\ m^{-1}\ K^{-1}$) $K_c = K_b - (K_b - K_s)\frac{\rho_b - \rho(z)}{\rho_b - \rho_s}$ | Lower $K_c$ | $K_s$ | 0.00005 | 0.1014 | 40.6 | 2243.4 |
| | | $K_b$ | 0.00009 | | | |
| | Upper $K_c$ | $K_s$ | 0.09 | 0.4171 | 317.9 | 202.0 |
| | | $K_b$ | 0.1 | | | |
| Heat Capacity ($J\ kg^{-1}\ K^{-1}$) $c_p = c0 + c1T + c2T^2 + c3T^3 + c4T^4$ | Lower $c_p$ confidence interval | $c0$ | -1.47E+02 | 0.1371 | 59.4 | 807.4 |
| | | $c1$ | 4.12E+00 | | | |
| | | $c2$ | -5.29E-03 | | | |
| | | $c3$ | 2.43E-06 | | | |
| | | $c4$ | -4.75E-10 | | | |
| | Upper $c_p$ confidence interval | $c0$ | -8.05E+01 | 0.1117 | 76.2 | 807.2 |
| | | $c1$ | 5.14E+00 | | | |
| | | $c2$ | -8.73E-03 | | | |
| | | $c3$ | 7.64E-06 | | | |
| | | $c4$ | -2.38E-09 | | | |

### 2.1.2 Boundary Conditions

#### 2.1.2.1 Surface Boundary Condition

The temperature at the surface is controlled by a balance between incident solar flux, thermal conduction of endogenic heat, and radiative energy loss, such that the surface boundary condition can be written as:

**Eq. 6**

$$\rho c_p \frac{\partial T_s}{\partial t} = \frac{(1-A) * F_{\odot}}{R^2}\cos(\Theta_z) - k\frac{\partial^2 T}{\partial z^2}\bigg|_s - \sigma\epsilon(T_s^4)$$

The subscript $s$, again, denotes evaluation at the surface. $A$ is the bolometric albedo and $\epsilon$ is the infrared emissivity of the surface. $R$ is the orbital radius of Mercury in AU, which varies from 0.308 to 0.467 AU. The solar constant, $F_{\odot}$, is 1367 W m$^{-2}$ and is multiplied by the cosine of the solar incidence angle $(\Theta_z)$. $\Theta_z$ depends on the latitude and longitude of a horizontal surface facet and the subsolar latitude and longitude. We set $\cos(\Theta_z)$ to zero on the night side of the planet.

#### 2.1.2.2 Bottom Boundary Condition

A profile of 10 m depth measured from the surface was chosen to ensure our model encompassed the depth at which the equilibrium temperature is achieved (~1 m in our model). The boundary condition at depth can be written:

**Eq. 7**

$$\rho c_p \frac{\partial T_b}{\partial z} = \frac{\partial}{\partial z}\left(k_b \frac{\partial T}{\partial z}\right)\bigg|_b + \frac{Q}{z} dz_b$$

The subscript $b$ denotes evaluation at the bottom boundary, $dz_b$ is the depth step at the bottom boundary, and $Q$ is the internal thermal heat flux. $Q$ has been calculated at 20 mW m$^{-2}$ by Schubert et al. (1988) and Hauck et al. (2004), and contributes relatively little to the subsurface temperature at the depths explored by our model.

With the described boundary conditions, the implicit method of Crank and Nicolson (1947) is used to solve **Eq. 1**. We split our model into three grid layers to sample more finely near the surface and more coarsely with depth. The thickness of the first grid layer is calculated as 10% of the diurnal thermal skin depth ($d_w$, typically ~11 cm in our model, Table 2). The first layer samples the first ~cm of the surface at ~100 µm, approximately the scale of a regolith grain (c.f., Hale and Hapke, 2002). The second grid layer extends from 1 cm to 1 m depth and is sampled at twice the depth interval as used in the first layer. The third grid layer extends from 1 to 10 m in depth and samples at a depth interval that is 10 times larger than that used for the first layer. We used the maximum time step for explicit stability, typically ~5 to 7 minutes in our model (see, e.g., Hayne et al., 2017). Solutions were said to reach equilibrium when the change in integrated surface temperature between consecutive model days became < 0.005 K.

Table 3: "Best Fit" Parameters

| Parameter | | Value |
|---|---|---|
| $A$ | | 0.06 |
| $\epsilon$ | | 0.95 |
| l (µm) | | 75 |
| $\rho_b$ $(kg\ m^{-3})$ | | 3100 |
| $K_c$ $(W\ m^{-1}\ K^{-1})$ | $K_s$ | 0.004 |
| | $K_b$ | 0.007 |
| $C_p$ $(J\ kg^{-1}\ K^{-1})$ | $c0$ | -8.05E+01 |
| | $c1$ | 5.14E+00 |
| | $c2$ | -8.73E-03 |
| | $c3$ | 7.64E-06 |
| | $c4$ | -2.38E-09 |
| $d_w$ (m) | | 0.1066 |
| $dt$ (s) | | 807.1 |
| Time-averaged $TI$ | | 92.7 |
| $kE_t$ | | 53.3 |

2.1.3 Model Sensitivity, Comparison to Mariner 10 IRR Data and to Other Thermal Models

We explored a range of physically plausible values for $A, \epsilon, l, \rho_b, k_s, k_b,$ and $c_p$ to test the sensitivity of our model to variations in these parameters (c.f., BeipColombo-STR, 2000; Clark, 2015). Table 2 shows upper and lower parameter values and some model outputs ($d_w, TI,$ and $dt$) that are useful for evaluating the behavior of the model. When exploring the extremes of an individual parameter, all other values were set to median values to isolate the effect of the parameter of interest on the model. The diurnal surface temperatures and noon and pre-dawn temperature profiles for these model runs are presented in Figures S3 and S4. We compare our model results to the model results produced by Yan et al. (2006) (hereafter referred to as Y06) and Hale and Hapke (2002) (hereafter referred to as HH02). The model produced by Y06 is a relatively recent effort with a similar approach to that presented here, and the model by HH02 is more detailed and formally considers the effects of radiative heat transfer. We also compare our surface temperatures to Mariner 10 Infrared Radiometer (IRR) nighttime surface temperatures (Chase et al., 1976). The parameter values ultimately used to calculate temperature profiles used for sublimation and diffusion rate calculations are presented in Table 3.

Albedo, $A$, controls how much insolation is absorbed by the surface. We considered a range from 0.06 to 0.14, which encompasses values commonly considered for Mercury (e.g., Clark, 2015; Rothery, 2015). Infrared emissivity, $\epsilon$, modulates the amount of energy emitted by the surface according to the Stefan-Boltzmann law. We considered a range for $\epsilon$ from 0.9 to 1, values typical for silicate-dominated planetary surface materials. While variations in $A$ and $\epsilon$ values can produce large effects on surface temperatures, the ranges of values plausible for Mercury are relatively small and as such these parameters do not greatly affect temperatures in our model (Figs. S3 and S4).

The inter-grain length scale, $l$, is used to calculate $\beta$ in **Eq. 4**. $l$ is typically on the order of a grain size, so we used values ranging from 20 to 130 µm (e.g., Carrier, 1973).

The value for density at depth, $\rho_b$, controls the density profile as described by **Eq. 2**. Values for $\rho_b$ ranged from 1920 kg m$^{-3}$ to 3300 kg m$^{-3}$, which correspond to surface densities of 1301 kg m$^{-3}$ and 2236 kg m$^{-3}$, respectively (see Fig. S5, c.f., Clark, 2015). The lower value was chosen because it is typical for lunar density profiles (Carrier et al., 1991) and the upper value was chosen to encompass an upper end density of gabbro (a reasonable upper limit estimate for hermean density at depth).

Thermal conductivity values can range over orders of magnitude; consequently, thermal conductivity has the largest effect on both surface and subsurface temperatures and, therefore, on uncertainty within our model (Figs. S3 and S4). We used a range that encompasses more than the variability in thermal conductivity found in lunar regolith (e.g., Heiken et al., 1991; Schreiner et al., 2016) to account for possible differences between the properties of hermean regolith and lunar regolith. Values for $k_s$ range from 5 x 10$^{-5}$ to 9 x 10$^{-2}$ W m$^{-1}$ K$^{-1}$ and values for $k_b$ range from 9 x 10$^{-5}$ to 1 x 10$^{-1}$ W m$^{-1}$ K$^{-1}$. Y06 calculated a value for $k_c$ by fitting their model to a model run from HH02 and determined 5 x 10$^{-5}$ W m$^{-1}$ K$^{-1}$ as the best-fit value. Our model results for the lower values of $k_{s,b}$ match well with the results of Y06 (diamond symbols in Fig. S4). The HH02 temperature-depth profile we use for comparison (squares, Fig. S4) is that which the authors suggested best represented the hermean regolith. Note that increasing values of $k_{s,b}$ has the expected effect of decreasing the equilibrium temperature and increasing the penetration depth of the thermal wave.

For $T$-dependent $c_p$, a fourth-degree polynomial (**Eq. 3**) was fit to lunar regolith data (see Fig. S2 for data and citations). To incorporate possible differences in $c_p$ between hermean and lunar regolith, we added an uncertainty of ±75 J kg$^{-1}$ K$^{1}$ to the data and calculated ±85% confidence intervals ("upper" and "lower" $c_p$, Table 2). Our fits are compared to those of Hemingway et al. (1973); Ledlow et al. (1992); Hayne et al. (2017) in Figure S2.

Table 4: Comparison to HH02 free parameters

| Parameter | This Work | HH02 |
|---|---|---|
| Lower $kE_t$ | 0.7 | 1 |
| Upper $kE_t$ | 1200.0 | 1000 |
| Median $kE_t$ | 40.0 | 50 |
| Lower $TI$ | 40.6 | 43 |
| Upper $TI$ | 317.9 | 67 |
| Median $TI$ | 70.1 | 65 |

As a check for the reasonableness of the range of parameter space explored, we compared our range to the range investigated by HH02. The free parameters in HH02's model were $TI$ and the "conductivity ratio" ($kE_t \equiv k_c/L_{Ex}$, where $L_{Ex}$ is the thermal extinction length). We calculated a value for the conductivity ratio in our model using $l$, the inter-grain length, as an estimate of $L_{Ex}$, which was suggested to be comparable for hermean regolith by HH02. Our lower limit of $kE_t$ is 0.7 and our upper limit is 1200, compared to 1 and 1000 by HH02. Our min, max, and median time-averaged $TI$ values were 40.6, 317.9, and 70.1 J m$^{-2}$ K$^{-1}$ s$^{-1/2}$ compared to 43, 67, and 65 J m$^{-2}$ K$^{-1}$ s$^{-1/2}$ by HH02 (Table 4).

## 2.2 Sublimation Rate Equations and Choice of Volatile Phases

### 2.2.1 Sublimation at the Surface

We considered volatiles that are likely to exist on Mercury from impact delivery (Pierazzo and Chyba, 1999; Botta and Bada, 2002), emplacement from volcanic processes (Kerber et al., 2009), or from space weathering (Domingue et al., 2014). Table S1 lists common volatiles and their concentrations found in meteorites (Botta and Bada, 2002), comets (Ferris et al., 2007), and hermean magmas (Kerber et al., 2009, 2011). For the behavior of volatiles at the surface, we are interested in the rate at which these volatiles will sublimate as a function of temperature in vacuum. To calculate the sublimation rate, we us the Hertz-Knudsen equation (e.g., Watson et al., 1961):
**Eq. 8**

$$E_i = P_v\left(\frac{\mu_i}{2\pi \mathrm{R}T}\right)^{\frac{1}{2}}$$

Where $E_i$ is the mass sublimation rate of volatile $i$ (kg m$^{-2}$ s$^{-1}$); $\mu_i$ is the molar mass of volatile $i$ (kg mol$^{-1}$); R is the gas constant (m$^2$ kg s$^{-2}$ K$^{-1}$ mol$^{-1}$); $T$ is the temperature (K); and $P_v$ is the temperature-dependent solid-vapor equilibrium pressure (Pa). $P_v$ is calculated using the standard vapor pressure equation (e.g., Lodders and Fegley, 1998):
**Eq. 9**

$$\ln\big(P_v(T)\big) = A_i + \frac{B_i}{T}$$

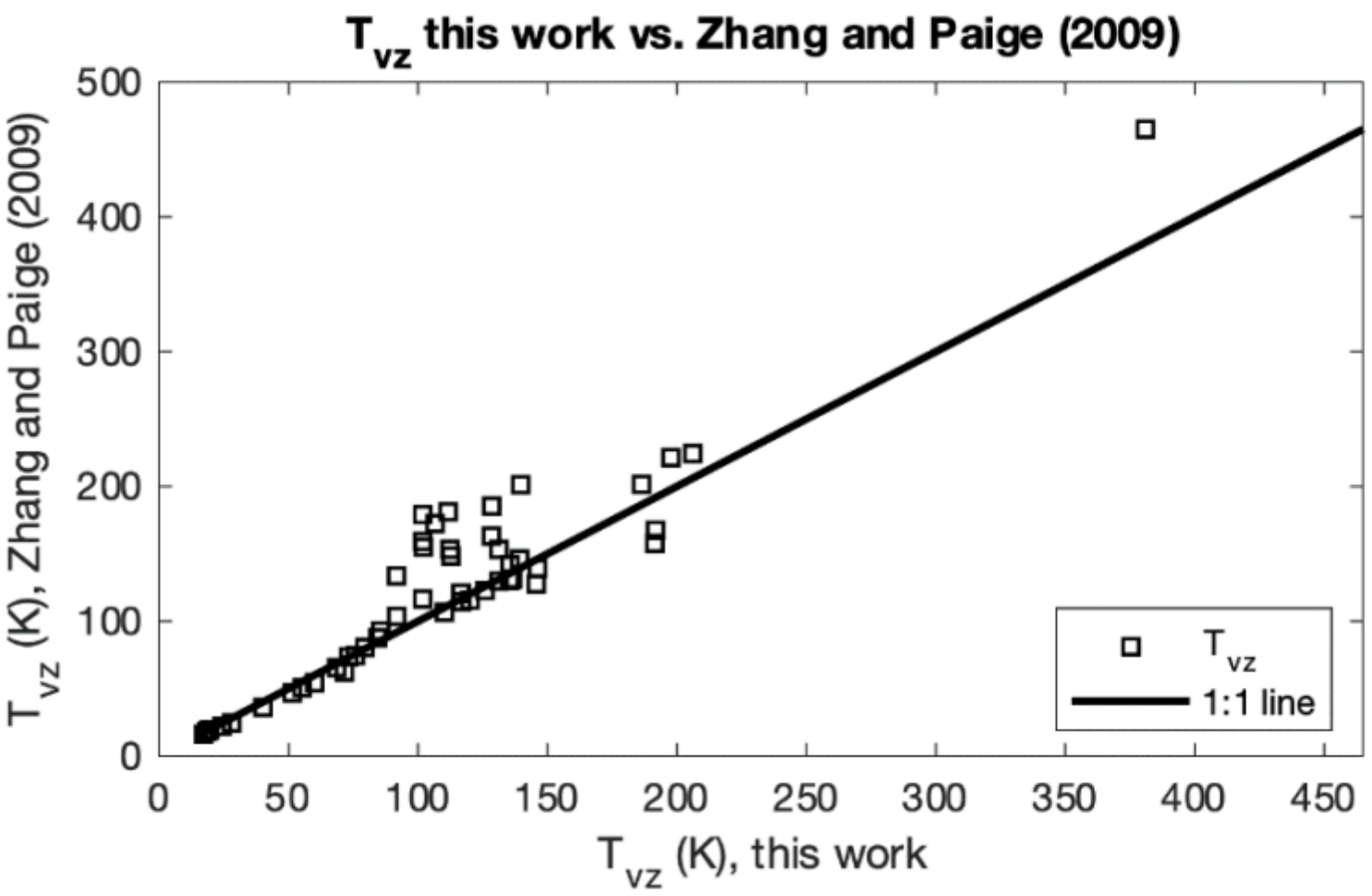

**Figure 4:** Comparison between the volatility temperatures of Zhang and Paige (2009) and those calculated in this work. See text for details on differences.

Where $A_i$ and $B_i$ are constants for volatile phase $i$ derived by fitting **Eq. 9** to laboratory-derived pressure and temperature data from Rumble (2018) and Rankin (2009). The coefficients derived from our fitting are presented in Table 1.

We convert $E_i$ into a sublimation rate in m $Gyr^{-1}$ by dividing by the density of each volatile phase (from Yaws, 2014) and converting seconds to Gyr. For each volatile, one can calculate a "volatility temperature" defined as the temperature at which $E_i$ equals some desired rate. Zhang and Paige (2009) calculated a "volatility temperature", $T_{vz}$, using $E_i$ = 1 mm $Gyr^{-1}$. We calculated values for $T_{vz}$ and compared to those of Zhang and Paige (2009) (Fig. 4). Differences between our calculated temperatures and those of Zhang and Paige (2009) arise from differences in the P-T data used to derive constants $A_i$ and $B_i$ (**Eq. 9**) and in our choices for volatile densities. Our calculation uses the density of each volatile phase ($\rho_i$) whereas Zhang and Paige (2009) used 1000 kg $m^{-3}$ in all calculations. The results of our sublimation rate calculations are reported in Figure 5. These sublimation rates are for pure volatile phases and do not incorporate possible chemical and physical interactions with other constituents of the regolith and should be thought of as ideal upper limits (i.e., phases may be stable on the surface for longer than modelled here).

Sulfides must be treated differently because they do not simply sublimate as do the other volatile phases considered here. Sulfides will decompose into their elemental constituents at temperatures below their melting point through a process called reactive sublimation or thermal decomposition. We use reactive sublimation to refer to surface reactions because it encapsulates both the decomposition of the metal-sulfide bond and the subsequent sublimation to vacuum of the gas phase products, and we use thermal decomposition to refer more generally to the decomposition reaction. Where this work refers to the "sublimation" of sulfides, reactive sublimation is implied. The thermal decomposition of $Na_2S$, MnS, MgS, and/or CaS is a proposed mechanism for hollow formation (Vaughan et al., 2012; Helbert et al., 2013; Thomas et al., 2016). Therefore, we consider these sulfides along with other sulfides that plausibly exist on or within Mercury as potential hollow-forming materials (Table 1). A general equation for the decomposition reaction of a sulfide can be written as:

**Eq. 10**

$$\mathrm{M_xS_{y(s)}} \rightarrow \mathrm{xM_{(g)}} + \frac{y}{2} S_{2(g)}$$

Vapor pressure data are not available for the decomposition reactions we consider. However, an ideal maximum rate as a function of temperature can be calculated from analysis of the Gibb's free

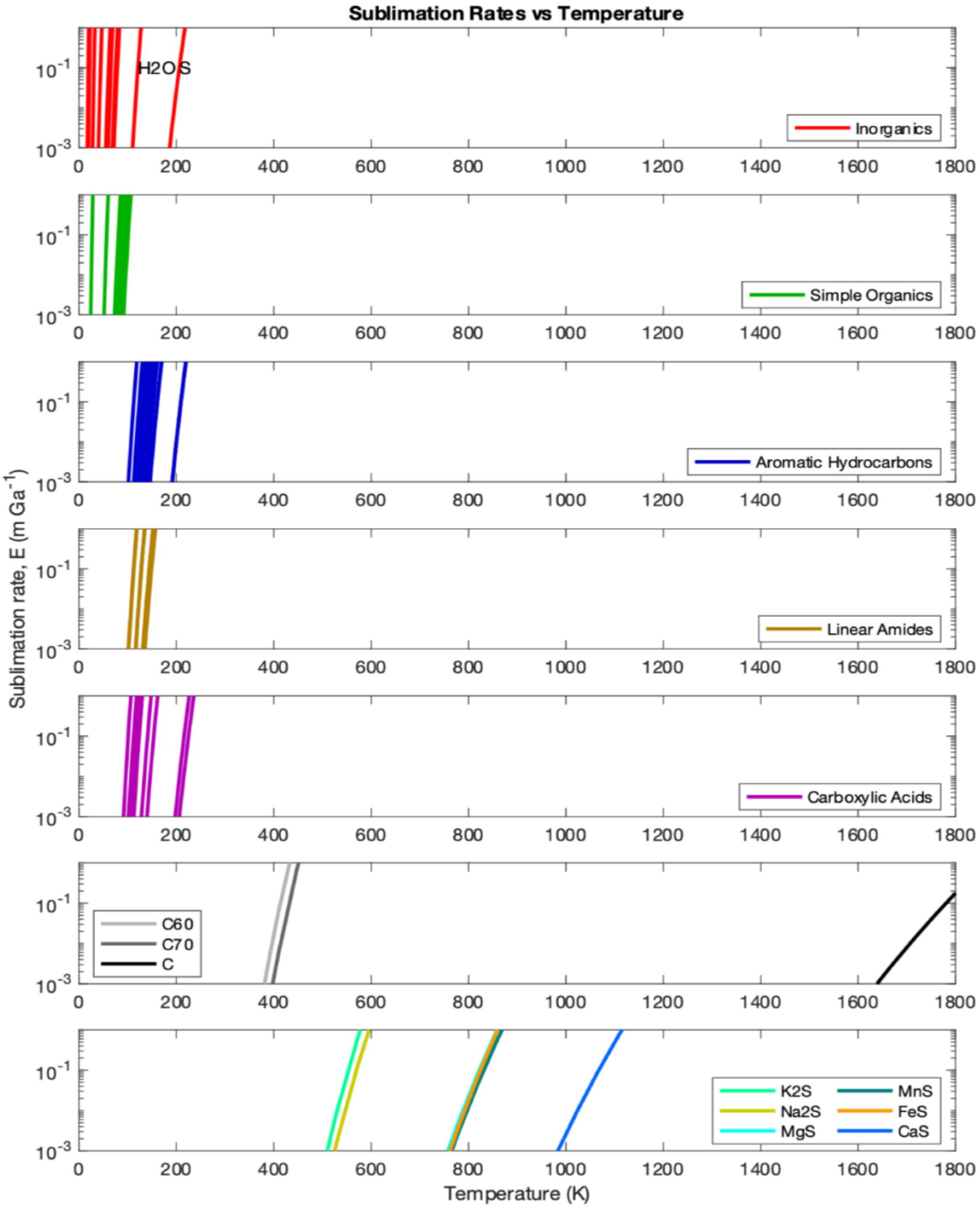

**Figure 5:** Sublimation rates vs. temperature for the volatiles in this study. Volatiles are grouped into different plots by compound classes. Legends for the bottom two plots list phases from left to right. Colors correspond to Table 1. Values listed in Table 1 were used to calculate rates using **Eq. 9**.

energy of the reactions (assuming the change in heat capacity between the products and reactants is independent of temperature) according to:

**Eq. 11**

$$\Delta G = \Delta G^{\circ} + RT\ln(K_{eq})$$

Where $\Delta G$ is the Gibbs free energy of the reaction and $\Delta G^{\circ}$ is the standard state Gibb's free energy of the reaction given by:

**Eq. 12**

$$\Delta G^{\circ} = \Delta H^{\circ} - T\Delta S^{\circ}$$

$\Delta H^{\circ}$ and $\Delta S^{\circ}$ are the change in enthalpy and entropy, respectively, between products and reactants at standard state. The equilibrium constant, $K_{eq}$, is given by the expression:

**Eq. 13**

$$K_{eq} = P_M^x P_{S_2}^{\frac{y}{2}}$$

Where, $P_M^x$ and $P_{S_2}^{\frac{y}{2}}$ are the partial pressures of the gas-phase products of the reaction, that is, the sulfide metal cation and disulfur, exponentiated by their molar concentrations, x and $\frac{y}{2}$. Substituting **Eq. 13** and **Eq. 12** into **Eq. 11** and noting that $P_{S_2} = \frac{y}{2x} P_M$, we solved for the vapor pressure (in atmospheres) of the reaction in terms of the vapor pressure of the metal, M (e.g., Leckey and Nulf, 1994):

**Eq. 14**

$$\ln(P_M(T)) = \frac{y}{2x + y}\ln\left(\frac{2x}{y}\right) + \frac{\Delta S^{\circ}}{\left(x + \frac{y}{2}\right)R} - \frac{\Delta H^{\circ}}{\left(x + \frac{y}{2}\right)RT}$$

In this form, we can see that constants $A_i$ and $B_i$ from **Eq. 9** are equal to:

$$A_i = \frac{y}{2x + y}\ln\left(\frac{2x}{y}\right) + \frac{\Delta S^{\circ}}{\left(x + \frac{y}{2}\right)R}$$

$$B_i = -\frac{\Delta H^{\circ}}{\left(x + \frac{y}{2}\right)R}$$

With an equation for vapor pressure, the sublimation rate is calculated as before with **Eq. 8**. Note that converting the vapor pressure from atmospheres to pascals requires an addition of $\ln(101{,}325)$ to the constant $A_i$. Using the method delineated above, we calculated constants $A_{PbS}$ and $B_{PbS}$ for the thermal decomposition of PbS, for which vapor pressure data do exist in the CRC handbook (and references therein), to compare to the constants derived by fitting **Eq. 9** to experimental data. Table 1 shows the coefficients derived from both methods ($A_{PbS,theory} = 27.6, B_{PbS,theory} = -28{,}801$; $A_{PbS,experiment} = 21.9, B_{PbS,experiment} = -26{,}851.9$).

For comparison on the effect each coefficient has on the vapor pressure, we calculated the percent error between the theoretically and experimentally derived vapor pressure curves (Fig. S6). The percent error is a function of temperature and is highest at low temperatures because the vapor pressure values are very low. The error between the curves is zero at 340 K and is ~94% at 710 K. The vapor pressure calculated from theory is higher (at temperatures > 340 K) than those calculated from experimental data, as expected, which results in calculated sublimation rates that are faster than those determined experimentally. We emphasize that our thermal decomposition rate calculations for sulfides are theoretical upper limits.

Lastly, we derived a surface sublimation rate for each volatile phase over one hermean day (i.e., two hermean years) by integrating **Eq. 8** over a diurnal surface temperature profile produced by our thermal model. The model-derived rates are compared to rates estimated for hollow formation based on observations of hollows on Mercury in *Section 3*.

2.2.2 Diffusion Through a Regolith Cover

To assess the rate of volatile loss through a regolith cover we use Fick's law:

**Eq. 15**

$$F_i = D_k(T)\frac{\phi}{\tau}\frac{\delta N_i(T)}{\delta z}$$

Where, $F_i$ is the loss flux (kg m$^{-2}$ s$^{-1}$) of volatile $i$, $\phi$ is porosity, $\tau$ is tortuosity, and $\frac{\delta N_i(T)}{\delta z}$ is the gradient in vapor density (mol m$^{-3}$) from the volatile layer to the surface, assuming the surface is at vacuum. We calculate vapor density from vapor pressure (**Eq. 9**) using the ideal gas law. The diffusion coefficient, $D_k(T)$, is equal to $\frac{4}{3}r\left(\frac{2RT}{\mu_i\pi}\right)^{\frac{1}{2}}$, and is calculated from Knudsen diffusion theory considering the mean molecular speed within a cylindrical pore of radius $r$ (e.g., Pollard and Present, 1948; Moore et al., 1996). Knudsen diffusion is valid when $r$ is comparable to or smaller than the mean free molecular path of the volatile phase, that is, when molecular collisions with the pore wall dominate over molecule-molecule collisions. This condition is true for the volatile phases and conditions explored here with the exception of simple organic molecules and the most volatile inorganics (Table S2). Molecules for which this condition is not true have regolith diffusion rates that are likely over estimated.

We define a column integrated porosity after Carrier et al. (1991) as $\phi = 1 - \frac{\bar{\rho}(z)}{G_h\rho_w}$, where $z$ is the thickness of the regolith cover, $\bar{\rho}(z)$ is the mean column density calculated from **Eq. 2**, $G_h$ is the specific gravity of hermean regolith and $\rho_w$ is the density of water. Values for $\phi$ range from 52.1% to 29.7% for 1 mm and 100 m regolith covers respectively. Tortuosity, $\tau$, captures the degree to which the diffusive path deviates from a straight line. Typical values of $\tau$ for various regolith and regolith-simulant materials range from ~1.5 to 7, but values can be as low as 1 to as high as 12.5 (Reiss, 2019 and references therein). In the absence of experimental data describing $\tau$ for hermean regolith, we use the Bruggeman model describing tortuosity as a function of porosity for cylinders in which $\tau = \phi^{-1}$ (Bruggeman, 1935; Tjaden et al., 2016). As with our values for the surface sublimation rate, $E_i$, we convert $F_i$ into a sublimation rate in m Gyr$^{-1}$ by dividing by the density of each sulfide phase (from Yaws, 2014) and converting seconds to Gyr.

$F_i$ was calculated for each volatile phase using time-varying temperature versus depth profiles of volatile-free regolith derived from our thermal model as inputs into **Eq. 15**. Rates were integrated over a hermean day (two hermean years) to derive loss rates as a function of depth along the hot and warm meridians. If the effects of volatile species in the pore space of regolith were included in our temperature profile calculations, it would enhance heat transfer (by enhancing contact conductivity), increase the temperature in volatile-bearing zones, and therefore increase sublimation rates. We comment on the implications of neglecting the effects of pore-filling volatiles on the thermal profile in our results (*Section 3*).

2.3 Metrics for Comparison: Hollow Formation Rates, Timescales, and Latitudinal Extent

In *Section 3*, we compare our model-derived hollow-formation rates and timescales for each volatile phase to rates and timescales derived from observations of hollows on the surface made in Blewett et al. (2016) because that study offers an analysis of over 500 images of hollows at pixel scales of < 20 m, which is ~10 times higher resolution than other large-scale studies of hollows. To calculate hollow formation rates, we require hollow depths and an estimate for the timescale over which hollows formed. As shown in Figure 2d, the average depth of hollows is latitude-dependent, so we chose to calculate latitude-dependent rates by averaging the depth of hollows between 0°N and 30°N, 30°N and 60°N, and 60°N and 90°N (Table 5). The tightest observationally derived constraint for the timescale over which hollows form is 300 Myr, and comes from hollows in the rayed crater Balanchine at 38.6°N, 175.3°E (Blewett et al., 2016). Blewett et al. (2016) estimated the lateral rate of hollow growth for hollows on the floor of Balanchine by taking the lateral extent of the hollows (300 m) and dividing that by the estimated age of the crater (300 Ma). Because we are interested in the rate of downward growth, we instead estimate a vertical-growth rate by using the depth of hollows rather than their lateral extent. The 300 Myr timescale for hollow formation in Balanchine crater is an upper limit for how long the hollows in Balanchine may have taken to form. The hollows in Balanchine are the only hollows for which a published timescale estimate exists, so we used this timescale along with the average hollow depths over each latitudinal range to calculate "characteristic" vertical-growth rates of hollows (characteristic vertical-growth rate, Table 5). The possibility that hollows in Balanchine formed over a faster timescale cannot be precluded, and therefore our characteristic rates are lower limit estimates of actual hollow formation rates.

| Latitude | Average Depth (m) | Characteristic vertical growth rate depth/0.3 Gyr | Minimum effective vertical growth rates depth/4 Gyr |
|---|---|---|---|
| 0°N to 30°N | 25.7 | 85.6 | 6.4 |
| 30°N to 60°N | 24.5 | 81.8 | 6.1 |
| 60°N to 90°N | 13.3 | 44.2 | 3.3 |

**Table 5**: Estimated hollow-formation rates based on the average depth of hollows within 30° latitudinal bins with two estimates for the timescale of hollow formation (300 Myr and 4 Gyr). Rates are presented in m $Gyr^{-1}$.

In addition to the characteristic rates of hollow formation, it is also necessary to estimate a rate below which hollow growth can be considered effectively inactive. We call these rates the "minimum effective" vertical-growth rates (Table 5). These rates were calculated for each latitudinal range based on an estimate for the absolute maximum timescale over which hollows could form coupled with average hollow depth values over each latitudinal range. For the upper limit timescale estimate, we used 4 Gyr based on an age estimate for heavily cratered terrain (the oldest terrain where hollows are commonly found) from Marchi et al. (2013)

Another metric for comparison between our model-derived results and observations of hollows is the range of latitudes over which hollows are observed on Mercury. Hollows are found at the hot poles. Therefore, a candidate hollow-forming volatile must be able to be sequestered beneath a regolith lag deposit of reasonable thickness (i.e., < 100 m). That is, the loss rate of a candidate hollow-forming phase must be suppressed to a slow enough rate when buried beneath a regolith lag deposit at the hot pole such that hollow formation can be considered effectively inactive. We call this constraint the "hot limit" of hollow formation. The hot limit can be considered a soft constraint because in addition to being limited by a lag deposit, hollow growth could be limited by exhaustion of the volatile phase. However, volatile exhaustion is considered a less favorable mechanism for hollow growth cessation than lag development because the thickness of the volatile-sourcing LRM is several km-thick, whereas hollows are only 10s to ~100 m-deep (Blewett et al., 2016). The second constraint is imposed by the location of the northernmost observed hollows and is twofold: 1) whether a phase can sublimate quickly enough at that latitude to account for hollow formation; and 2) whether a phase is refractory enough that it is not expected to form hollows at that latitude. We call this constraint the "cold limit" of hollow formation, which we discuss in more detail below.

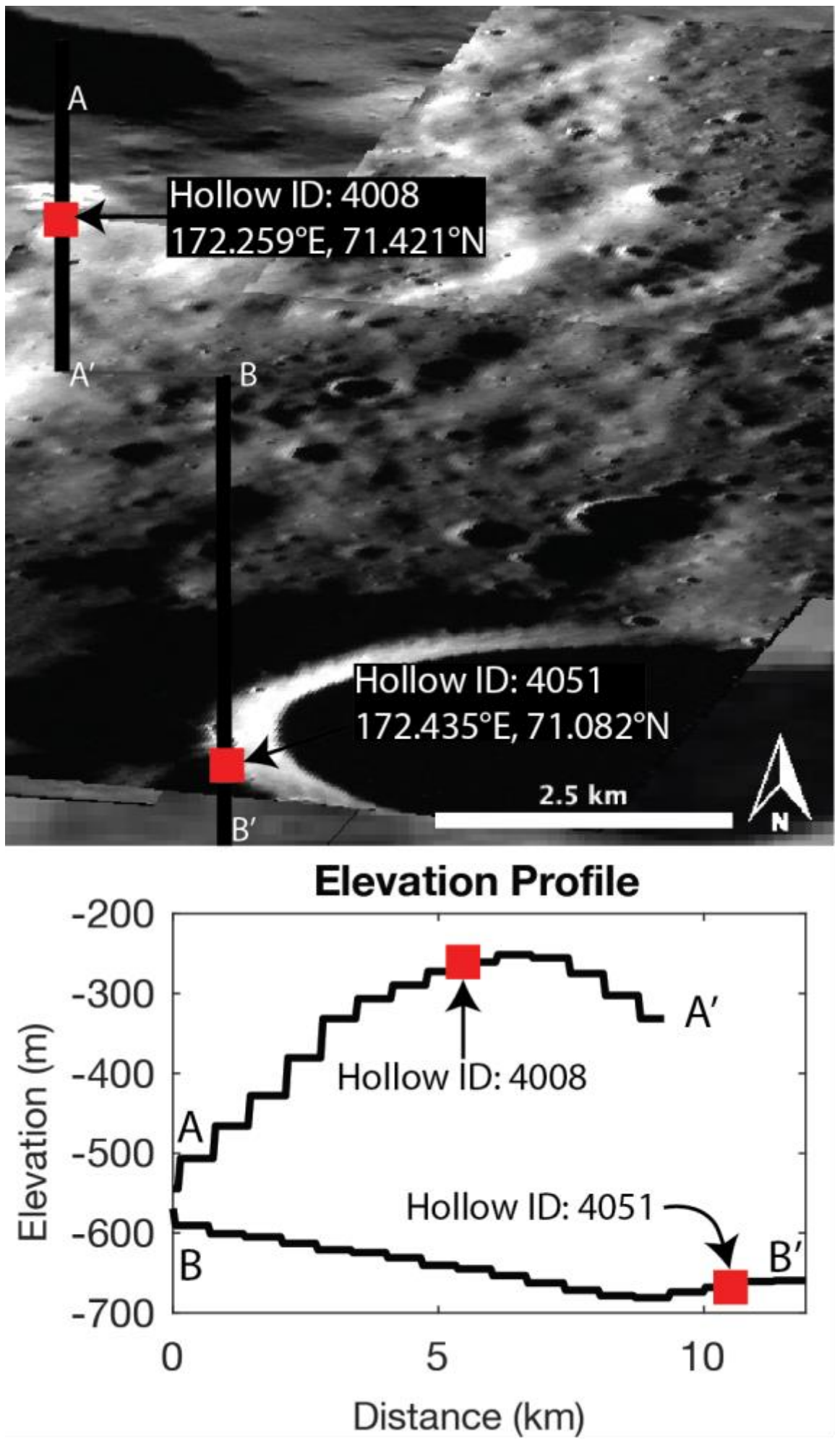

**Figure 6:** Hollows at the cold limit of hollow formation. North-south oriented elevation profiles showing the slopes for hollow 4008 at 71.4°N (A to A') and 4051 at 71.1°N (B to B'). Vertical black bars in the elevation profile show the locations of each hollow along the profiles. Hollow ID numbers are from the global hollow database of Thomas et al. (2016). The elevation profiles show that the slopes on which these hollows have formed are slightly north facing (~1.2° for hollow 4008 and ~0.6° for hollow 4051). Analysis was done with MDIS images <20 m/pixel and with the USGS digital topography map (665 m/pixel resolution) in JMARS.

The solar insolation received by a surface facet depends on its location on Mercury, its albedo, its orientation, and whether the facet suffers from target adjacency effects (i.e., if adjacent topography fills any portion of its "sky", contributing radiation to, and thereby warming, the target). Our model takes into account the location and physical properties of a surface facet, but we assume a horizontal surface freely radiating to space; which means our calculations do not include the effects that surface facet orientation or that target adjacency might have on the local thermal environment of a hollow. Therefore, we set the cold limit of hollow formation for our model using the latitude of the northernmost hollows that form on a horizontal surface and have no obvious local topography that might produce target adjacency effects. We found that this occurred at ~71°N with hollows 4008 and 4051 of the Thomas et al. (2016) database. All 20 hollows north of 4008 and 4051 occur along the south-facing walls of their host craters (Fig. 6).

Hollow 4008 is on a regional north-dipping slope (~1.2° as measured using a global MESSENGER-derived 665 m/pixel elevation raster from Denevi et al., 2018), and hollow 4051 occurs on the crest of the western rim of an approximately 5 km-diameter crater, which gently dips northward at ~0.6° (Fig. 6). Both hollows are near the apexes of their local topography, satisfying the target adjacency condition, and are on near-horizontal surfaces (slopes < ~1.2°) that do not dip toward the equator. From these two hollows, we will set the cold limit of hollow formation at 71°N.

## 3. Results

We have calculated hollow-formation rates along the hot and warm meridians at 1° latitudinal increments at the surface and as a function of depth for 57 different candidate hollow-forming volatile phases (listed in Table 1). Our mass-loss rate calculations are upper limits for pure volatiles that are assumed to be heated to the ambient temperatures of surrounding hermean regolith at the surface and in the subsurface (see *Section*s *2.2.1* and *2.2.2* for details). The assumption of "pure volatiles" means that we ignore possible chemical and physical interactions between volatiles and regolith grains that could inhibit sublimation, such as local increases in vapor pressure or adsorption of volatiles onto regolith grains; however, for volatiles covered by regolith, we explicitly model volatile diffusion through a porous regolith cover (*Section 2.2.2*). For the input temperature profiles that we use, our results should be interpreted as ideal upper limit sublimation rates. We also assume that hollow depth is equal to the amount of volatile substance lost, i.e., the volatile mass-loss rate equals the hollow formation rate. For each volatile phase, we used our model-derived hollow-formation rates to 1) estimate hollow-formation timescales at various latitudes assuming typical hollow depths at those latitudes (see Table 5); 2) estimate the thickness of a regolith lag deposit necessary to suppress hollow formation rates to inactive levels at the hot and cold limits of hollow formation; 3) predict a range of latitudes over which hollows would be observed on Mercury. We compare these model results to observations of hollows on the surface (see *Section 2.3* for an explanation of comparison metrics).

### 3.1 Hollow-Formation Timescales

Figure 7 shows the model-derived hollow-formation timescales at mid-latitude and at the hot and cold limits of hollow formation for each of the 57 phases in our study. Along the hot meridian (Fig. 7 top panel), most substances would form hollows in less than a year even at the cold limit of 71°N. Sulfur would form hollows in ~0.1 and 10 yrs at the hot and cold limits, respectively. Fullerenes would form hollows in ~1 kyr and 100 Myr at the hot and cold limits, respectively. The only substances that do not sublimate quickly enough to explain hollow formation at the cold limit are sulfides. However, the mid-latitude rate for $K_2S$ is the closest match to the characteristic timescale estimate for hollow formation of 300 Myr (Table 5), which was derived from a hollow at mid-latitude (~39°N, Blewett et al., 2016).

Along the warm meridian, most substances would form hollows in less than 100 yrs even at 71°N (Fig. 7 bottom panel). Sulfur would form hollows in ~10 and 100 yrs at the hot and cold limits, respectively. The mid-latitude sublimation timescale for fullerenes is the closest match to the fast timescale estimate of 300 Myr, but fullerenes cannot form hollows in less than 4 Gyr at 71°N. Sulfides cannot form hollows in less than 4 Gyr even at 0°N along the warm meridian.

There are no volatile phases with mid-latitude rates that closely match the characteristic timescale of 300 Myr and yet can also form hollows faster than the minimum necessary timescale of 4 Gyr at the cold limit of 71°N. Therefore, hollows most likely form over a timescale shorter than 300 Myr, consistent with recent results from Wang et al. (2020).

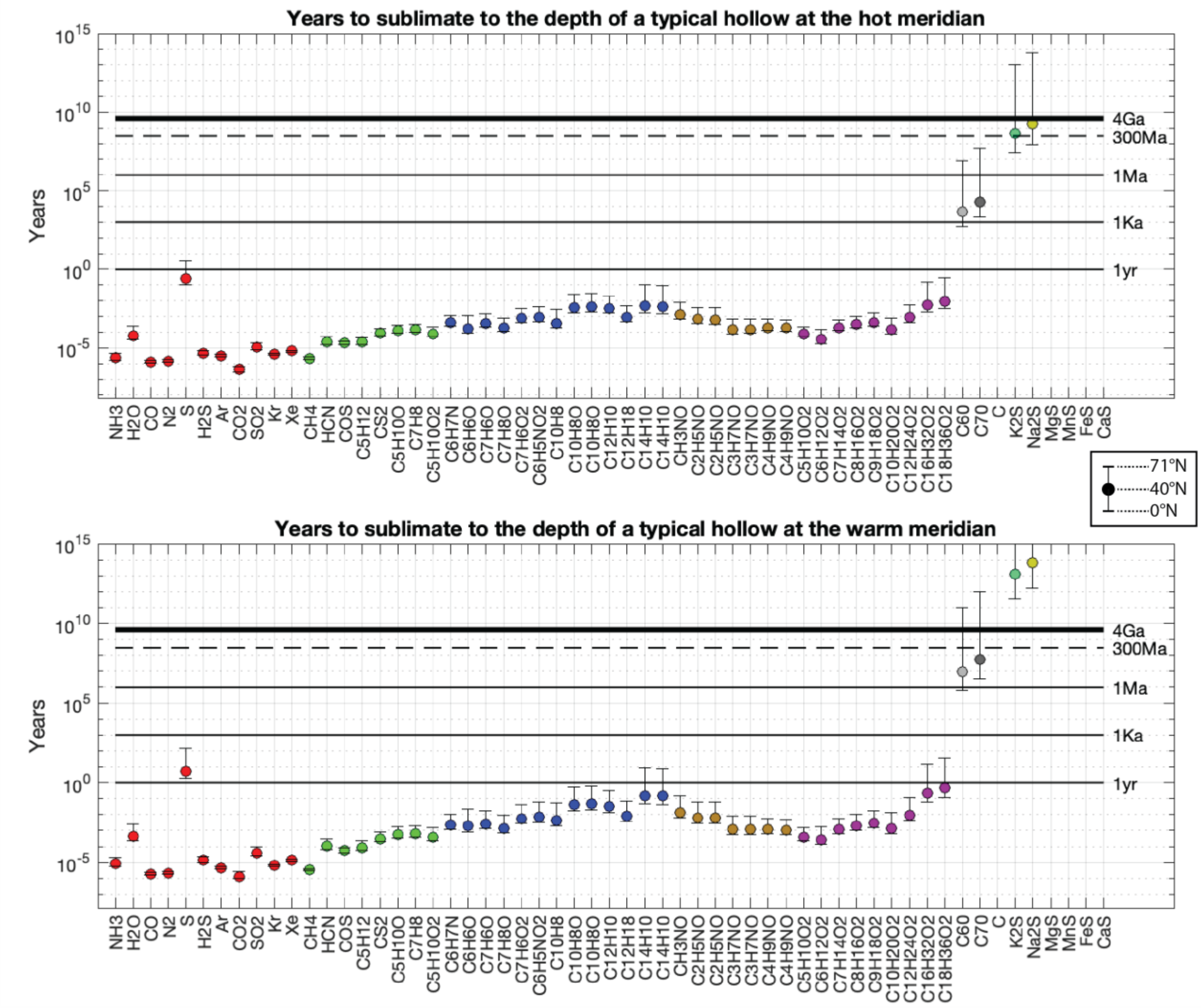

**Figure 7**: The vertical axis shows the time in years it would take to sublimate each substance to the average hollow depth for each latitudinal range (Table 5). Colors correspond to the compound classes in Table 1. Horizontal lines are drawn at 1 yr, 1 kyr, 1 Myr, 300 Myr, 4 Gyr to facilitate interpretation. The estimated timescale for hollow formation (derived from hollows in Balanchine crater, Blewett et al., 2016) corresponds to the dashed line at 300 Myr. The circle markers represent the hollow formation timescale for each volatile phase at latitude 40°N, the upper bar is for 71°N, and the lower bar is for 0°N (see legend). If a candidate hollow-forming phase plots below the dashed line at 40°N along the warm meridian and below the solid line at 4 Gyr along the warm meridian, it is considered a viable candidate. Sulfides and graphite (C) do not meet this requirement. The mid-latitude sublimation timescale for fullerenes are the closest match to the mid-latitude timescale of 300 Myr, but fullerenes cannot form hollows quickly enough at the cold limit of 71°N. Note that graphite and sulfides more refractory than $Na_2S$ plot off the top of the chart; values for these species are provided in Table S3 of the supplementary materials.

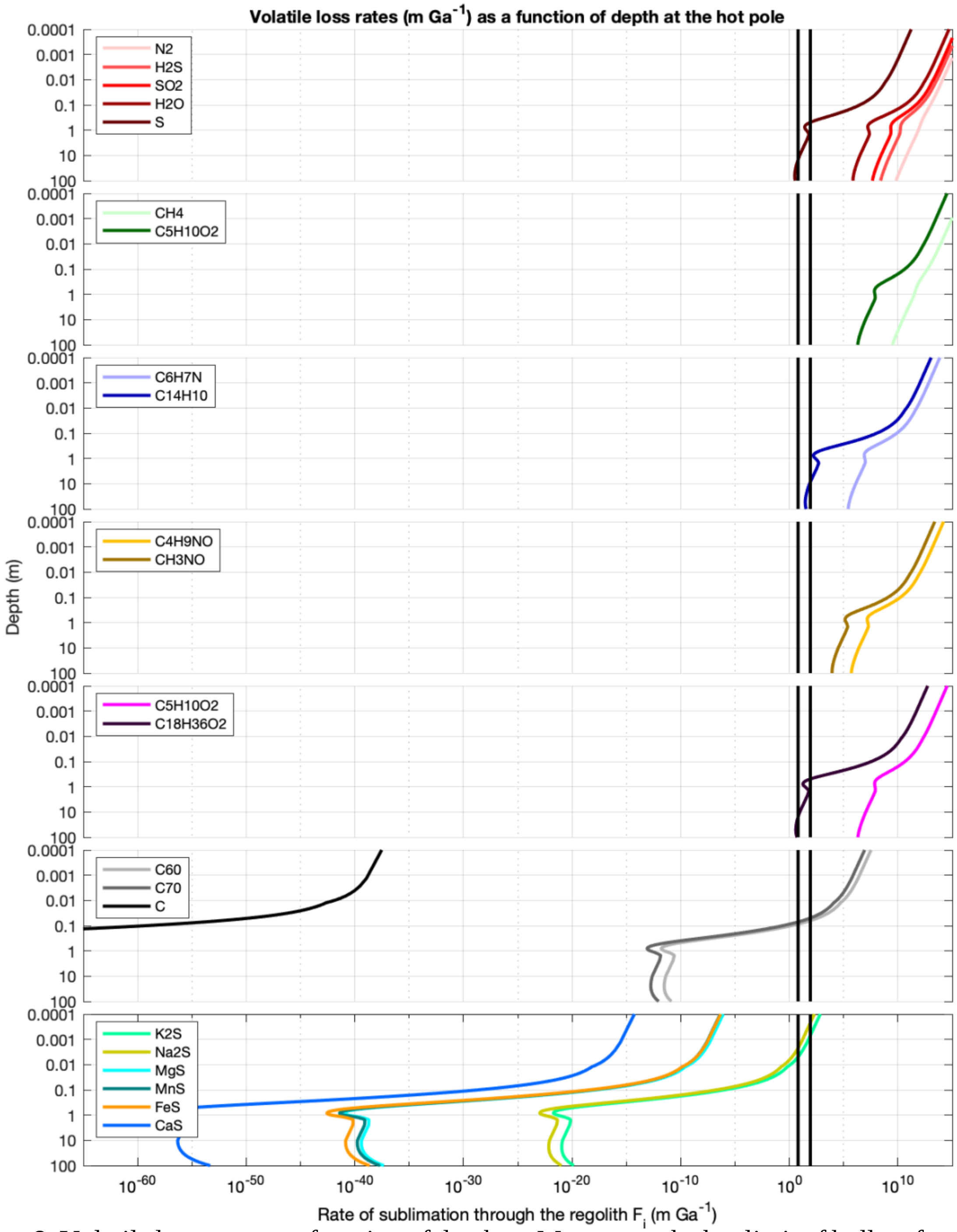

**Figure 8**: Volatile loss rates as a function of depth on Mercury at the hot limit of hollow formation. Colors correspond to the compound classes in Table 1. For simple organics (green), aromatic hydrocarbons (blue), linear amides (yellow), and carboxylic acids (magenta) only the minimum and maximum of the compound classes are plotted for simplicity. Vertical lines are the minimum effective (6.4 m $Gyr^{-1}$) and the characteristic rate (85.6 m $Gyr^{-1}$) for hollow formation at 0°N (Table 5). A viable candidate volatile is expected to plot to the right of the rightmost vertical black line at the surface (depth approaches 0) and, at minimum, to the left of the leftmost vertical black line in the subsurface. Thus, a candidate volatile crosses both vertical black lines.

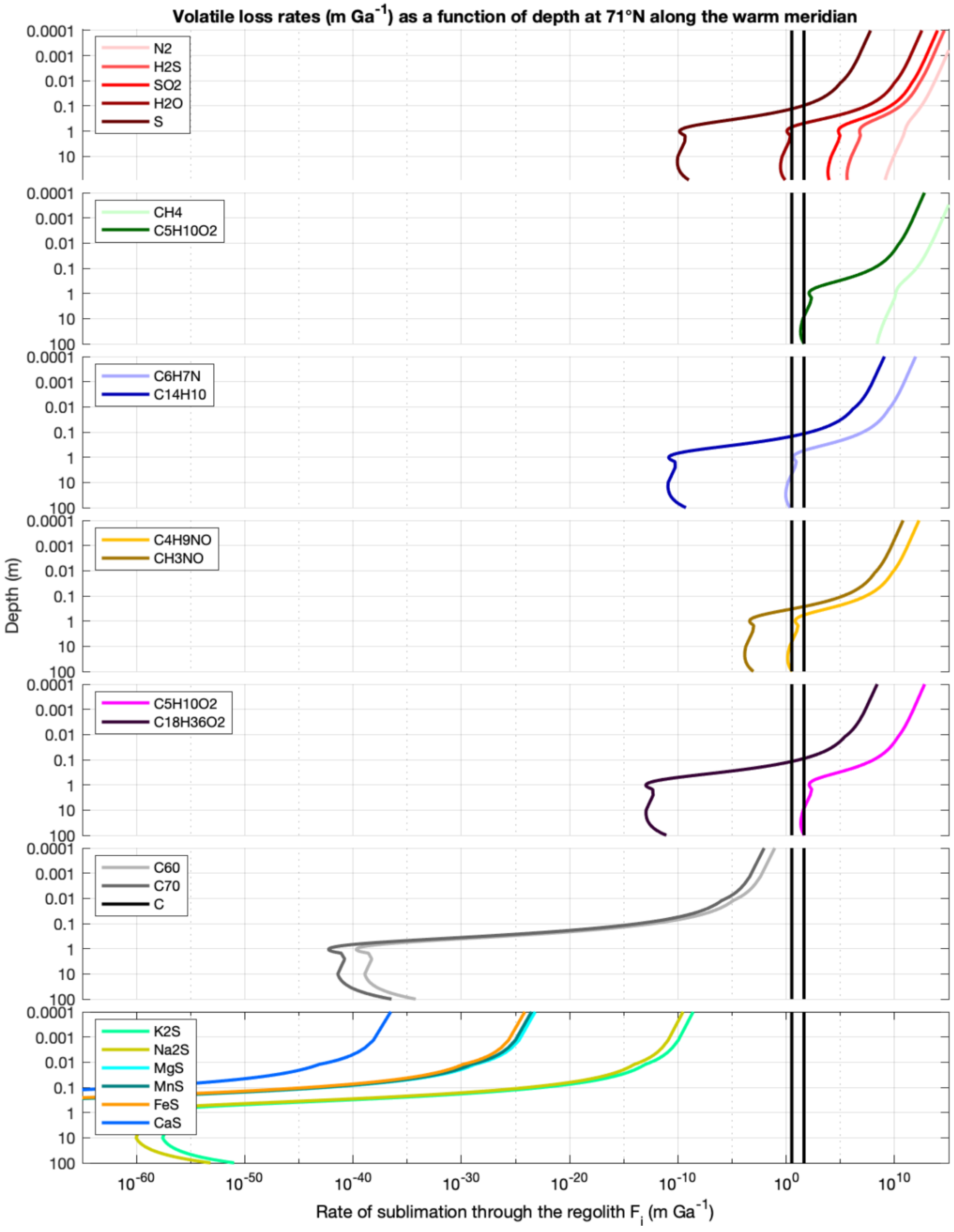

**Figure 9**: Volatile loss rates as a function of depth on Mercury at the cold limit (71°N along the warm meridian). Colors correspond to the compound classes in Table 1. Vertical lines are the minimum effective rate (3.3 m Gyr$^{-1}$) and the characteristic rate (44.2 m Gyr$^{-1}$). In a SEALS hollow formation model, a viable candidate volatile is expected to plot to the right of the rightmost vertical black line at the surface (depth approaches 0) and to the left of the leftmost vertical black line at some point in the subsurface. Thus, a candidate volatile crosses both vertical black lines.

### 3.2 Regolith Lag Deposits

Loss rates as a function of depth (see *Section 2.2.2* for details on calculation methods) were used to calculate the thicknesses of regolith lag deposits necessary to suppress volatile loss to less than relevant threshold values at the hot and cold limits of hollow formation (Figs. 8, 9, and 10). We estimate a range of lag deposit thicknesses for each volatile phase by calculating lag thicknesses at the hot (thickest lag) and cold (thinnest lag) limits of hollow formation. The threshold rate values we use are the minimum effective rate values listed in Table 5 (see *Section 2.3* for how these rates were calculated) because hollow formation that proceeds slower than these rates is considered effectively inactive. Regolith lag deposit thicknesses were calculated as the point in the subsurface where the mass-loss rate is equal to the minimum effective hollow-formation rate value (Table 5). In Figures 8 and 9, this is the depth at which the left vertical black line intersects the mass-loss rate curves. For some volatile phases, the mass-loss rate is never suppressed to below the minimum effective rate value, which indicates that hollow formation would not be suppressed to less than the minimum effective rate value even under a 100 m-thick regolith lag deposit.

In Figure 10, we summarize the volatile phases for which: a) sublimation at the surface is quick enough to account for hollow formation, and b) hollow formation becomes effectively inactive after development of a lag deposit. These are the volatile phases with mass-loss rates that plot to the right of the rightmost vertical black line and to the left of the leftmost vertical black in Figures 8 and 9. If condition a) is not met, then no hollow forms and, therefore, no lag deposit forms. If condition b) is not met, then hollow depth is not limited by development of a lag deposit and, thus, a lag deposit thickness cannot be calculated. Under the assumption that a single volatile phase is responsible for hollow formation, then the volatile phases that satisfy above conditions a) and b) at both the hot and cold limits could be responsible for hollow formation in a SEALS model framework. The two phases for which conditions a) and b) are met at both the hot and cold limits are S and $C_{18}H_{36}O_2$ (i.e., the two phases in both the right and left panels of Fig. 10). If pore-filling volatiles enhance subsurface temperatures in volatile-rich regions (*Section 2.2.2*) then the cold limit of formation would be moved to higher latitudes, perhaps allowing fullerenes to span the full latitudinal range of hollow formation.

Assuming that the depth of a hollow represents the original thickness of the volatile-rich layer minus the thickness of the lag, a percent purity of the original volatile-rich layer can be estimated. For the volatile phases that have a nonzero lag deposit thickness ($L$), we calculated the percent purity of the original volatile layer as: $\%purity = \frac{d_h}{d_h+L} * 100$ for a hollow of depth $d_h$ (see Table 5 for values of $d_h$ used at each latitude). These values are reported in Figure 10. For example, at the hot limit of hollow formation (Fig. 10 left panel), S requires a 15 m-thick lag deposit to suppress its sublimation rate to below the limit of effective inactivity, which would be an original S-bearing layer of ~62% purity. At the cold limit of hollow formation (Fig. 10 right panel), S requires a 0.13 m-thick lag deposit to suppress its sublimation rate to below the limit of effective inactivity, which would be an original S-bearing layer of ~99.5% purity.

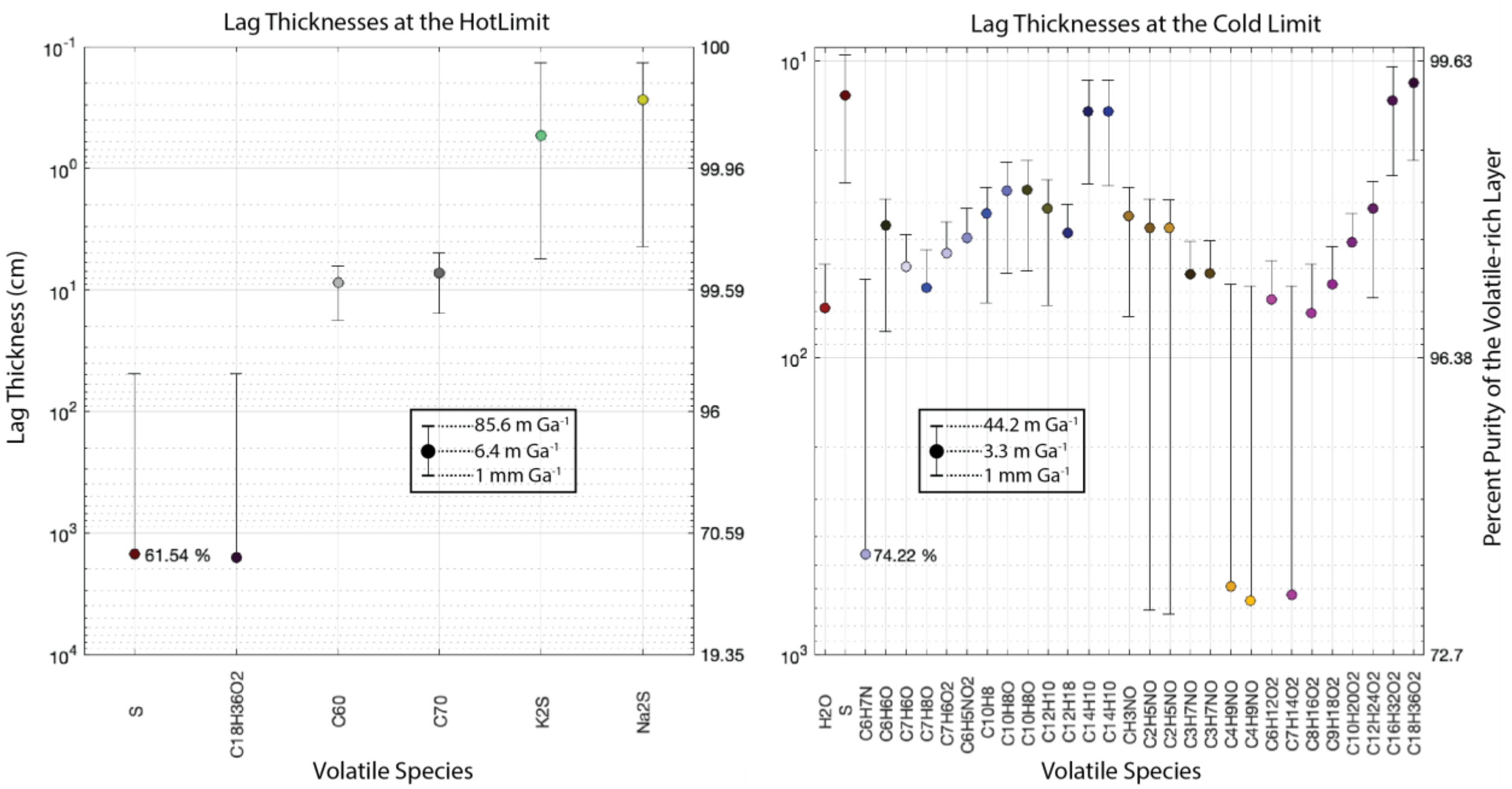

**Figure 10**: Lag thickness at which volatile loss rate is equal to specified threshold loss rate values (see legends for values), which correspond to the vertical lines in Figs. 8 and 9. Thicker lag deposits are necessary to suppress volatile loss rates to lower values. Percent purity of the volatile layer for a given lag thickness is plotted on the right axes. Values without lower bars indicate that a lag deposit 100 m-thick is not sufficient to suppress the loss rate to < 1 mm $Gyr^{-1}$. The two volatiles that are viable at both the hot and cold limits are S and $C_{18}H_{36}O_2$. Volatiles that either do not sublimate fast enough at the surface or cannot be suppressed beneath a volatile lag are not plotted.

### 3.3 Model-Predicted Latitudinal Ranges

In a SEALS hollow formation model, a candidate volatile phase must sublimate quickly enough at the cold limit of hollow formation (see *Section 2*.3 for an explanation of the hot and cold limits of hollow formation) to account for the typical depth of hollows (Table 5); the candidate volatile phase must also be refractory enough at the hot limit of hollow formation that a lag deposit can suppress the loss rate to less than the rate of effective inactivity (minimum effective rate, Table 5). With these two constraints, we derived an expected latitudinal range over which hollows should form for each of the 57 candidate volatile phases in our study. One can think of these latitudinal ranges as representing model-predicted hot and cold limits of hollow formation specific to each volatile phase in our study.

A phase-specific hot limit (low-latitude limit) was calculated as the latitude at which a lag deposit > 100 m-thick would be necessary to suppress the sublimation rate of a phase to less than the rate of effective inactivity (minimum effective rates, Table 5). 100 m was chosen as an extreme upper limit of possible regolith lag deposit thickness. The low latitude limit is represented as the bottom of each colored bar in Figure 11. A phase-specific cold limit (high-latitude limit) was calculated as the latitude at which the hollow-formation timescale became longer than a) 300 Myr (top of the colored portion of the bars in Fig. 11), and b) 4 Gyr (top of the black portion of the bars in Fig. 11). Because both of these timescales are conservative estimates for the temperature profiles we use (see *Sections 2.2 and 2.3*), our phase-specific high-latitude limits are likely lenient, and the true high-latitude limits may be at lower latitudes.

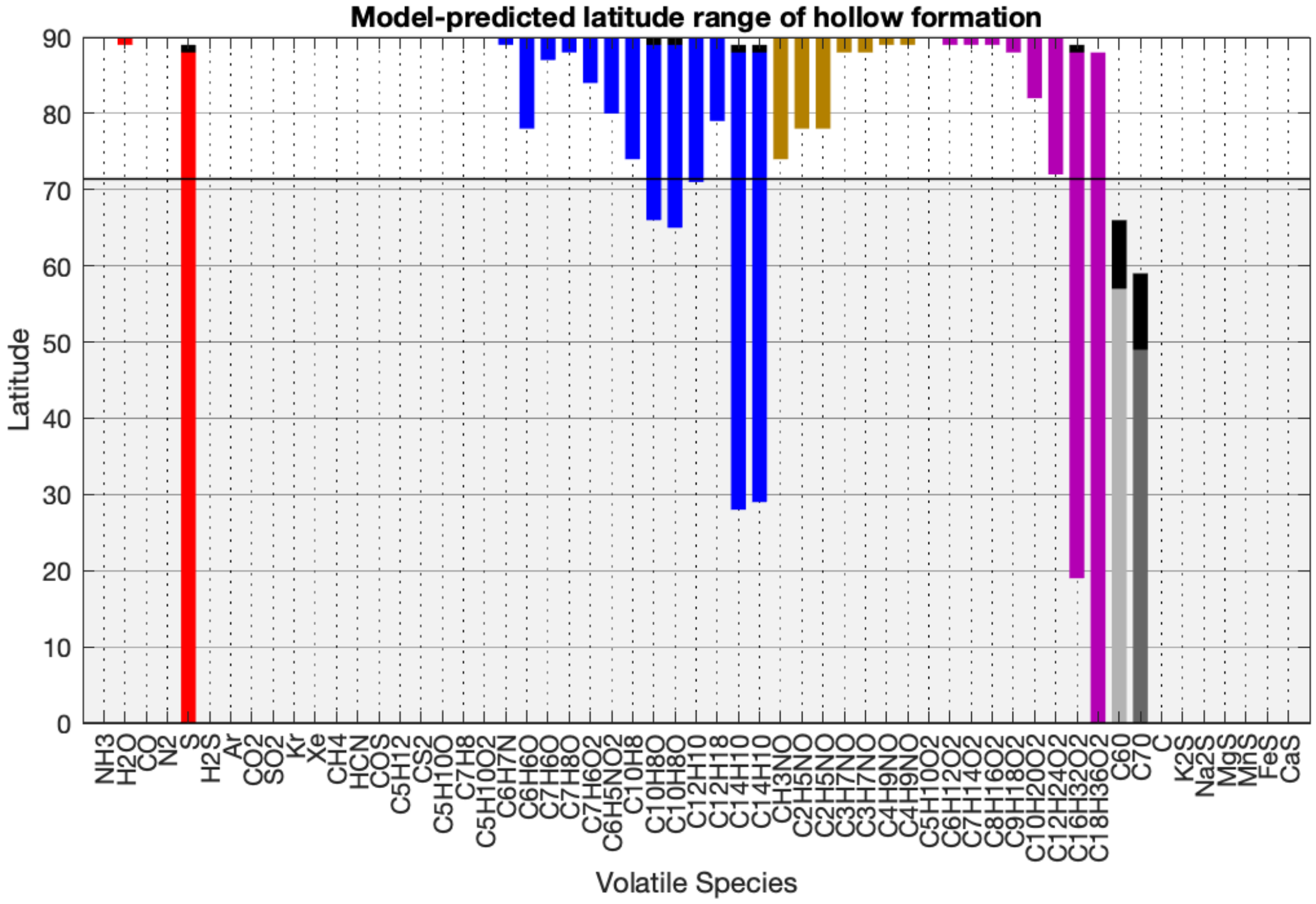

**Figure 11**: Model predicted range of latitudes over which each volatile phase is expected to form hollows on horizontal surfaces in a SEALS model framework. The highest model-predicted latitude depends on the timescale used for hollow formation. If the estimate from Blewett et al. (2016) of 300 Myr is used, then the highest latitude predicted for hollow formation is at the top of the solid colored bars. If a conservative estimate of 4 Gyr is used, then the highest latitude predicted for hollow formation is extended as indicated by the black portion of the colored bars. The light gray region between 0°N (hot limit for hollow formation) and 71.4°N (cold limit for hollow formation) indicates the observed limits of hollow formation on a horizontal surface. See *Section 2.3* for details on the hot and cold limits of hollow formation.

To summarize our results described above, volatile phases that do not match the observed limits of hollow formation are considered less likely to be the hollow-forming phase. For the more-volatile classes of compounds (i.e., inorganics, simple organics, aromatic hydrocarbons, linear amides, carboxylic acids), the hot limit imposes a stricter constraint on their viability as hollow-forming phases than the cold limit. This is because, for many volatile phases, even a 100-m thick regolith lag cannot inhibit hollow growth at low- to mid-latitudes and so the constraint imposed by the hot limit is not met. Conversely, for more-refractory phases (i.e., fullerenes, graphite, sulfides), the stricter constraint on their viability as the hollow-forming phase is the cold limit. Given that our calculated phase-specific high-latitude limits are likely lenient (see above), S and $C_{18}H_{36}O_2$ are the two phases that have the best match with the hot and cold limits of hollow formation derived from observation of hollows on the surface. Fullerenes satisfy the hot limit constraint, but do not satisfy the cold limit constraint even if 4 Gyr is the assumed timescale for hollow formation. Even the most volatile sulfides ($K_2S$ and $Na_2S$) cannot form hollows at 0°N along the warm meridian, and, therefore, do not have a high-latitude limit and do not plot in Figure 11.

## 4. Discussion

Based on our model results, we have narrowed a list of 57 candidate hollow-forming volatile phases down to 3 volatile phases that are viable based on their thermophysical characteristics: S, $C_{18}H_{36}O_2$, and, to a lesser extent, fullerenes. These three volatile phases are the most consistent with the hypothesized behavior of a hollow-forming volatile phase in a SEALS hollow-formation model (as described in our hypothesis in *Section 1.5*). Below we consider broader geological constraints on the viability of these phases as the hollow-forming volatile phase. Specifically, we explore how Stages 1 and 2 of the Blewett et al. (2013) model for hollow formation (see Fig. 3) might be accomplished. Elemental sulfur, S, is discussed because it is, perhaps, the best match to the thermophysical properties expected of a hollow-forming volatile phase as proposed in our hypothesis. We discuss fullerenes because their mid-latitude hollow-formation rate and their expected latitudinal range both agree moderately well with those of observed hollows. We also discuss possible mechanisms for the delivery of fullerenes to and/or the synthesis of fullerenes on the hermean surface in the Pre-Tolstojan Period (~4.6-3.9 Ga, Spudis, 2001; Wagner et al., 2001). Although it is possible that fullerenes may have been generated via space weathering of the hermean graphite crust, we conclude that the volume of fullerenes generated in this way would not have been sufficient to account for hollows. We do not consider $C_{18}H_{36}O_2$ in our discussion because $C_{18}H_{36}O_2$ alone (i.e., without contribution from other comet/asteroid-derived organic compounds) could not have been delivered and sequestered in a large enough volume to be a global hollow-forming layer on Mercury (see Table S1 for reference).

After discussing S and fullerenes, we consider the role sulfides might play in hollow formation. Although sulfides do not exhibit the expected behavior for a hollow-forming volatile in a SEALS model, they have been suggested most frequently in the literature as the species potentially responsible for hollow-formation (Vaughan et al., 2012; Blewett et al., 2013; Helbert et al., 2013; Xiao et al., 2013; Thomas et al., 2014; Thomas et al., 2016; Vilas et al., 2016). Thomas et al. (2014) first suggested that endogenic heat sources may drive volatiles to the surface, contributing to hollow formation. Here, we explore this possibility in detail. Finally, we offer an alternative hollow formation model to SEALS models, wherein thermal decomposition of sulfides in the subsurface via (primarily) impact-related heating drives S-rich fumarolic systems that generate hollows through a diurnal cycle of sublimation and deposition.

### 4.1 Elemental Sulfur

Out of the 57 volatile phases tested in our model, the volatile behavior of S on Mercury is in best agreement with the expected behavior of a hollow-forming phase based on observations of hollows on the surface. If S were the hollow-forming volatile, our model indicates that hollow formation could occur at the observed cold limit at 71°N and potentially as far north as 89°N. Additionally, one would expect a latitudinal dependence on the extent of hollowing, with larger areal extent and deeper hollows closer to the equator, which is consistent with observations (e.g., Fig. 2). Hollows generated via sublimation of S are expected to form in a matter of days (~38 days) at the hot pole to ~145 years at 71°N along the warm meridian (measured in Earth days and Earth years, Fig. 7). These rates are fast compared to previous rate estimates (Blewett et al., 2016; Blewett et al., 2018), but are in keeping with recent results that suggest hollows could form quickly (Wang et al., 2020). Additionally, the rate estimates made by Blewett et al. (2016, 2018) are for lateral growth of hollows, whereas our growth estimates are for downward growth. It could be that an initial rapid phase of downward hollow growth occurs, followed by an extended period of lateral hollow growth by scarp retreat (e.g., Blewett et al., 2018).

#### 4.1.1 Possible Formation of a Near-Globally Distributed, Hollow-Sourcing, S-rich Layer within Mercury's Crust

If S were the phase responsible for hollow formation in a SEALS model, then a near-globally distributed S-rich "layer" on the order of 20 to 110 m-thick must exist within the hermean crust, and specifically the LRM. This S-rich layer may have been emplaced during the Pre-Tolstojan and Tolstojan Periods when voluminous secondary crustal production, and associated eruption of volcanic gases, including S and S-bearing phases) covered the primary crust (i.e., what is observed today, presumably, as LRM) (Denevi et al., 2009; Denevi et al., 2013b; Marchi et al., 2013; Vander Kaaden and McCubbin, 2015; Klima et al., 2018). Our model results suggest that a hollow-sourcing S-rich layer would need to be ~62% pure at equatorial latitudes and ~99% pure at high latitudes (Fig. 10). Sulfur species are a likely product of magmatic activity on Mercury and could have been emplaced through volcanic outgassing and eruptions or magmatic intrusions. Kerber et al. (2009) conservatively estimated the fraction of S in hermean magmas to be 0.05% by weight, and the estimate is much higher, up to 3.68%, for magmas driving pyroclastic deposits. Using 0.05% as an estimate for sulfur in hermean magmas from Kerber et al. (2009) and a lower estimate of 15 km for the thickness of the crust from Sori (2018), a conservative estimate for the volume of sulfur in the hermean crust is 5.6 x $10^{14}$ $m^3$. This volume is equivalent to a ~7.5 m-thick global layer of pure S. If instead we use 3.68% as the estimate for the fraction of S in hermean magmas, then an estimate for the amount of S in the crust of Mercury would correspond to a global layer of pure S ~550 m thick. Thomas et al.'s (2014) nearly global survey of hollows found that hollows cover 5.74 x $10^{10}$ $m^2$ and have an average depth of 47 m. Because hollows have steep sides and flat floors, we can estimate the total volume of hollows, and, therefore, the volume of volatile substance that must have sublimated to generate hollows that are presently observed on the surface, by multiplying hollow area by depth. In this way, we estimate that the total volume of volatiles necessary to be released to form hollows on Mercury is 2.7 x $10^{12}$ $m^3$ (or 1.4 x $10^{12}$ $m^3$ if instead we use 24 m for the average depth of hollows as determined by Blewett et al., 2016; c.f., Wang et al., 2020). These volumes expressed as global layers of pure volatile substance correspond to layers ~1.87 and 3.61 cm thick. Therefore, it seems plausible that enough sulfur was generated via magmatism to account for the volume of material lost to form hollows. A key issue with this estimate for sulfur volume, however, is that most magmatic sulfur will not be in elemental form, but rather in the form $H_2S$ or $SO_2$ (Kerber et al., 2009). Nevertheless, even if < 1% of magmatic sulfur was deposited as elemental sulfur, it would be sufficient to account for the volume of hollows observed. Below we consider possible scenarios for the emplacement of S in the hermean subsurface as a global hollow-forming layer through intrusive magmatism or extrusive magmatism and subsequent burial.

The volume of intrusive magmatism is plausibly higher than extrusive magmatism, and intrusive emplacement of S does not require the extra step of burial. While there are no reliable estimates for the ratio of intrusive to extrusive magmatism (I:E) on Mercury (Solomon, 2018), Earth is estimated to have an I:E ~5:1 for oceanic crust (White et al., 2006) and Mars is estimated to have an I:E ~5-12:1 (Greeley and Schneid, 1991). The I:E value for the Moon is thought to be even higher, perhaps 10-30:1 (Kirk and Stevenson, 1989). It may be that the I:E value is smaller for Mercury because of its dense crust (e.g., Sori, 2018), which would increase the buoyancy differences between magmas and the crust enhancing their ability to rise through the crust. On the other hand, Mercury has also experienced global contraction of its crust that would restrict the rise of magmas to the surface (Byrne et al., 2014). If other rocky bodies in the solar system can be used as a guide, then the majority of magma on Mercury may have been emplaced intrusively (i.e., I:E

> 1). LRM exposed at the surface appears to have been derived from the lower crust/upper mantle (Murchie et al., 2015). Magmatism at the base of the crust may be a mechanism by which magmatic sulfur species are intruded into the LRM. However, an intrusively emplaced hollow-forming layer of S would require extraction of magmatic sulfur from the melt into relatively pure (~63 – 99% pure) layers of S that could then be exhumed by impacts to form hollows according to Stages 1 and 2 of the hollow-formation model of Blewett et al. (2013) (Fig. 3a). Intrusive magmatic sulfur would not likely be emplaced predominantly as S, but rather as sulfides or outgas as $H_2S$ or $SO_2$.

The sulfur content of extrusive magmatism may be considerably higher than intrusive magma bodies because volatile-bearing magmas would be buoyant and preferentially erupt on the surface. Features that have been interpreted as pyroclastic vents provide evidence that volcanism on Mercury can be volatile-rich and even explosive in the fashion of Hawaiian-type explosive eruptions (Kerber et al. 2011; Goudge et al., 2014). Goudge et al. (2014) showed that ~78% of Mercury's pyroclastic deposits are Calorian in age or older (i.e., ~3.3 Ga) as determined by the degradation state of the craters hosting such deposits. Sulfur species are likely contributing gases for these eruptions (e.g., Kerber et al., 2009, 2011), and, per Blewett et al. (2013), may have condensed on the surface during nighttime eruptions. The eruption of S-rich volatiles early in Mercury's history and their subsequent rapid sequestration is also consistent with the apparent global volcanic resurfacing of Mercury ~4 Ga (e.g., Marchi et al., 2013).

In light of the voluminous and S-rich volcanism early on Mercury, it is worth considering how this S might concentrate into thick, nearly global layers. It is possible that volcanically erupted/outgassed sulfur species would sublimate on the sunlit side of the planet and undergo deposition on the nightside of the planet (e.g., Sprague et al., 1995) in a diurnal sublimation/deposition cycle that could potentially concentrate volatile phases into thick layers (e.g., Stages 1 and 2 of Fig. 3a). Similar high-volume, global, volatile transport by means of sublimation and deposition cycles takes place on other planetary bodies, for example on Mars with the seasonal sublimation/deposition of $CO_2$ ice (Hess et al., 1979), on Iapetus imparting its global albedo dichotomy (Mendis and Axford, 1974; Spencer and Denk, 2010), on the uranian satellites affecting asymmetrical distributions of $H_2O$ and $CO_2$ ices (Grundy et al., 2006), as well as on Pluto and Triton perhaps inducing polar wander (Rubincam, 2003; Keane et al., 2016). The distribution of a diurnally induced volatile-rich layer across Mercury would depend on the ancestral spin-orbit resonance state of Mercury. Based on the distribution of large craters on the surface, it has been inferred that the planet was likely in either a 1:1 or 2:1 spin-orbit resonance before being captured into its present 3:2 spin-orbit state, likely at the end of the LHB (Wieczorek et al., 2012; Knibbe and van Westrenen, 2017). If the ancestral spin-orbit resonance of Mercury was synchronous (i.e., 1:1 as suggested by Wieczorek et al., 2012), then one side of the planet would always face the sun and the other hemisphere would be in perpetual darkness. In this state, a hypothetical hollow-sourcing layer would develop on the dark hemisphere whereas no volatiles would be deposited on the sunlit hemisphere; hollows would be expected to have a strong longitudinal dependence in this case, occurring only on one hemisphere. Conversely, in a 2:1 resonance Mercury would have one solar day per year, and a hypothetical volatile-rich layer would develop globally. If a hollow-forming volatile layer developed in this manner, it would imply that Mercury was not in a synchronous orbit for any significant duration and that a 2:1 resonance more likely dominated throughout the Pre-Tolstojan (Knibbe and van Westrenen, 2017).

Using **Eq. 9** and assuming a 2:1 spin-orbit resonance with Mercury's present-day semimajor axis and orbital eccentricity, and a solar flux of 75% to account for the faint young sun (Newman and Rood, 1977), we estimate that S would sublimate at a rate of 120 m $yr^{-1}$ at 0°E, 0°N (the point where solar noon is at perihelion) and a rate of ~4 m $yr^{-1}$ at 180°E, 0°N (the point where solar noon is at aphelion). These rates constitute an upper estimate for the rate at which S could be sublimated from each hemisphere. For the hemisphere that experiences solar noon at aphelion, it would take ~30 years to sublimate a 110 m-thick layer of sulfur. For the hemisphere that experiences solar noon at perihelion it would only take ~1 year to sublimate a sufficiently thick layer of sulfur. S on the hermean nightside is very stable with a nighttime sublimation rate of only $1 \times 10^{-22}$ m $yr^{-1}$. Sulfur-bearing gases would have similar differences in stability between the day- and nightsides of the planet. For example, $H_2S$, which is more volatile than $SO_2$, sublimates at 2600 km $yr^{-1}$ at 180°E, 0°N, and at a hemispheric nighttime average rate of ~0.2 mm $yr^{-1}$. The differences in stability between the day- and nightsides could concentrate S and S-bearing gases (and other volcanic gases) on the nightside of the planet as layers of frost. Some portion of the frost layer would sublimate and then migrate from the sunlit hemisphere to the ephemeral cold trap of the dark hemisphere on a diurnal cycle. Loss of more-volatile S-bearing gases to space may contribute to the purification of the volatile layers into mostly elemental S-bearing. The purity of such a volatile layer would depend on latitude, with more volatile-rich layers deposited at more poleward latitudes. The thicknesses of the frost layers would depend on the eruption rate of volcanic sulfur, which would likely be lower than the daytime sublimation rates, the local topography and temperature of the deposition surfaces, and the probability that a given eruption buries a S-frost layer. To be consistent with the hypothesized hollow-forming volatile layer, S-frost deposits would have to be between ~10 and 110 m thick and ~99% pure at high latitudes (Fig. 10). Due to the complexity of this process, it is difficult to say with the simple calculations performed here how likely it is that a hollow-sourcing S layer could have been deposited in this manner in the Pre-Tolstojan to Tolstojan Periods (~4.6 – 3.8 Ga). The dependence of the process on the eruption rate of S could mean that if S-frost layers developed on Mercury, they may have been covered by eruptions before developing into thick layers, and would therefore be thin relative to the expected thickness of the hypothesized hollow-forming layer, and patchy rather than nearly global. This conceptual model for concentration and sequestration of a S-rich volatile layer through global volatile cycling requires detailed modeling of the physical processes involved to better assess its viability.

Assuming that the aforementioned processes successfully concentrated S-rich layers sufficiently thick to account for hollows on the nightside of the planet, these layers would then have to be sequestered by lava flows, as suggested by Blewett et al. (2013). However, burial by lava flows presents another challenge. If S (and sulfur-bearing gases) are vulnerable to volatilization when heated, would lava flows bury the materials or sublimate them? Using terrestrial lava flows over ice as an analog can be illuminating in this regard. Edwards et al. (2012) describe field observations and measurements of the 2010 Fimmvörðuháls lava flows over ice and snow fields. Their observations reveal that, in most cases, ice/snow is not completely, or in many cases even significantly, melted beneath lava flows. The highest rates of melting were ~1 – 4 m $day^{-1}$ and occurred via heat conduction between the base of fast-moving lava flows in contact with ice. Rates of ~1 – 4 m $day^{-1}$ are broadly consistent with rates calculated by Wilson and Head (2007) for subaerial lava flows over ice. The bases of slower moving lava flows observed at Fimmvörðuháls cooled down relatively quickly and did not cause as much melting (only ~cm $day^{-1}$) as faster moving flows. Ice/snow loss via melting by radiative heating was slow compared to lava-advance rates, and therefore lavas flowed overtop of snow/ice fields rather than melting them as they advanced. In cases where a layer of tephra/breccia covered snow/ice, melting was suppressed. In the case of lava flowing over S-frost layers on Mercury, we might expect even less volatile loss because the latent heat of vaporization for S is 22,120 J mol$^{-1}$ whereas the latent heat of fusion for water is 334 J mol$^{-1}$. However, $H_2S$ and $SO_2$ are more volatile than water, and may experience more loss. Based on the ability of lava to bury deposits of ice on Earth, it seems plausible that this process could have sequestered S-frost deposits on Mercury if such S-frost deposits existed.

Once buried, it is important to note that the volcanic units must be low porosity/permeability, or in other words, they must form an effective capping unit. Otherwise, the buried volatiles would diffuse through the cover and not remain in the crust as the hypothesized hollow-forming layer. Local instances of venting may occur, perhaps producing rootless cones, and some evidence of pyroclastic vents associated with the LRM have been noted (Kerber et al., 2009; Thomas et al., 2014). Venting could drive off more-volatile S-bearing gases, leaving behind relatively pure S deposits. Other considerations exist for the burial and sequestration of sulfur, such as its stratigraphic position and its stability within the subsurface. Such a layer would likely be stratigraphically above the LRM rather than a constituent of the LRM. Regolith and megaregolith forming processes may also lead to removal and mixing of the layer such that it does not maintain the necessary purity required of the hollow-forming layer.

#### 4.1.2 Excavation of a Hollow-Sourcing, S-Rich Layer, by Impact Processes

Another potential hurdle for hollow formation in a SEALS model framework is the mechanism by which the hypothetical volatile layer is exhumed. Exhumation by the impact cratering process is most often the mechanism proposed because hollows are closely associated with impacts craters. If impacts exhume a S-rich layer, it is not clear whether the S would survive the impact process because temperatures would far exceed its volatilization temperature. It could be that the duration over which the S-rich material is exposed to elevated temperatures is short compared to the volatilization rate at those temperatures, and therefore a majority of the S-rich materials would still be emplaced on the crater floor, walls, and ejecta. But, the modeling necessary to test this hypothesis is outside of the scope of this study. Additionally, the distribution of an impact-exhumed S-rich layer within and surrounding a crater may not be consistent with the distributions of hollows observed within and surrounding craters. Specifically, in complex craters, hollows can occur on the central complexes, floors, walls, and rims of a single crater (e.g., Blewett et al., 2011; Thomas et al., 2014). The materials at each of these locations within and surrounding a crater are excavated from different depths (e.g., Croft, 1980), which is potentially inconsistent with excavation of a single S-rich layer. Lastly, the purity of the layer is unlikely to be maintained during the excavation and emplacement process when mixing with other materials is probable.

### 4.2 Fullerenes

The model-predicted sublimation rates, hollow-formation timescales, and latitudinal extent of fullerenes match moderately well with the rates, timescales, and latitudinal extent of hollows as observed on the surface. The expected warm-meridian mid-latitude (40°N) sublimation timescales for fullerenes range from 5 to 90 Myr for C60 and C70, which are within the estimated (near warm-meridian) mid-latitude timescale of 300 Myr from Blewett et al. (2016). The model-predicted latitudinal extent of fullerene-generated hollows in the northern hemisphere (0°N to 66°N, Fig. 11) is slightly less than the observed extent of hollows in the northern hemisphere (0°N to 71°N, Figs. 6, 11), but could be extended if heating in the subsurface were enhanced by pore-filling volatiles (*Sections 2.2.2* and *3.2*).

#### 4.2.1 Production of Fullerenes on Mercury

Fullerenes condense spontaneously in carbon vapors, and the $C_{60}$ molecule has exceptional photochemical and thermal stability (Kroto et al., 1985). As such, fullerenes have been hypothesized as ubiquitous in the stellar medium and in chondrites (Krätschmer et al., 1990; Kroto, 1990); although, fullerenes are typically not as abundant as expected, or even present at all, in meteorites (e.g., Kroto, 1992; De Vries et al., 1993; Becker et al., 1994; Heymann et al., 1995; Luann et al., 1999; Pizzarello et al., 2001). Fullerenes are found in several geologic settings on Earth, perhaps most notably in association with the carbonaceous Karelian minerals of Russia, such as shungite (Buseck et al., 1992; Heymann, 1995; Melezhik et al., 2004).

Fullerenes associated with impacts and the interstellar medium are more relevant to Mercury. The $C_{60}$ and $C_{70}$ molecules were found in material related to the Sudbury impact (Becker and Bada, 1994) and in impact-related material from the Cretaceous-Tertiary boundary (Heymann et al., 1994). Fullerenes have been shown to form during impact processes in an experimental setting as well (Radicati Di Brozolo et al., 1994). If fullerenes were ubiquitous in the early solar system (Buseck et al., 1992; Kroto, 1992; De Vries et al., 1993; Heymann et al., 1995), then it is likely they were delivered to Mercury during accretion and through asteroidal and cometary impacts. However, there has been little evidence for fullerenes in lunar rocks and meteorites (Heymann et al., 2003), so an exogeneous source may be unlikely to contribute significantly to a putative fullerene population on Mercury.

Another scenario in which fullerenes could have come to exist on Mercury is via space weathering of a graphite crust. It has been suggested that Mercury accreted from carbon-rich material (e.g., Ebel et al., 2011) and would have formed a graphitic-primary crust through magma ocean processes (Brown and Elkins-Tanton, 2009; Vander Kaaden and McCubbin, 2015). Petrological modeling by Vander Kaaden and McCubbin (2015) predicted that graphite would be stable as a floatation crust. Other workers have shown that carbon (possibly in the form of graphite or nano- to microphase amorphous carbon) could be responsible for spectral darkening of Mercury's surface, and that this carbon could be sourced from a primary graphite crust (Denevi et al., 2009; Riner et al., 2009; Rothery et al., 2010; Riner, 2012; Murchie et al., 2015; Peplowski et al., 2016; Thomas et al., 2016; Trang et al., 2017).

A global, or near global, primary crust of graphite on the surface of Mercury would have been subject to intense space-weathering (e.g., Cintala, 1992; Domingue et al., 2014) until buried by a secondary volcanic crust (e.g., Marchi et al., 2013) and mixed through impact gardening. Trang et al. (2018) reported space-weathering experiments on graphite powders at the University of Hawaiʻi space-weathering laboratory using a 20 Hz, 1064-nm Nd:YAG pulse laser in which multiple submicroscopic carbon phases and amorphous carbon were produced. However, no explicit mention of the crystalline form of the submicroscopic carbon was made and it is unclear whether fullerenes were produced in this manner. One mechanism for fullerene synthesis in a laboratory setting is through condensation from a carbon vapor generated from a graphite source (e.g., Kroto et al., 1985; Mintmire, 1996; Kozlov et al., 1997). Mercury's graphite crust, combined with its space-weathering environment, may have been a natural laboratory for synthesis of fullerenes on a global scale.

A simple upper estimate of the total volume of fullerenes that could have been produced in the first 200 – 800 Myr of the solar system (when Mercury's graphite crust was likely exposed based on volcanic smooth plains age-estimates, Denevi et al., 2013a; Marchi et al., 2013) can be made using reasonable values for the vaporization rate due to space-weathering at Mercury and the efficiency with which fullerenes can condense from graphite vapor. Cintala (1992) gives a mean rate for space-weathering-induced vaporization of regolith at Mercury of 1.6 x $10^{-14}$ kg $m^{-2}$ $s^{-1}$. If the total surface area of Mercury is presumed as graphite and 20% of vaporized graphite precipitates to fullerenes (a high end of efficiency for laboratory fullerene synthesis using graphite vaporization, e.g., Mintmire, 1996), then an upper estimate of fullerenes produced over an 800 Myr time span would be 6.01 x $10^{12}$ kg. If a constant rate of secondary crustal production covered the graphite primary crust over a 200 Myr time span and a 2% efficiency of fullerene synthesis from carbon vapor is used (Mintmire, 1996), then a lower estimate of 7.51 x $10^{10}$ kg of fullerenes would form. Using the density of C60 (1650 kg $m^{-3}$) and the range of masses calculated above, an estimate for the total volume of fullerenes produced on Mercury is ~4. 6 x $10^7$ to 3.6 x $10^9$ $m^3$. These volumes are equivalent to a global fullerene layer ~ 0.6 to 48 µm thick. If the above estimates of volumes of fullerenes produced on Mercury encompass the actual upper limit of fullerene volume produced by space-weathering of a hermean graphite crust, then fullerenes would be volumetrically insufficient by themselves to account for the total volume of hermean hollows, equal to ~2.7 x $10^{12}$ $m^3$ (*Section 4.1*).

If fullerenes did form globally on Mercury, they would have been a component of the primary crust in which they formed, which was rapidly buried and later exhumed, perhaps as the LRM observed today (Denevi et al., 2009; Murchie et al., 2015). There would, therefore, be a plausible connection between hypothesized fullerenes and the hollow-sourcing LRM.

If fullerenes are involved in hollow formation, they could be present in the hermean exosphere or within permanently shadowed craters at the poles as possible constituents of dark lag deposits covering ice (e.g., Chabot et al., 2014). Fullerenes exhibit UV/VIS spectral features at 213, 257, and 329 nm (Hare et al., 1991; Ehrenfreund and Foing, 1997), which might be detectable with the Ultraviolet and Visible Spectrometer (UVVS) and Visible and Infrared Spectrograph (VIRS) instruments, which are part of the Mercury Atmospheric and Surface Composition Spectrometer (MASCS) system onboard the MESSENGER spacecraft (McClintock and Lankton, 2007). However, fullerene absorptions features are too narrow (e.g., Ehrenfreund and Foing, 1997) to be observed in published spectra of hollowed terrain (e.g., Thomas et al., 2016; Trang et al., 2017). Fullerenes also display weak IR spectral features near 7, 8.5, 17 and 19 µm (Krätschmer et al., 1990), which may be detectable with the Mercury Radiometer and Thermal Infrared Spectrometer (MERTIS) on the BepiColombo spacecraft (Hiesinger and Helbert, 2010). Detection of fullerene spectral features in the exosphere and/or in permanently shadowed regions at the poles could be considered indirect evidence that fullerenes play a role in hollow formation.

### 4.3 Sulfides

Sulfides are present, and relatively abundant on the surface of Mercury (Sprague et al., 1995; Boynton et al., 2007; Denevi et al., 2009; Zolotov et al., 2013; Goudge et al., 2014; Murchie et al., 2015). Sulfides have recently been considered as products of a hermean magma ocean (Parman et al., 2016; Boukaré et al., 2019) and have been considered as candidate hollow-forming materials by several workers (Blewett et al., 2013; Helbert et al., 2013; Thomas et al., 2014; Thomas et al., 2016; Vilas et al., 2016). Sulfides may have formed in relatively pure layers in the upper mantle during the early Mercury magma ocean, or they may have risen to the surface as a sulfide lid (Parman et al., 2016) or sulfide-rich plumes (Boukaré et al., 2019). The thickness of the sulfide lid has an expected lower limit of ~5 km, which would comprise a volume of ~3.7 x $10^{17}$ $m^3$. This volume is orders of magnitudes more than necessary to account for the total volume of hollows (*Section 4.1*). There is, therefore, a plausible mechanism to emplace a sufficient volume of sulfide-rich deposits subjacent to, or as a constituent of, the LRM.

The species of sulfides that are suggested to have formed from the Mercury magma ocean are FeS, MnS CaS, MgS, and $Na_2S$. However, MnS and FeS are expected in smaller proportions due to the relative scarcity of Fe and Mn on Mercury, and $Na_2S$ is expected to crystallize out of the magma after CaS and MgS and in smaller proportions (Boukaré et al., 2019). The elements U and Th will likely partition into CaS and MgS, and K will partition into $Na_2S$ (Boukaré et al., 2019). Although pure $K_2S$ is not expected to form, it is useful to consider it as an "extreme endmember" of K substitution into $Na_2S$.

Our model results suggest that most sulfides are stable on the hermean surface, even at the hot pole, and can persist for billions of years without substantial loss (*Section 3*, Fig. 7). For the most volatile sulfide, $K_2S$, the predicted time necessary to form a typical hollow at the hot pole is ~25 Myr, increasing to ~230 Myr at 40°N, and would form over a time period much longer than the age of Mercury at 71°N. The times necessary to form a typical hollow are similar for $Na_2S$, taking ~86 Myr at the hot pole, ~887 Myr at 40°N, and orders of magnitude longer than the age of Mercury at 71°N. At the warm pole, the rate of reactive sublimation, even for $K_2S$, is not fast enough to account for hollow formation. Our results, therefore, show that sulfides cannot be the hollow-forming volatile in a SEALS model framework.

#### 4.3.1 Impact-Induced "Hermeothermal" Systems

If sulfides are responsible for hollows, then solar heating alone is not sufficient to account for hollow formation and additional heat sources are necessary to drive hollowing. In addition to solar heating, some combination of local increases in temperature caused by impact cratering or subsurface magma bodies, lava flows, faulting, or other sources may provide the heat necessary to thermally decompose sulfides. Volatile phases generated through subsurface heating of sulfides could be driven to the surface, forming hollows in the process. Non-crater related hollows are commonly associated with lava flows and pyroclastic deposits (Xiao et al., 2013; Thomas et al., 2014; Goudge et al., 2014). For example, hollows at higher latitudes (> 50°N) northwest of the Caloris basin are associated with a lava flow front, perhaps indicating a genetic link between the two as suggested by Thomas et al. (2014). Almost all hollowed surface area is associated with impact craters that could provide the heat necessary to decompose sulfides. The local temperature increase in the vicinity of an impact can persist for $10^5$ – $10^6$ years (e.g., Kring, 2000) depending on the size of the impact and the amount of melt produced. Elevated post-impact temperatures may enable the majority of hollowing to occur within the first several thousands of years after the impact when the local temperature, especially under the central complex and crater floor, is elevated (Kring, 2000; Osinski et al., 2001; Abramov and Kring, 2004).

At present, the only impact for which subsurface temperature has been modeled as a function of time after impact is the Caloris basin-forming impact (e.g., Potter and Head, 2017). Although no models presently exist for the evolution of temperature with time beneath median-sized (~15 km diameter) impacts on Mercury, we can compare to models for the Earth, Moon, and Mars. Abramov et al. (2012) calculated, for oblique impacts, how melt volume scales with impactor size, velocity, and planetary target body (Earth, Moon, and Mars). While they do not include Mercury in their evaluation, it can be inferred that Mercury would experience more melting per impact than any of the three bodies included in their study because: 1) impact velocity strongly controls melt volume production (second only to impactor diameter), and 2) the thermal environment of the target rock significantly affects the total melt volume (Abramov et al., 2012). For both factors Mercury has an environment conducive to the production of melt because impact velocities are high at Mercury due to its proximity to the Sun (Marchi et al., 2005; Domingue et al., 2014) and subsurface temperatures at Mercury are hot relative to the Earth, Moon, and Mars. For impacts on Earth that produce craters with a final diameter of ~12 km, Abramov et al. (2012) predict ~1 $km^3$ of melt would be produced (their Fig. 4). If this volume is considered a minimum estimate for a similar-sized crater on Mercury, then such an impact would generate > 1 $km^3$ of melt. Best estimates for melt produced by the Sudbury impact (~180 km diameter), for example, are 8,000 – 10,000 $km^3$ (Grieve and Cintala, 1992; Wu et al., 1995). A typical melting temperature for basalt can be taken as ~1400 K (Frost and Frost, 2014). Such a volume of melted rock would keep surrounding areas in the subsurface at elevated temperatures for > $10^4$ years and drive volatilization from the bottom up (e.g., Rathbun and Squyres, 2002; Abramov and Kring, 2004). Hydrothermal springs around the Sudbury impact, for example, were driven in this manner, and hydrothermal systems on Mars were likely powered by impact heat as well (Rathbun and Squyres, 2002; Schwenzer and Kring, 2009; Ruff et al., 2011).

Similar to hydrothermal systems on Earth and Mars that are driven by impact-generated heat, "hydrothermal-like" systems could be driven by impact-generated heat and melt on Mercury. A more general term for hydrothermal systems that is agnostic to the surface expression of the system and the fluids circulating within them is "geothermal system" or "areothermal system" (for Earth and Mars, respectively). We will call analogous systems on Mercury "hermeothermal systems" to avoid the implication that water is involved in these systems and to avoid invoking particular surface features that are commonly associated with geothermal systems, such as constructional vents, sinter deposits, hot springs, outwash aprons, etc. The fluids in the case of hermeothermal systems would be different from geo- and areothermal fluids because the starting compositions (including presence of water) and oxygen fugacities are significantly different on Mercury (e.g., McCubbin et al., 2017). Sulfides within LRM source material may decompose and deliver S and S-bearing species to the surface. Because of the sulfur-rich composition and gas-dominated rather than liquid-dominated nature of hermeothermal systems, the surface expression would likely be most similar to terrestrial solfataras (e.g., Allard et al., 1991; Caliro et al., 2007), albeit with many differences remaining between them. In the proposed impact-generated hermeothermal systems, sulfur and sulfur-bearing phases would undergo deposition on the surface in the immediate vicinity of the vent sites, accumulating within the regolith at night, and would sublimate during the day. This diurnal pattern of volatile saturation at night and sublimation during the day at fumarole vent locations combined with the volume lost from thermally decomposing material in the subsurface may drive hollow formation (see *Section 4.4*).

Could sulfides decompose and source sulfur-bearing volatiles (e.g., S, $H_2S$, $SO_2$) in an elevated post-impact temperature environment rapidly enough to explain hollow formation? To address this question, we calculated the thermal decomposition rates of sulfides ($K_2$-, $Na_2$-, Mg-, and Ca-sulfides) as a function of temperature from 600 to 1500 K at the surface (Fig. 12, solid lines) and at 1 km depth (Fig. 12, dashed lines). At the surface, even CaS sublimates quickly enough to account for hollow formation at temperatures above ~1100 K. If impact-exhumed LRM is emplaced at high temperatures, the more volatile sulfides (i.e., $Na_2S$ with possible K substitution) will reactively sublimate quickly relative to expected hollow formation rates. $Na_2S$ will sublimate meters per day at temperatures above ~1230 K (Fig. 12). If sulfides were heated conductively by a subsurface impact melt body, our model results indicate that $K_2S$ and $Na_2S$ decompose rapidly enough to account for hollow formation at temperatures above ~950 K. MgS and CaS require much longer to decompose when buried, even at basalt melting temperatures. However, our assumption about porosity (i.e., $\phi = 1 - \frac{\bar{\rho}(z)}{G_h \rho_w}$) likely does not hold because fractures may allow volatiles to travel to the surface more easily. In this regard, our calculations are conservative estimates. Nevertheless, $Na_2S$ (with possible K substitution) will more readily decompose directly from the surface after an impact and will more readily deliver sulfur-bearing volatiles to the surface through decomposition in the subsurface than more refractory Mg- and Ca-sulfides.

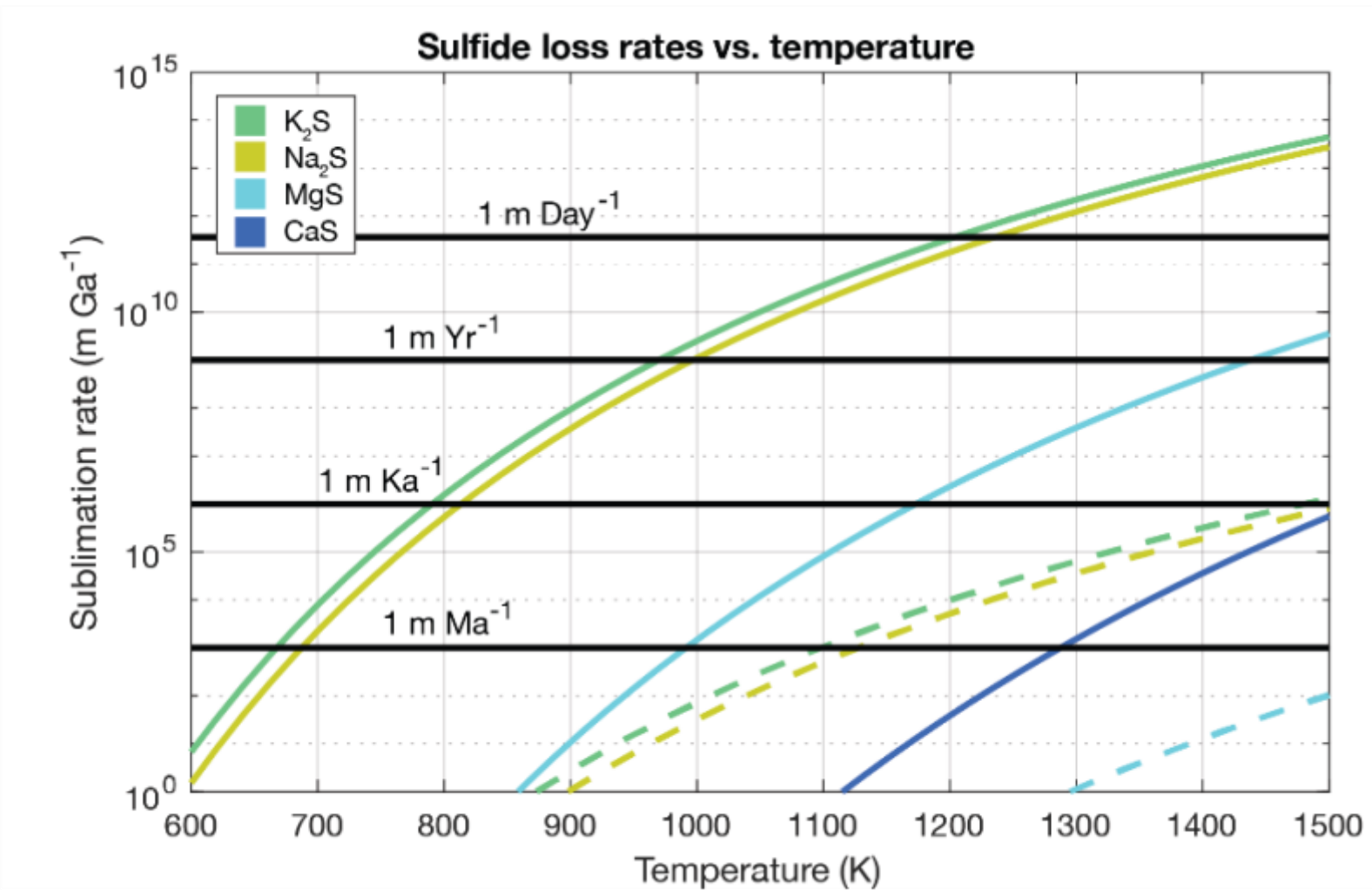

**Figure 12:** Sulfide loss rates as a function of temperature between 600 K and 1500 K. Dashed lines are loss rates for sulfides buried at 1 km. At temperatures relevant to a post-impact environment, $Na_2S$ can be thermally decomposed rapidly at the surface, and above the estimated hollow formation rate at temperatures above ~1100 K buried at 1 km. CaS buried at 1 km falls off the plot.

If sulfides contribute to hollow formation, then it could be a source for exospheric Na, K, and Mg (Potter and Morgan, 1985). The reactive sublimation of sulfides within hot LRM on the surface immediately following an impact or thermal decomposition of sulfides within the subsurface and delivery to the surface in a vent system could produce Na, K, and Mg ions that could adsorb to the surface to be launched into the exosphere via solar wind particles, photons, or micrometeorite impacts (Killen et al., 2010; Mangano et al., 2013). Na is lost from the hermean exosphere and some surface mechanism replenishes it (Potter and Morgan, 1985); thermal decomposition of sulfides associated with hollow-formation could be a mechanism for replenishment. S would be produced in this manner as well and should be observable in the hermean exosphere depending on the age of hollows and the expected lifetime of S in the exosphere. A S absorption feature centered at 1813.7 Å would be observable with the ESA BepiColombo PHEBUS UV instrument (see, e.g., Killen et al., 2007).

### 4.4 A Novel Model for Hollow Formation: Sublimation Cycling Around Fumarole Systems (SCArFS)

Sulfides have been the phase suggested most often in the literature as possibly responsible for hollow formation, but they do not have the appropriate thermophysical characteristics to be responsible for hollow formation in a SEALS model framework in which solar heating is the principal heat source that drives hollow formation. Although the model-predicted latitudinal extent of hollows formed through sublimation of elemental S agrees well with the observed latitudinal extent of hollows, the mechanism by which a global layer of S up to 110 m-thick and 62% to 99% pure could come to exist within Mercury's crust presents several challenges. These challenges include the production of enough volume of S, the concentration of S into thick, nearly pure, layers, the sequestration of those layers through burial by lava flows, and the survival of that layer through the impact exhumation process (*Section 4.1*). While the volatile behaviors of fullerenes match moderately well with those expected from a hollow-forming volatile in a SEALS model, their synthesis on and/or delivery to Mercury and subsequent sequestration within the crust at high enough volume to be responsible for hollows may be unlikely (*Section 4.2.1*).

Any viable alternative model for hollow formation must account for certain key observables, which include: 1) the close association of hollows with impact craters and the distribution of hollows within and surrounding both simple and complex impact craters, including their occurrence on the steep slopes of central peak structures, crater walls, and the Caloris knobs; 2) the association of hollows with LRM; 3) the possibility for some hollows to form outside the vicinity of impact craters; 4) the trend of decreasing hollow areal extent and depth with increasing latitude; 5) the morphology of hollows (i.e., their flat floors, steep walls, amorphous shape, and depth); 6) the apparent youth of hollows.

Because of the challenges inherent within a SEALS hollow-formation model, we propose an alternative model for hollow formation (Fig. 13). Our hollow-formation model involves deposition of S and S-bearing phases, such as chalcogenides[1] and sulfosalts[2], in the vicinity of fumarole vents at night and sublimation during the day in a diurnal cycle. Therefore, we call our model the Sublimation Cycling Around Fumarole Systems model for hollow formation (SCArFS). The SCArFS hollow-formation model is summarized in the following four stages. It should be noted that the events described in Stage 1 would occur throughout the hollow-forming process (i.e., throughout Stages 2 and 3) until the processes that terminate (or slow) hollow formation, described in Stage 4, take place.

*Stage 1: Embedded heat sources produce volatile phases, such as S, $S_2$, and $H_2S$, through thermal decomposition of sulfides within LRM. Sulfur-bearing phases are driven along fractures to the surface in hermean fumarolic systems (Fig. 13, Stage 1).* Because of the association of hollows with impact craters, we propose that impact melt and related heating are the principal heat sources that would drive thermal decomposition reactions. It is possible that other mechanisms, in addition to thermal decomposition, could liberate S, such as oxidation reactions between sulfides and impact-generated melt (e.g., Weider et al., 2016). It is important to note that other heat sources could drive volatile production in a SCArFS model, including subsurface magma bodies that intrude sulfide-rich LRM (e.g., Xiao et al., 2013; Weider et al., 2016) and heating related to faults (Thomas et al., 2014). Based on the apparent abundance of sulfides within LRM and the higher volatility of $Na_2S$ compared to MgS, CaS, and graphite, our results suggest that the fundamental species involved in Stage 1 of the SCArFS model would be $Na_2S$ (with possible K substitution), but other sulfides are expected to contribute to a lesser degree as well. The volume lost through the destruction of sulfides in the subsurface would, ultimately, be accommodated at the surface as a hollow according to the processes described in Stages 2 and 3.

The thermal decomposition of sulfides would produce gaseous sulfur at the surface immediately following an impact, as well as in the subsurface for longer time periods feeding fumarolic systems. Hot LRM emplaced on the surface as ejecta may produce incipient hollows as sulfides reactively sublimate from the surface. The products of sulfide thermal decomposition in the subsurface would likely react with other constituents of the subsurface and regolith during their ascent, producing sulfur allotropes, $H_2S$ or other sulfur-bearing gases, minerals such as chalcogenides and sulfosalts, as well as altering the composition of subsurface materials. Additionally, it is possible that methane and other C-bearing phases may form by subsurface

[1] Chalcogenides are minerals comprised of a chalcogen (group 16 elements, typically excluding oxygen) and an electropositive element, such as the alkali, alkaline earth, or transition metals. Common chalcogenides include pyrite and galena.

[2] Sulfosalts are minerals with the general formula $A_mB_nS_p$, where A is a metal, B is a metalloid, and S is sulfur (or, rarely, a heavier chalcogen).

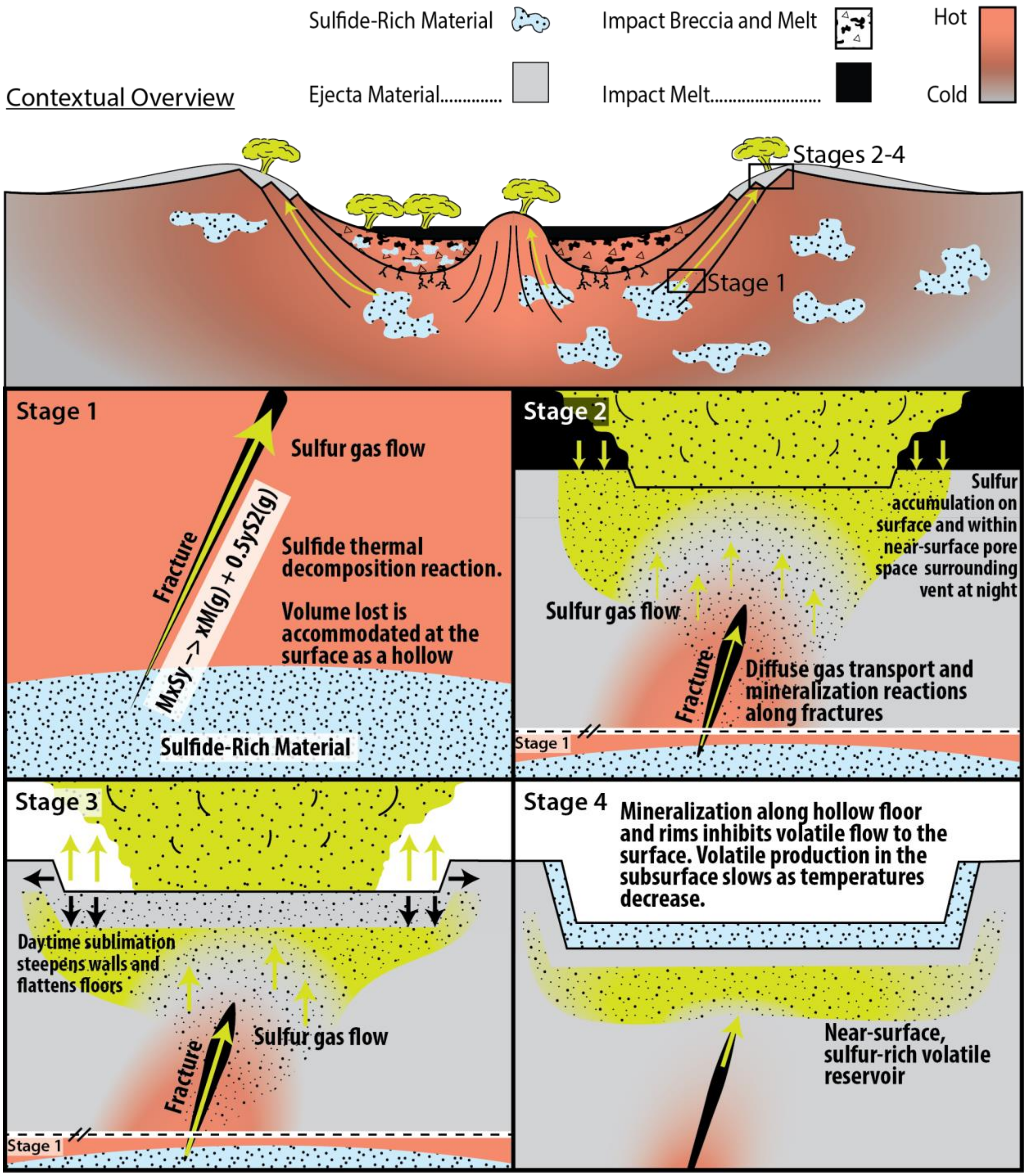

**Figure 13**: Sublimation Cycling Around Fumarole Systems (SCArFS) model for hollow formation. Top panel shows a complex crater in the first several thousand years after impact into sulfide-bearing LRM and gives context for Stages 1-4 of the model. *Stage 1:* Heating of LRM produces volatile phases, for instance by thermal decomposition of sulfides. Volatile phases are driven toward the surface along fractures in a fumarole system. The processes of Stage 1 occur throughout hollow formation until hollow formation ceases in Stage 4. To show this, a small panel

**(Fig. 13 cont.)** displaying the processes of Stage 1 is located at the bottom of the Stage 2 and 3 panels beneath a dashed line that indicates a break in space. *Stage 2:* Sulfur and other phases accumulate in the vicinity of fumarole vents on the surface and in near-surface pore spaces at night, becoming structurally integral members of the surface in a “sulfur-rich permafrost” layer. Some amount of gases, especially highly volatile species such as $H_2S$, would likely continue to vent at night. *Stage 3:* Phases that precipitated on the surface and within the near-surface at night sublimate during the day, widening and deepening hollows. Stages 2 and 3 would repeat on a diurnal cycle. *Stage 4:* Fresh sulfides may precipitate along hollow floors, walls, and rims developing a low porosity capping unit that impedes hollow growth. Hollow formation may also slow when the subsurface heat source cools and thermal decomposition of sulfides terminates. Retention of a S-rich reservoir is anticipated below the influence of diurnal temperature swings and above the influence of subsurface heating.

heating of carbon-bearing material within the LRM (especially if the LRM is H-rich from solar wind saturation, per Blewett et al., 2016; and McCubbin et al., 2017) and contribute to the volatile phases in Stage 1 of the SCArFS hollow-formation model.

*Stage 2: Volatile phases accumulate within surface materials in the vicinity of fumarole vents at night where S and S-bearing minerals precipitate (Fig. 13, Stage 2).* The extremely low nighttime temperatures on Mercury would generate a zone of thermal stability for S extending from the surface into the subsurface. This zone can be thought of as analogous to terrestrial permafrost zones (regions in the subsurface where water ice is thermally stable and that can either host water ice or not). The hermean “permafrost zones” we suggest are with respect to S, not water, and would be ephemeral on a diurnal cycle. Solid S would accumulate on the surface and within near-surface material within the “permafrost zone” at night. The result would be a sulfur-rich surface/near-surface layer containing solid S and, likely, other species.

In the SCArFS model, crystallization of S and other species would have to disrupt the surface layer in order for hollows to form, which could occur by several mechanisms. One such mechanisms would be if the crystalizing phases became structurally integral components of the near-surface layer. This could occur if precipitating volatile species saturate pore spaces and displace other constituents of the near-surface layer. That is, if the total volume of the crystalizing phases is higher than the locally available pore-space volume, then disruption of the pre-existing near-surface layer is expected. Such a process may result in a nightly “inflation” of the surface surrounding a vent as the volume in the area is increased by the crystalizing species.

Some highly volatile phases may continue to vent at night, such as $H_2S$, but deposition of phases would advance as the night progresses and the surface and near-surface grow colder. Elemental S and S-bearing minerals would preferentially accumulate in the near-surface over other, more-volatile, S-bearing phases (e.g., $H_2S$, $SO_2$). A gradient in volatile concentration in the near-surface radiating outward from vent locations might be expected because gas fluxes would be highest near vent sites.

The distribution of fumarole vents, and therefore hollows, would depend on the subsurface distribution of sulfide-rich materials and the connection of those materials to the surface via fracture networks (see e.g., the upper panel contextual overview of Fig. 13). Because hollows do not occur everywhere that there are impacts in the LRM, it is likely that the distribution of sulfide-rich material within the LRM is patchy. Sulfide-rich patches may be distributed within the LRM on a scale smaller than the size of a hollow-hosting impact crater (as shown in Fig. 13), or on a scale larger than the hollow-hosting crater. If sulfide-rich patches are small relative to the impact, hollows may only be expected in isolated sections of the crater where fractures intersect a sulfide-rich region. If sulfide-rich areas are large compared to the size of the impact crater, then most fractures might be expected to intersect sulfide-rich material and, therefore, anywhere fracture networks intersect the surface might be expected to host hollows. In either case, in a SCArFS model for hollow formation, hollows are predominantly expected to form where fractures vent gases produced from sulfide decomposition to the surface; for simple craters this corresponds to the rims (French, 1998), and for complex craters this corresponds to central peaks, peak rings, terraced walls and rims (Melosh, 1989; French, 1998). It is possible that sulfide-rich materials could be entrained in the impact melt and breccia lens of a complex crater (Fig. 13, contextual overview); in this case, extensive venting, and therefore hollowing, would also be expected on the crater floor, as is observed in, for example, Tyagaraja and Sander craters.

*Stage 3: Solar heating drives sublimation weathering during the day that reworks the surface and results in steep walls and flat floors characteristic of hollows (Fig. 13, Stage 3).* In the SCArFS model, a subsurface heat source drives volatile phases to the surface/near-surface where they accumulate (Stage 2), but solar insolation is the heat source that drives sublimation of the accumulated phases from the surface/near-surface. The nightly process of volatile deposition through upward transport of fumarolic volatiles recharges the "sulfur permafrost zone" with new material that sublimates during the day furthering hollow growth in a diurnal cycle. Recharging of the surface is expected while the fumarole system is active, which would depend on the longevity of the heat source. The surface would be reworked by sublimation weathering and other physical and chemical weathering processes. We suggest that processes applicable to thermokarst landscape development (Czudek and Demek, 1970; Kargel, 2013) and in salt weathering (e.g., Melosh, 2011 Ch. 7) may operate in hollow formation. Analogy to permafrost landscapes (molards) has also been drawn to explain the morphology of circum-Calrois knobs on Mercury (Wright et al., 2020), some of which contain hollows. We propose that in the SCArFS model, solid phase S and S-bearing minerals become structurally integral members of the surface (Stage 2), as is the case with terrestrial ice-rich permafrost, and that removal of these phases results in collapse, flat floors, and steep walls, analogous to *alas* formation in thermokarst landscapes (Czudek and Demek, 1970). In the SCArFS model, broad areas of the surface would synchronously deflate, whereas other models propose a central downward-growing focal point followed by lateral growth (e.g., Blewett et al., 2018). Just how "broad" the area is that deflates would depend on the flux of volatiles at a specific vent location, the architecture of the vent and its subsurface plumbing, and the diffuse nature of volatile phase emplacement.

Because our model results favor elemental S as the hollow-forming phase (*Sections 3* and *4.1*), we suggest that elemental S would have been the predominant phase that precipitated on and within the surface surrounding fumarole vents at night and subsequently sublimated during the day to generate hollows. However, the thermal stability of other phases, such as chalcogenides and sulfosalts, should be explored in the future. Where solar insolation is not sufficient to sublimate S (e.g., at high latitudes and in shadowed regions), hollows would not be expected to form because there would be negligible net loss of material from the area. However, S-rich regions on the surface and near-surface would be expected in locations around high-latitude craters where hollows are typically found at lower latitudes. Similarly, S is expected to be relatively stable at depths where the amplitude of surface diurnal temperature swings is sufficiently muted. The depths at which S is relatively stable would vary with both latitude and longitude on Mercury (see Figs. 8 – 10, *Section 3.2*), but typical depths of a S-rich reservoir would be decimeters near the cold limit of hollow formation to ~15 m near the hot limit of hollow formation (Fig. 13 Stages 3 and 4). The expectation of a relatively stable, S-rich volatile reservoir, below the reach of solar heating and above the influence of subsurface heating, is potentially key to the maintenance of fresh hollow surfaces in the SCArFS model (c.f., *Section 4.4.1*).

*Stage 4: Hollow growth ceases (or becomes nearly dormant) when one of three conditions are met: i) the heat source responsible for volatile-phase production from the LRM cools down and, therefore, volatile-phase production stops; ii) the sulfide-rich source is exhausted and no more volatile phases are produced; iii) mineralization along fractures and of the surface creates a capping unit that precludes continued upward transport of volatiles to the surface (Fig. 13, Stage 4).* The first two conditions involve lowering the volatile production rate to a sufficiently low value that hollow formation can be considered effectively inactive. It could be that above conditions *i* and *ii* are never achieved for some hollows. In such cases, slow and continued hollow growth may be happening today. The third condition requires mineralization of pathways to the surface and of the surface itself with phases that do not sublimate at hermean daytime temperatures (e.g., sulfides). As the pore spaces of hollow floors (and walls) mineralize with refractory phases, rising volatiles would take paths of least resistance to the surface in a process of vent relocation. The vent relocation process could give rise to the irregular shape of hollows because relocation would happen in a quasi-random manner. Clusters of hollows may also form in this way if the vent relocates outside the extent of the original hollow. This process could also lead to hollow clusters aggregating into larger hollows. If the development of a mineralized capping unit inhibits hollow formation, then downward hollow growth would be moderated from the bottom up in a SCArFS model, whereas in a SEALS model hollow growth is terminated from the top down by development of a lag deposit. We suggest that the minerals that act as the "capping" unit may likely be sulfides that (re-)precipitated in the fumarole systems; sulfide precipitation is known to happen in impact-generated geothermal systems (e.g., Grieve, 2005).

Fresh sulfide precipitation along hollow rims and floors may contribute to the relatively bright floors and halos characteristic of some hollows. Recent spectral analyses of hollows on Mercury and sulfides in the lab indicate that sulfides are present in hollow floors and in their bright halos (Helbert et al., 2013; Thomas et al., 2016; Vilas et al., 2016; Lucchetti et al., 2018; Pajola et al., 2021). MgS has an absorption centered at ~0.62 µm but CaS is spectrally featureless (Helbert et al., 2013). Analyses of $Na_2S$ spectra show that this mineral has a broad absorption similar to that of MgS, but centered at ~0.56 µm (Gradie and Moses, 1984). Although the 0.62-µm absorption of MgS is consistent with some MESSENGER 8-band WAC spectra of the wall and rim hollows of Dominici crater, other spectra display a shorter wavelength absorption feature more consistent with the spectrum of $Na_2S$ (Vilas et al., 2016). Additionally, hollow floors and rims are smooth and/or fine-grained in texture as determined by phase-ratio images (Blewett et al., 2014). It might be expected that fumarole-emplaced minerals would be fine-grained.

In a SCArFS model, it is likely that the majority of hollow formation would occur shortly after the impact ($<10^3$ – $10^6$ yrs, depending on how long the subsurface stays hot). However, hollows would appear young through processes that freshen hollow floors and rims over longer timescales, which we describe in further detail in *Section 4.4.1.*

#### 4.4.1 Reasons to Favor a SCArFS Hollow-Formation Model

*1) Distribution of hollows in and around impacts.* In the SCArFS model, hollows are expected to be predominantly associated with impact craters because they provide the heat and fracture networks necessary to drive hermeothermal systems. The locations of hollows in simple and complex craters agrees well with the expected locations of fumarole vents in both simple and complex craters (see *Section 4.4,* Stage 2). In simple craters, hollows typically are found along crater rims and are found less often on crater floors (Thomas et al., 2014), which would be consistent with where gases are expected to vent at the surface in simple craters (e.g., French, 1998; Oehler and Etiope, 2017). In complex craters, the preference of hollows to form on terraced walls, central peaks, and central peak rings (Thomas et al., 2014) is consistent with where fractures are expected to vent the subsurface and with where hydrothermal systems have been preserved in terrestrial impact structures (Ames, 1999; Osinski et al., 2001; Osinski et al., 2013). Conversely, in a model for hollow formation where a volatile-rich hollow-forming source material is exhumed by impacts, hollows would be expected anywhere along the impact structure that corresponds to the depth of excavation of the volatile-rich material instead of predominantly along central complexes and terraced walls.

*2) Association of hollows with LRM.* A SCArFS model does not require a near-globally distributed layer of hollow-forming volatiles, as does the model suggested by Blewett et al. (2013). Instead, hollow-forming volatiles are produced from the sulfide-rich portions of LRM through thermal decomposition or other interactions (e.g., oxidation reactions) with impact melt or magma bodies. We suggest that the production of S (and S-bearing phases) through impact-generated hermeothermal systems is a more direct mechanism for the involvement of S in hollow formation than the concentration and sequestration of near-globally distributed S-rich layers as described in *Section 4.1.1* (see also Blewett et al. 2013, 2016). Our model results indicate that temperatures achieved adjacent to magma bodies or impact generated melt could decompose $Na_2S$ at a rate sufficient to generate hollows, even at a depth of 1 km (Fig. 12). MgS and CaS would decompose at lower rates, but could still contribute to hollow formation, especially immediately following an impact when temperatures are highest.

*3) Hollows not associated with impacts.* There are few examples of hollows separate from an impact heat source. Hollows that do occur apart from impact craters, such as hollows northwest of Caloris basin, are spatially associated with lava flows, pyroclastic vents, red pitted terrain, and/or faults, which implies an association with subsurface heat (Thomas et al., 2014). Fumarole systems could have been stimulated by dikes and sills related to the lava flows, pyroclastic vents, and/or faults, that intruded subsurface sulfide-rich LRM and generated hollow-forming fumarole systems. Another heat source for the decomposition of sulfides could be radiogenic decay of K, Th, and U, which are expected to strongly partition into sulfides. K is expected to partition into $Na_2S$ and Th and U are expected to partition into MgS and CaS in Mercury's magma ocean (Boukaré et al., 2019). Hollows within small dark patches known as "dark spots" are observed in a variety of terrains and have been suggested to form via high velocity outgassing events (Xiao et al., 2013). Radiogenic self-heating of sulfides and their subsequent decomposition may play a role in small isolated hollows, such as dark-spot hollows.

*4) Decreasing area and depth of hollows with increasing latitude.* Hollow formation in the SCArFS model requires sublimation from solar heating, otherwise minerals would precipitate on and within the surface, perhaps with some vent formation as is typical in fumarolic systems, but hollows would not form because there would be negligible net loss of material from the system. The dependence of hollow formation on solar heating would impart the observed trend of decreasing depth and areal extent with increasing latitude.

*5) Hollow morphology.* The steep, rimless walls, flat floors, and amorphous shapes of hollows that are indicative of formation by sublimation (Blewett et al., 2011) are accounted for in a SCArFS hollow-formation model because sublimation is the principal mechanism for forming hollows at the surface (*Section 4.4, Stage 3*). We draw analogy to development of features in thermokarst landscapes, such as alases (Czudek and Demek, 1970). In the SCArFS model, hollows form primarily through downward growth over a broad area. Vent relocation (described above in *Stage 4*) could allow clusters of hollows to form, hollows to widen, and adjacent hollows to merge. Additionally, the likely diffuse nature of volatile emplacement surrounding vents (see *Stage 2* above) could contribute to widening of hollows through slow-acting scarp retreat as volatiles more slowly accumulate further from the vent. Other models for hollow formation suggest that most hollow growth happens laterally through scarp retreat after an initial downward phase of hollow growth (see, e.g., Blewett et al., 2018).

*6) The apparent youth of hollows.* In the SCArFS model, the bulk of hollowing is expected to occur shortly after the hollow-hosting impact while the subsurface of the impact site is still hot. Thus, the apparent youth of hollows does not obviously follow within the SCArFS model. We propose two mechanisms, which are natural extensions of processes proposed within the SCArFS model, that could maintain the fresh appearance of hollows. In this way, the lifecycle of hollows in a SCArFS model has two primary phases: 1) a primary phase initiated by introduction of a (typically impact-generated) heat source in which the majority of hollow growth occurs followed by; 2) a maintenance phase of slower hollow growth that is long-lived and possibly currently active.

The first mechanism that could support a long-lived is gradual loss of S from a S-rich, near-surface, reservoir retained beneath hollows (see Fig. 13, Stages 3 and 4). A region where S is relatively stable would exist beneath hollows at depths of decimeters to decameters, depending on hollow location (see *Section 4.4*, *Stage 3)*. The reservoirs would be shallower at colder locations and deeper at warmer locations. Mass loss rates of S from such reservoirs would also depend on hollow location, ranging from 3.3 m $Gyr^{-1}$ from a 13 cm-deep reservoir at the cold limit of hollow formation to 6.4 m $Gyr^{-1}$ from a 15 m-deep reservoir at the hot limit of hollow formation (see Figs. 8 – 10, *Section 3.2*).

The second mechanism we propose for maintaining the youthful appearance of hollows within the SCArFS model is slow, but continual, hollow growth if conditions *i – iii* of Stage 4 of the SCArFS model are not (completely) achieved. Continued transport of volatiles to the surface throughout the tail-end of impact-induced heating and by the hermeothermal gradient within the megaregolith may be expected. Temperatures within the upper ~1 – 5 km of the hermean crust are influenced by megaregolith properties and have been estimated at ~500 – 720 K (Mohit et al., 2009; Egea-González and Ruiz, 2014). Such temperatures are not sufficient to account for the full depth of hollows in a SCArFS model but may be sufficient to impart a veneer of youth to hollows. Volatiles produced at depth would have an effective path to the surface provided by the fumarole vent systems, and hollow formation could proceed (per Stages 1 – 3 of the SCArFS model) at a rate on the order of ~1 m $Gyr^{-1}$ (c.f., Fig. 12) – so long as the volatile source is not depleted. This second mechanism would also augment the first mechanism through replenishment of the S-rich reservoir beneath hollows.

Any mechanism that maintains a fresh appearance for hollows must compete with mechanisms that degrade hollows, such as space weathering and macro-scale erosional processes. The bright floors and halos of hollows might be expected to diminish over a similar time frame as the bright rays of rayed craters (≤ ~270 Ma, Xiao et al., 2012). If this is the case, then the mechanisms proposed here to maintain a fresh appearance to the surface hollows would act sufficiently quickly (freshening 1 to 6 mm of surface every Myr) to combat surface darkening by space weathering. Erasure of decameter-scale topography is unlikely to occur through surficial space weathering processes, but hollows would be affected by macro-scale erosional processes, such as small impactors that cause "topographic diffusion" (e.g., Fassett and Thomson, 2014). Topographic diffusion is the process of progressive erasure of topography via small impactors (Soderblom, 1970; Hirabayashi et al., 2018). On the Moon, a characteristic erosion rate for features > ~500 m in extent (appropriate for hollows) over the past 3 Ga has been estimated at between 0.2 m $Gyr^{-1}$ (Craddock and Howard, 2000) and 0.4 m $Gyr^{-1}$ (Fassett and Thomson, 2014). Fassett et al. (2017) conservatively estimated that the rate of erosion due to topographic diffusion is a factor of 2 higher on Mercury than on the Moon. Because impact flux scales with distance to the sun (R) as $\frac{1}{R^2}$, the rate of topographic diffusion for Mercury could be as much as 6.7 times that of the Moon's. Therefore, we estimate a characteristic erosion rate for hollows at between 0.4 and 2.68 m $Gyr^{-1}$. These rate estimates are slower than our estimated freshening rates for hollows, so it is plausible that hollows could maintain a fresh appearance through geologic time on Mercury via the mechanisms proposed here. It is interesting to note that our upper estimate for erosion by topographic diffusion is not sufficient to completely erase a typical 24 m-deep hollow over a 4 Ga period, even in the absence of mechanisms to refresh hollows. The process of erosion via topographic diffusion is scale-dependent, however, and smaller features are, counterintuitively, expected to experience *lower* rates of erosion via topographic diffusion than larger features (Minton et al., 2019). More rigorous estimates of hollow erosion rates should be made, as well as more detailed modeling of the SCArFS processes to test whether these estimates hold under more detailed scrutiny.

In the SCArFS model, hollow formation could eventually cease if all volatile sources are depleted. Therefore, two endmember classes of hollows are predicted: **1) "degraded" hollows** that no longer have a volatile source, are morphologically and topographically muted, and are dark; and **2) "fresh" hollows** that have a volatile source (either actively sourced from the subsurface or passively sourced from a remnant, near-surface, S-rich reservoir), are morphologically crisp, and are bright. Both classes of hollows would begin their growth directly after introduction of a heat source, typically impact-generated. If this is true, one might expect older craters to have a relatively higher proportion of "degraded" to "fresh" hollows than younger craters because there would have been more time to deplete the volatile source. Thomas et al. (2014) looked at the percentage of crater floor surface area covered by hollows as a function of crater degradation state as a proxy for age (Barnouin et al., 2012). Thomas et al. (2014) found that younger craters have a higher percentage of their floors covered by hollows than do older craters. One interpretation for this trend is that erosional processes mute dormant hollows, obscuring their observation (Thomas et al., 2014). This explanation would be consistent with the SCArFS model in which "degraded" and "fresh" hollows are expected.

*7) Some other considerations*. The SCArFS model for hollow formation does not require space weathering processes to account for hollow formation. We do not favor models that require space weathering processes because they act on only the uppermost portion of the surface and downward growth would likely be self-limiting after the buildup of a thin lag deposit. Impact gardening rates are slow (Killen et al., 2007; Domingue et al., 2014; Costello et al., 2019) compared to estimated hollow formation rates (Blewett et al., 2016), and as such, regolith turnover could not replenish volatiles to the surface quickly enough to account for hollow formation. However, the role that space weathering processes may play in the surface appearance of hollows should be explored because space weathering will likely affect the lifetime and spectral properties of sulfides and other materials present on hollow floors and rims (e.g., Domingue et al., 2014). Additionally, Ca, Mg, K, and Na in Mercury's exosphere have a conceivable link to hollows (c.f., Bennett et al., 2016). Reactive sublimation of sulfides at the surface immediately following an impact and/or precipitation of fresh sulfides in fumarolic systems and subsequent space weathering could be replenishing sources for these exospheric species.

Lastly, the SCArFS model can account for hollows in all their settings. Other models for hollow formation can plausibly account for hollows in certain settings, but have difficulty explaining all instances of hollows (see, for example, Section 12.6 of Blewett et al., 2018). We suggest that it is more parsimonious for a single model to explain all instances of hollows rather than adopting several setting-specific explanations for hollow-formation.

### 4.5 Future Work

In this paper, we showed that elemental S is a viable hollow-forming volatile phase. However, more work is needed to test the hollow-formation model proposed by Blewett et al. (2013) and the SCArFS model proposed here. In the hollow-formation model proposed by Blewett et al. (2013), a near-globally distributed S-rich layer is necessary; whether such a S-rich layer could have been emplaced within the hermean crust is unknown. We discuss a conceptual model for how a hollow-sourcing layer of S-rich material could have been emplaced in *Section 4.1.1*, but the detailed modeling necessary to test the viability of this conceptual model is outside the scope of this work. If a S-rich hollow-forming layer exists within the hermean crust, then, according to the model of Blewett et al. (2013), hollows across Mercury are formed from material excavated from this layer. Therefore, calculating the depth of excavation associated with hollows both globally across Mercury and locally at individual craters could provide a useful test of the validity of that model. Thomas et al. (2014) presented a preliminary analysis of the depth of excavation of LRM and concluded that LRM is not globally present at a specific depth. However, a detailed study bracketing the range of excavation depths of hollow-sourcing material may prove useful. For example, the presence of a S-rich layer would be supported if hollows could be shown to be sourced from a subsection of the several kilometers-thick LRM.

To test the SCArFS model for hollow formation, more modeling and observational work should be done. We suggest several phases (most importantly, S) that could be present in hermean fumarolic systems (*Sections 4.3.1* and *4.4*), but detailed modeling of the phases that would be produced by heating, melting, and shocking sulfide-rich LRM should be done. The relative volume of volatile phases produced by thermal decomposition of sulfides versus alteration from heating other phases (e.g., C) in LRM is unconstrained. Other mechanisms for producing gases in a hermeothermal system could include degassing of magmas and of impact-melted LRM, as well as oxidation of sulfides. The degree and type of mineralization along fractures and in pore spaces is also unknown. We suggest sulfides may (re-)precipitate from the proposed fumarolic systems, but the mineral species that may form in such systems would be dependent on the fluid composition, the temperatures, and oxygen fugacities involved and may involve other sulfur-bearing species, such as chalcogenides and sulfosalts. The types of minerals deposited would likely evolve over time, as the temperatures surrounding the impact evolve and the fluid composition changes, and with latitude because the solar insolation differences would drive off different proportions of S and other mineral species. A detailed model of the fluid phases produced and circulated after an impact into sulfide-rich LRM on Mercury (similar to models of impact-generated hydrothermal systems on Earth and Mars, Rathbun and Squyres, 2002; Abramov and Kring, 2004) would be useful in addressing many of these questions.

Statistical analyses of the distribution of hollows within craters would prove useful for distinguishing between expected distributions of hollows in a SEALS model versus a SCArFS model for hollow formation. For example, in a SCArFS model, hollows are expected to be predominantly, but not exclusively, associated with terraced walls and central complexes; whereas in a SEALS model, hollows are expected wherever hollow-sourcing material is exhumed by the impact. Although our preliminary inspection suggests that hollows are predominantly associated with terraced walls and central complexes of complex craters (Fig. S7), this has not been rigorously shown through a statistical analysis.

Detailed observations of hollows are planned with the BepiColombo mission (Rothery et al., 2020). More detailed geomorphological analyses of hollow shapes, depths, and floor structures could be useful in (in)validating the SCArFS model. As part of the Spectrometer and Imagers for Mercury Planetary Orbiter (MPO) BepiColombo Integrated Observatory SYStem (SIMBIO-SYS, Poulet et al., 2015; Cremonese et al., 2020), the BepiColombo mission will image > 20% of the hermean surface with its high-resolution imager (HRIC, Poulet et al., 2015; Cremonese et al., 2020) at a ground sampling distance of 6 m at 480 km altitude (periherm). Targeted images of hollows may reveal that small vents pockmark hollow floors. Constructional vent features may not have formed if most of the phases that were present in the hermeothermal systems (e.g., S) sublimated away. However, some constructional vent features composed, perhaps, of sulfides may be expected. We suggest (in *Section 4.4.1*) that the SCArFS model predicts two endmember classes of hollows: “degraded” and “fresh”. Evidence for “degraded” hollows, while scarce, does exist (Blewett et al., 2018). If the SCArFS model accurately describes hollow formation, then a higher number of “degraded” hollows likely exist than have presently been observed. Therefore, a prediction from the SCArFS model is that many “degraded” hollows are yet undiscovered on Mercury. More complete high-resolution imaging of Mercury may reveal evidence for “degraded” hollows and allow for classification of hollows into categories of morphological degradation state, similar to that which exists for hermean craters (Barnouin et al., 2012; Kinczyk et al., 2020). Comparison between hollow degradation state and host crater degradation state might help address some of the questions concerning hollow age, timescales over which hollows are active, hollow modification and degradation rates and processes, and the potential for a “lag-time” between impact crater formation and hollow formation (which would be evidence against the SCArFS model). Additionally, the SIMBIO-SYS suite will have a moderate-resolution visible and near-infrared hyperspectral imager (VIHI) with a spectral range of 400 – 2000 nm, and with a spectral resolution of 6.25 nm/pixel. Hyperspectral images of hollow rims and floors could confirm the presence of sulfides within hollows (Helbert et al., 2013; Thomas et al., 2016; Lucchetti et al., 2018; Pajola et al., 2021) and the Mercury Imaging X-Ray Spectrometer (MIXS) will map S abundance across the planet (Bunce et al., 2020). Lastly, if the pore spaces of some sections of hollow floors and rims are mineralized, these units may have a higher thermal inertia than surrounding materials, which could be measured with the MERTIS instrument on BepiColombo (Hiesinger et al., 2020).

### 4.6 Possible Implications

If hollow formation is linked to hermeothermal systems, then the study of hollows may present a unique opportunity for comparative planetology between geothermal, aerothermal, and hermeothermal systems and their various surface expressions (e.g., hot springs, geysers, mud pots, fumaroles, etc.). On Earth and Mars, erosion and weathering can erase evidence of and preclude observation of deposits and landforms created within geothermal and areothermal systems. Although some processes, such as space weathering and mass wasting, could erase hollows over time, hollows are likely a more accurate representation of the full and relatively pristine extent of a hermeothermal system than are the surface deposits and landforms left by geothermal and areothermal systems. Therefore, studying hollow morphology, depth, and distribution within and surrounding craters could lead to a fuller understanding of the extent of thermal systems on Earth and Mars where orbit-based evidence of (extinct) thermal systems can be more difficult to detect. Informing the search for extinct thermal systems on Mars is of specific importance because these sites are of astrobiological interest (e.g., Schmidt et al., 2009; Ruff et al., 2011; Squyres et al., 2012; Ruff and Farmer, 2016).

## 5. Conclusions

We used a thermal model to calculate the loss rate of 57 candidate hollow-forming volatile phases as a function of depth, time, and location on Mercury. We found that only three volatile phases have the appropriate thermophysical characteristics to be plausible hollow-forming volatiles: stearic acid ($C_{18}H_{36}O_2$), fullerenes ($C_{60,70}$) , and elemental sulfur (S). Notably, sulfides are shown to be too refractory to be the phase responsible for hollow formation if solar heating drives hollow formation. We summarize our conclusions about the plausibility of $C_{18}H_{36}O_2$, fullerenes, and S as the hollow-forming phase below:

- $C_{18}H_{36}O_2$ is rejected as the volatile phase responsible for hollow formation on the basis that it could not have been delivered to Mercury and sequestered in sufficient volume to account for the observed volume of hollows.
- Fullerenes may have been generated through space weathering of a graphite crust in the Pre-Tolstojan Period, but it is unlikely that a sufficient volume of fullerenes was created in this way to account for hollows (*Section 4.2.1*). If fullerenes were generated on Mercury, it is possible that they contribute to dark lag deposits in permanently shadowed regions at the poles.
- S is the most viable candidate for the hollow-forming phase based on its thermophysical characteristics and on its relative abundance on Mercury.

If S is the hollow-forming volatile phase, then within the hollow-formation model framework proposed by Blewett et al. (2013) two conditions must be met: 1) a hollow-sourcing layer rich in S must exist within the hermean crust, and 2) this S-rich layer must survive the impact-excavation process and be emplaced on crater central complexes, floors, walls, and rims in a manner consistent with the observed distribution of hollows around impacts. In *Section 4.1.1*, we discuss the possible formation of a near-globally distributed, S-rich, hollow-sourcing layer through a diurnal sublimation/deposition cycle during global volcanic resurfacing of Mercury in the Pre-Tolstojan Period. We suggest that this mechanism cannot be ruled out, but that more detailed modeling is required to assess the viability of this conceptual model (see also *Section 4.5*). In *Section 4.1.2*, we discuss the possibility that a S-rich layer could survive the impact-excavation process and be distributed around hollows in a manner consistent with the observed distribution of hollows around impacts. We suggest that the distributions of hollows observed within and surrounding individual complex craters indicates that, if hollows form from excavated material, this material is likely sourced from a range of depths, which is potentially inconsistent with excavation from a single S-rich layer. However, more work should be done to calculate the depths of excavation for hollow-forming material (*Section 4.5*).

Because of the complications involved with the emplacement and excavation of a S-rich hollow-sourcing layer (*Section 4.1*), we propose an alternative hollow-formation model (*Section 4.4*). We suggest that S may have been produced through the thermal decomposition of sulfide-bearing LRM heated by impact-related, magmatic, or other subsurface heat sources. The production of sulfur-rich gases would generate fumarolic systems. The S-bearing phases within these fumarole systems would undergo deposition in the vicinity of vents at night, forming patchy sulfur-rich layers, and sublimation during the day to produce hollows in a process analogous to development of thermokarst landscapes (*Sections 4.3.1* and *4.4,* Fig. 13). We call our alternative model for hollow formation the Sublimation Cycling Around Fumarole Systems (SCArFS) model. We favor the SCArFS model over SEALS hollow-formation models for reasons delineated in *Section 4.4.1*.

**Acknowledgements**

We would like to thank two anonymous reviewers for edits, comments, and suggestions that greatly improved the quality of this manuscript.

**Supplementary Materials**

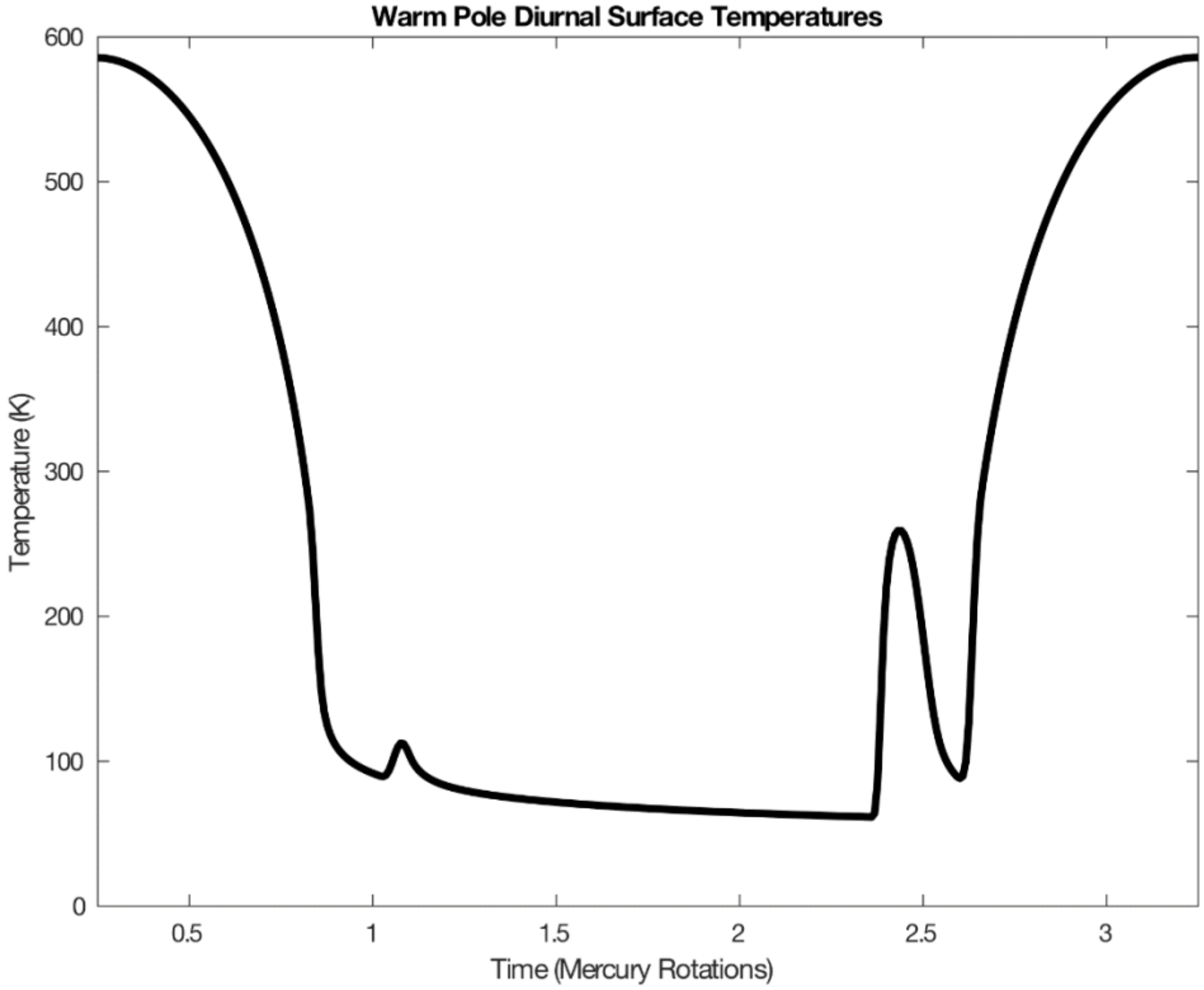

**Figure S1**: Diurnal surface temperature at 90°E 0°N. notice the anomalies on the nightside due to Mercury's unique orbital characteristics discussed in Section 1.1.

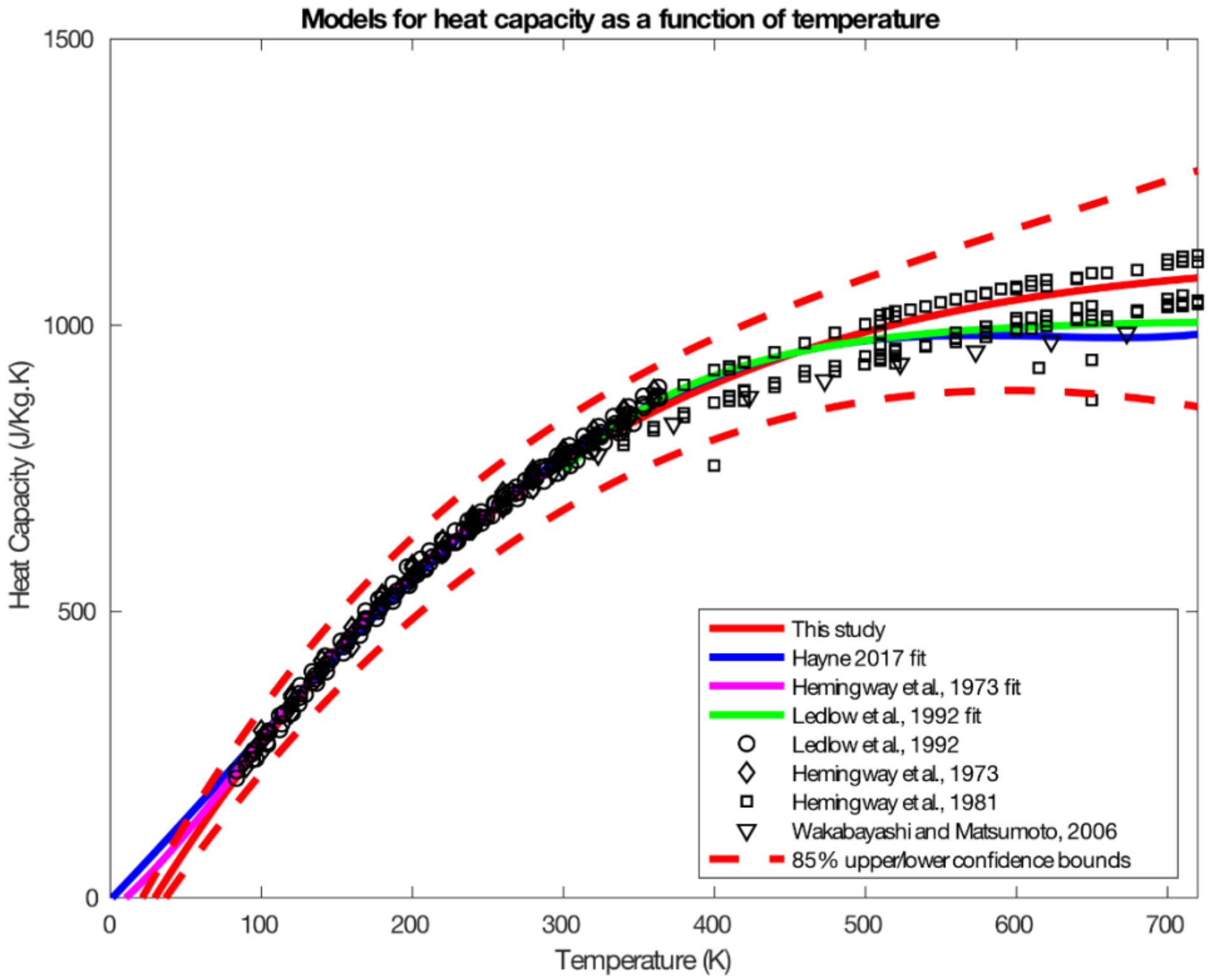

**Figure S2**: Polynomial fits to heat capacity data for lunar regolith. Our fit (red) is compared to previous fits.

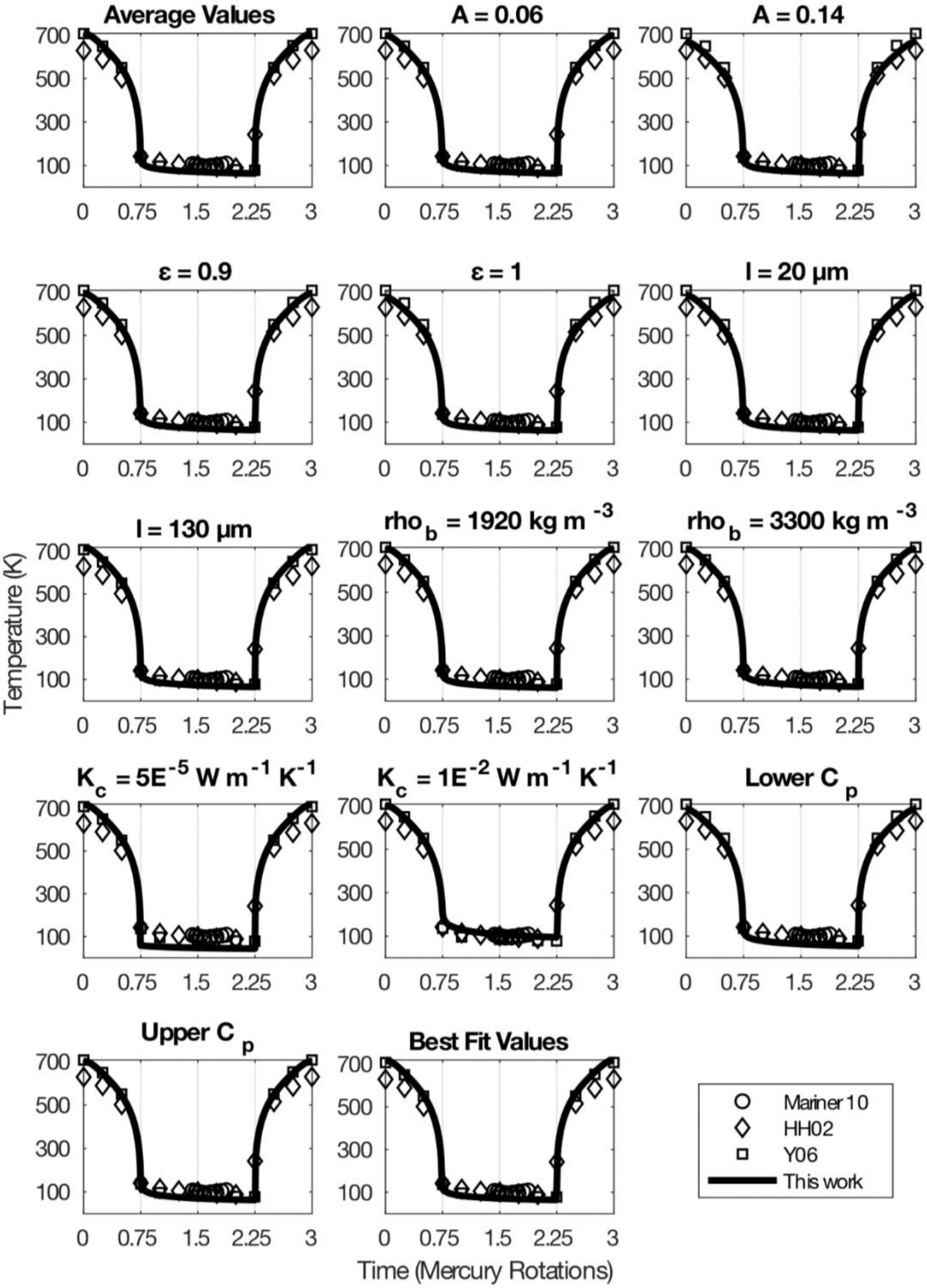

**Figure S3**: Surface diurnal temperatures for models run with various parameter configurations. See Table 2 and text for details. Values are compared to Mariner 10 IRR data from Chase et al. (1976), and to model data from Hale and Hapke (2002) (HH02) and Yan et al. (2006) (Y06).

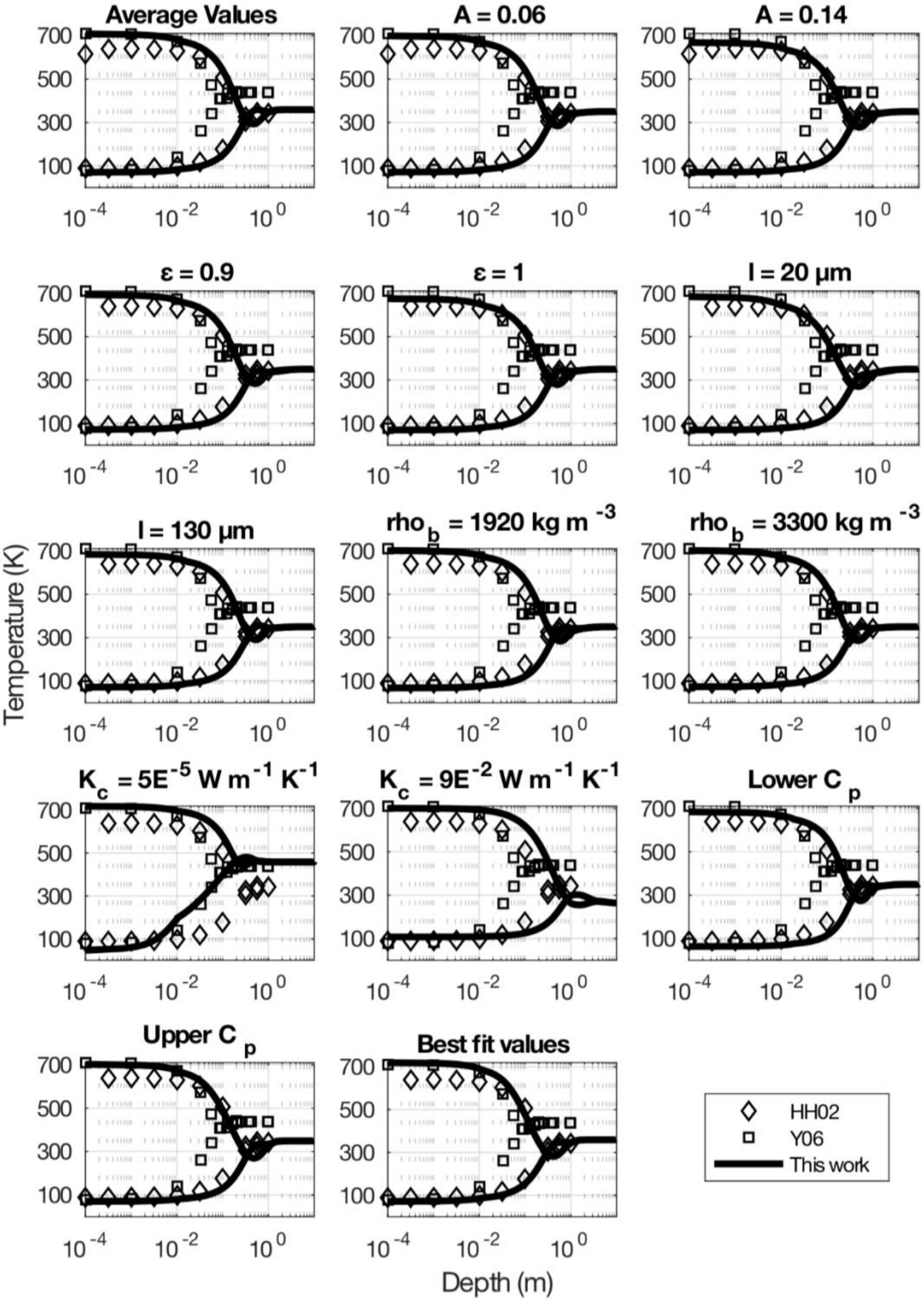

**Figure S4**: Noon and pre-dawn temperatures as a function of depth for various parameter configurations (same as those shown in Fig. S3). Values are compared to model runs from Hale and Hapke (2002) (HH02) and Yan et al. (2006) (Y06).

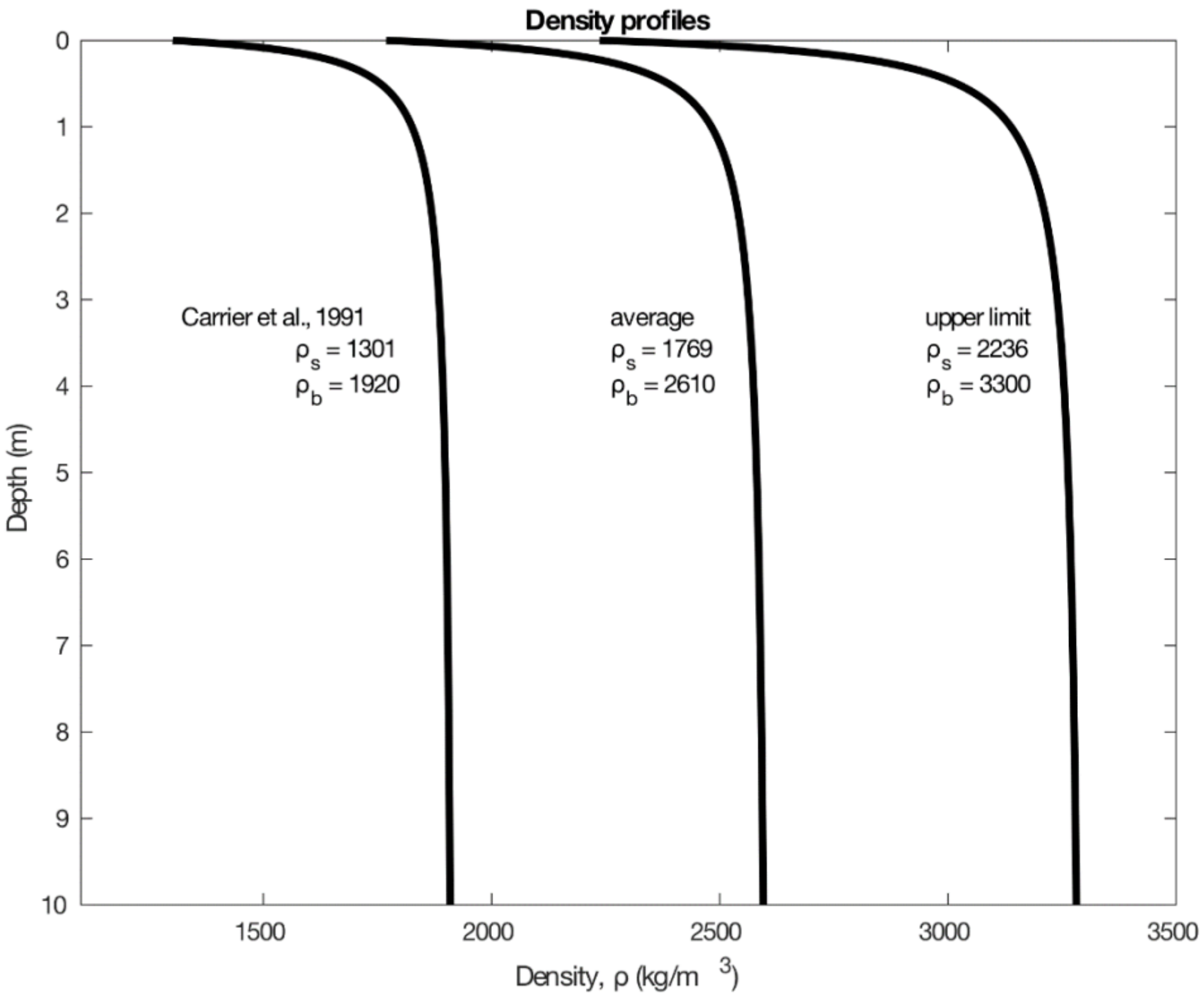

**Figure S5**: Upper and lower limit density profiles as a function of depth used in our thermal model.

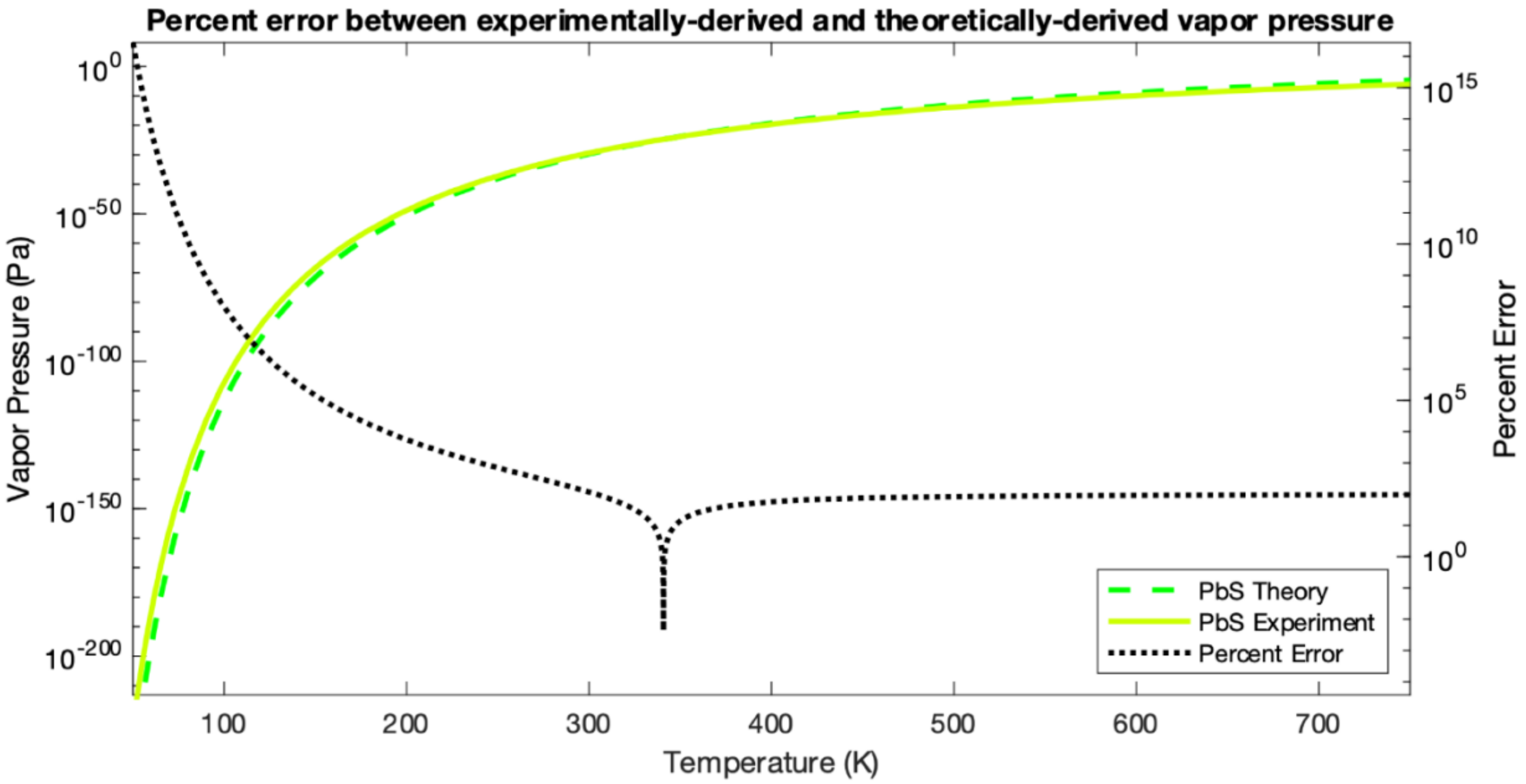

**Figure S6**: Percent error between theoretically and experimentally derived vapor pressure curves for PbS. At 340 K the two curves cross and percent error asymptotically approaches zero.

| Meteorites[a] | |
|---|---|
| **Compound Class** | **Concentration (ppm)** |
| Amino acids | 17– 60 |
| Aliphatic hydrocarbons | >35 |
| Aromatic hydrocarbons | 3319 |
| Fullerenes | >100 |
| Carboxylic acids | >300 |
| Hydroxycarboxylic acids | 15 |
| Dicarboxylic and Hydroxydicarboxylic acids | 14 |
| Purines and Pyrimidines | 1.3 |
| Basic N-heterocycles | 7 |
| Amines | 8 |
| Amides (linear) | >70 |
| Amides (cyclic) | >2 |
| Alcohols | 11 |
| Aldehydes and Ketones | 27 |
| Sulphonic acids | 68 |
| Phosphonic acids | 2 |

| Comets[b] | |
|---|---|
| **Compound** | **Concentration (% relative to water)** |
| $H_2O$ | 100 |
| CO | <1– 40 |
| $CO_2$ | 2–20 |
| $H_2CO$ | <1– 5 |
| $CH_3OH$ | 0.9–7 |
| HCOOH | <0.2 |
| $CH_4$ | 0.1– 0.5 |
| $C_2H_6$ | 0.2–2 |
| $C_{14}H_{10}$ | 0.15 |
| HCN | 0.007– 0.2 |
| $CH_3CN$ | 0.01– 0.02 |
| $NH_3$ | 0.1–2 |
| $H_2CS$ | <0.1 |
| $H_2S$ | 0.1– 1.6 |
| $N_2$ | <1.5 |
| Ar | <17 |

| Magma[c] | |
|---|---|
| **Compound** | **Concentration (ppm)** |
| CO | 5500 |
| $CO_2$ | 8700 |
| $H_2O$ | 3600 |
| $SO_2$ | 13000 |
| $H_2S$ | 500 |
| S | 500 |

**Table S1**: Phases likely delivered to or generated on Mercury. a: Botta and Bada (2002), b: Pierazzo and Chyba (1999), c: Kerber et al. (2009).

| Table S2 | Molecular form | molecular diameter (m) | mean free path (m) |
|---|---|---|---|
| Inorganics | NH3 | 2.60E-10 | 1.15E-06 |
| | H2O | 1.62E-10 | 0.000898 |
| | CO | 3.76E-10 | 1.38E-08 |
| | N2 | 3.64E-10 | 1.57E-08 |
| | S | 1.80E-10 | 134.293432 |
| | H2S | 3.60E-10 | 4.45E-07 |
| | Ar | 3.40E-10 | 3.05E-08 |
| | CO2 | 3.30E-10 | 6.90E-08 |
| | SO2 | 3.60E-10 | 2.39E-06 |
| | Kr | 3.60E-10 | 4.84E-08 |
| | Xe | 3.96E-10 | 1.25E-07 |
| Simple Organics | CH4 | 3.80E-10 | 4.80E-08 |
| | HCN | 2.19E-10 | 2.90E-05 |
| | COS | 2.76E-10 | 2.45E-06 |
| | C5H12 | 6.78E-10 | 3.39E-06 |
| | CS2 | 3.30E-10 | 3.20E-05 |
| | C5H10O | 6.78E-10 | 3.32E-05 |
| | C7H8 | 5.81E-10 | 4.96E-05 |
| | C5H10O2 | 7.92E-10 | 2.54E-05 |
| Aromatic Hydrocarbons | C6H7N | 5.79E-10 | 0.0003375 |
| | C6H6O | 5.21E-10 | 0.00318778 |
| | C7H6O | 6.02E-10 | 0.00112419 |
| | C7H8O | 6.94E-10 | 0.00034796 |
| | C7H6O2 | 6.02E-10 | 0.00210014 |
| | C6H5NO2 | 6.03E-10 | 0.00373439 |
| | C10H8 | 7.10E-10 | 0.00502455 |
| | C10H8O | 7.11E-10 | 0.05242604 |
| | C10H8O | 7.82E-10 | 0.04333839 |
| | C12H10 | 9.18E-10 | 0.01359902 |
| | C12H18 | 6.74E-10 | 0.00598434 |
| | C14H10 | 9.40E-10 | 2.33119844 |
| | C14H10 | 9.16E-10 | 2.45495765 |
| Linear Amides | CH3NO | 3.09E-10 | 0.02849364 |
| | C2H5NO | 4.20E-10 | 0.00752404 |
| | C2H5NO | 4.10E-10 | 0.00789555 |
| | C3H7NO | 4.28E-10 | 0.00091273 |
| | C3H7NO | 5.22E-10 | 0.0006136 |
| | C4H9NO | 5.38E-10 | 0.00023452 |
| | C4H9NO | 5.42E-10 | 0.00023107 |
| Carboxylic Acids | C5H10O2 | 7.87E-10 | 2.57E-05 |
| | C6H12O2 | 9.10E-10 | 5.23E-05 |
| | C7H14O2 | 1.04E-09 | 8.43E-05 |
| | C8H16O2 | 1.16E-09 | 0.0001438 |
| | C9H18O2 | 1.29E-09 | 0.00022574 |
| | C10H20O2 | 1.41E-09 | 0.00036285 |
| | C12H24O2 | 1.66E-09 | 0.00252262 |
| | C16H32O2 | 2.16E-09 | 1.41301819 |
| | C18H36O2 | 2.41E-09 | 4.18537228 |
| Carbon | C60 | 7.10E-10 | 4.8559E+13 |
| | C70 | 7.96E-10 | 6.8516E+14 |
| | C | 3.00E-07 | 1.27E+98 |
| Sulfides | K2S | 1.73E-09 | 1383.342 |
| | Na2S | 1.73E-09 | 6.70E+22 |
| | MgS | 8.66E-10 | 1.32E+40 |
| | MnS | 8.66E-10 | 2.58E+40 |
| | FeS | 1.93E-10 | 4.62E+42 |
| | CaS | 2.27E-10 | 8.07E+57 |

**Table S2**: molecular diameter and mean free path of volatile phases explored in this study. Phases with mean free paths highlighted in red do not meet the assumptions outlined in *Section 2.2.2.*

| | Hot Meridian | | | Warm Meridian | | |
|---|---|---|---|---|---|---|
| Compound | 0°N | 40°N | 71°N | 0°N | 40°N | 71°N |
| C | 8.35211E+47 | 9.05145E+51 | 4.14023E+74 | 1.49897E+60 | 1.12886E+65 | 1.84652E+94 |
| MgS | 2.41196E+16 | 1.85951E+18 | 7.30607E+28 | 1.12862E+22 | 2.14599E+24 | 9.65575E+37 |
| MnS | 4.58345E+16 | 3.58374E+18 | 1.52394E+29 | 2.23908E+22 | 4.33056E+24 | 2.1578E+38 |
| FeS | 4.20661E+16 | 3.88499E+18 | 4.2098E+29 | 3.41975E+22 | 8.09019E+24 | 1.34732E+39 |
| CaS | 3.89343E+24 | 1.10256E+27 | 6.46933E+40 | 9.77554E+31 | 8.96894E+34 | 5.01331E+52 |

**Table S3:** Time in years for each phase to sublimate to the depth of a typical hollow at each latitude (c.f., Table 5).

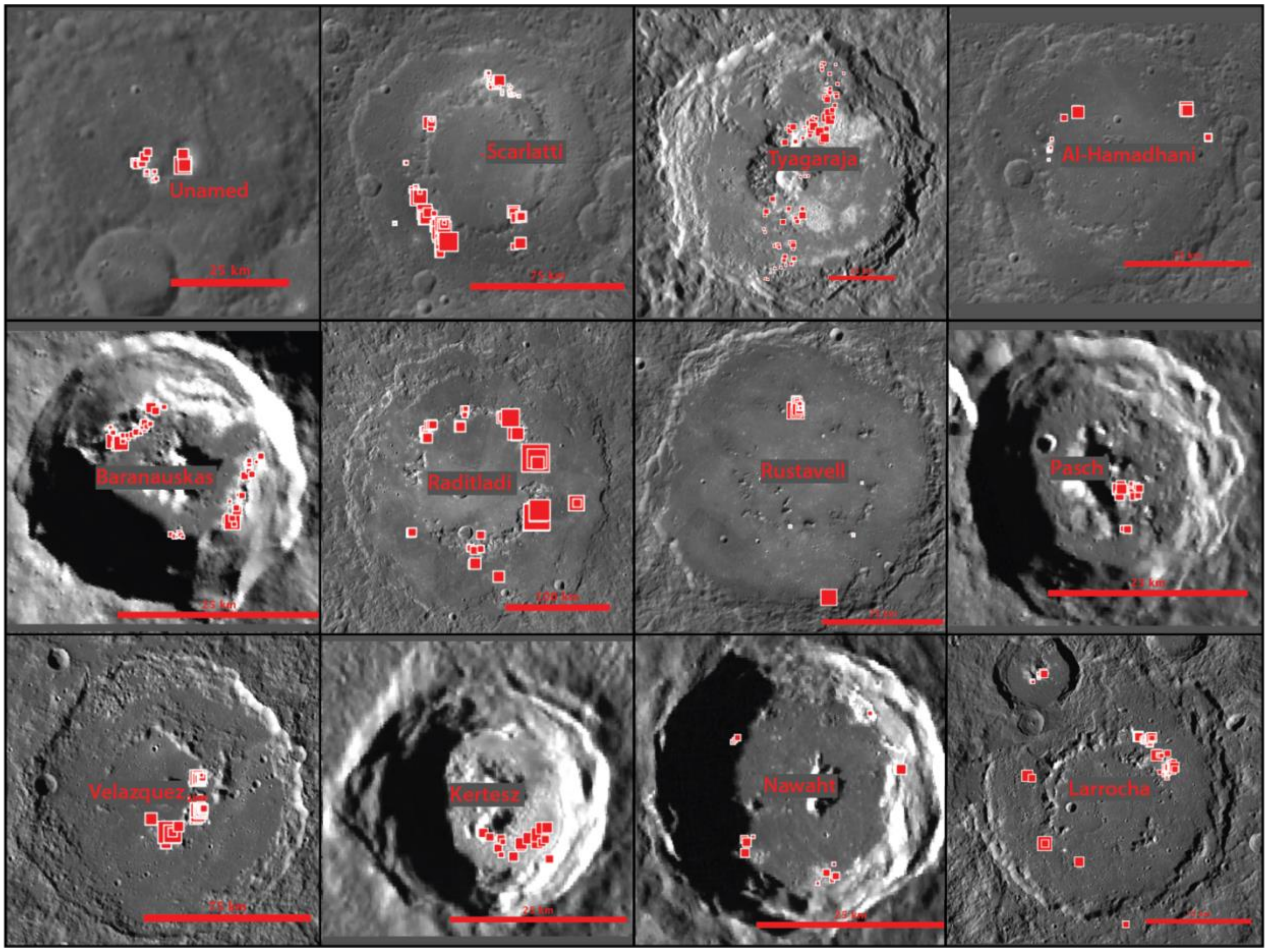

**Figure S7:** Collage of representative complex craters that host hollows. The crater name is presented in the center and north is up in all images. Red squares are locations where hollow depths were measured by Blewett et al. (2016). Larger squares represent a deeper value for hollow depth. Note that hollows occur predominantly along central peaks and peak rings and terraced walls. Hollows generally appear to be deeper closer to the central complexes than further from them.